\documentclass[11pt,oneside,letterpaper]{article}
\pdfoutput=1
\usepackage[lmargin=70pt,rmargin=70pt,footskip=50pt]{geometry}
\usepackage{setspace}
\usepackage{graphicx}
\usepackage{subcaption}
\usepackage{amsmath}
\usepackage{amssymb}
\usepackage{mathrsfs}
\usepackage{physics}
\usepackage{comment}
\usepackage{cite}
\usepackage{float}
\usepackage{units}
\usepackage{tikz}
\usepackage{cancel}
\usepackage{mathtools}
\usepackage{orcidlink}
\usepackage{empheq}
\usepackage{tcolorbox}
\usepackage{framed}
\usepackage{bm}
\usepackage{booktabs}
\usetikzlibrary{arrows.meta}
\definecolor{figblue}{RGB}{35,76,155}
\definecolor{figred}{RGB}{190,43,53}
\tikzset{
  fig boundary/.style={black, line width=0.5pt},
  fig hidden/.style={line width=0.4pt, dash pattern=on 2.3pt off 2pt},
  fig horizon/.style={line width=0.6pt, dash pattern=on 3pt off 2.4pt},
  fig worldline/.style={line width=0.95pt},
  fig time arrow/.style={black, line width=0.5pt, -{Latex[length=3.8pt]}},
  fig label/.style={inner sep=2pt, outer sep=1.5pt},
}

\numberwithin{equation}{section}
\newcommand{\om}{\omega}
\newcommand{\omA}{\omega_{_\text{AdS}}}
\newcommand{\gsc}{G_\text{scalar}}

\newcommand{\nn}{\nonumber}
\newcommand{\compi}{\mathfrak{i}}
\newcommand{\hds}{{\rm H}_{_\text{dS}}}

\begin{document}

\begin{center}
 \ \\
\vspace{1in}

{ \Large {\bf Post-Newtonian theory in de Sitter and anti-de Sitter spacetimes}}

\vspace{1cm}

R. Loganayagam$^{a}$\orcidlink{0000-0002-5240-3077}, Avi Wadhwa$^{a}$\orcidlink{0000-0002-6724-0872}

\vspace{1cm}

{\small{\it 
$^a$ International Centre for Theoretical Sciences (ICTS-TIFR), \\
Tata Institute of Fundamental Research,\\ Shivakote,  Hesaraghatta Hobli,  \\
    Bengaluru--560089,  Karnataka,  India.\\
\vspace{0.3cm}
 }}

\vspace{1.6cm}

\end{center}

\begin{abstract}
We develop the post-Newtonian framework for the near-zone dynamics of a binary in de Sitter and anti-de Sitter spacetime. We consider a binary interacting via a free scalar field in a fixed (A)dS background in $D+1$ spacetime dimensions, with $D\ge3$ and odd. Using multipole expansion and re-summation techniques, we solve for the field degrees of freedom to obtain an effective particle description in the form of a Fokker action. We derive the equations of motion for the binary up to second post-Newtonian (2PN) order, and obtain the corresponding 2PN Hamiltonian and conserved charges in phase space. We show that the conformal algebra, corresponding to the isometry algebra of the (A)dS background, is realised in phase space as a curvature-dependent deformation of the Poincare algebra at each PN order. Our results provide a systematic extension of post-Newtonian effective dynamics to finite-curvature backgrounds. This could serve as a useful framework for studying bound-state dynamics in (A)dS
as well as in the corresponding holographic CFT.
\end{abstract}

\pagebreak

\tableofcontents{}
\newpage

\section{Introduction}

The observation of gravitational waves from merging black holes and neutron stars has opened a new window into the dynamics of gravitating systems \cite{LIGOScientific:2016aoc}. As two compact objects orbit one another, they lose energy through gravitational radiation and gradually spiral inwards. The resulting gravitational-wave signal carries a remarkably detailed record of their orbital motion. Its frequency and amplitude increase as the bodies approach one another, while the accumulated phase is sensitive to even small corrections to their dynamics. Extracting this information requires accurate theoretical descriptions of the binary throughout its inspiral. The post-Newtonian (PN) expansion provides such a description when the orbital velocities are small compared with the speed of light, allowing relativistic corrections to the familiar Newtonian two-body problem to be calculated systematically \cite{Blanchet:2013haa,Goldberger:2007hy,Porto:2016pyg,Levi:2018nxp,Foffa:2013qca}. With gravitational-wave astronomy moving towards increasingly precise measurements, and future space-based observatories such as LISA opening a new range of frequencies, understanding the dynamics of relativistic binaries remains an important theoretical problem.

Most calculations of binary dynamics assume that the spacetime far from the system is approximately flat. This is an excellent approximation for astrophysical binaries, but it leaves open a natural question: how does the orbital motion change when the binary is placed in a curved background? Our Universe has a positive cosmological constant, and its late-time geometry is approximately de Sitter (dS). Although the corresponding curvature is far too small to produce appreciable local corrections to the motion of observed compact binaries, it is interesting to understand how cosmological curvature enters their dynamics \cite{Carrera:2008pi,Iorio:2012wva} and, ultimately, gravitational-wave generation \cite{Ashtekar:2015lxa}. For sufficiently large and weakly bound systems, however, the relative importance of these corrections can change \cite{Merafina:2014ysa,Benisty:2025mfv}. On megaparsec scales, such as those characteristic of the Coma galaxy cluster \cite{Chernin:2013rpa,Benisty:2025tct}, cosmological-curvature effects can become comparable to post-Newtonian corrections in suitable orbital regimes, even though both remain small relative to the leading Newtonian dynamics \cite{Benisty:2023clf}. Galaxy-cluster outskirts also provide observational probes of dark energy \cite{Morandi:2016cet}. Earlier studies have investigated the effect of a cosmological constant on the two-body problem \cite{Bisnovatyi-Kogan:2023aqf}, including corrections to orbital precession and gravitational-wave signals \cite{Jetzer:2006gn,Escobedo:2023exh}. A systematic treatment of binary motion on curved backgrounds also provides a useful setting in which to study the interplay between relativistic effects, field propagation and spacetime geometry.

The simplest curved backgrounds in which to address these questions are de Sitter (dS) and anti-de Sitter (AdS) spacetime. Their effects on non-relativistic motion are particularly intuitive. In AdS, the curvature acts as a confining harmonic potential: a freely moving particle is drawn towards the centre of the spacetime. In dS, the corresponding potential is inverted, producing an outward acceleration \cite{Adkins:2006kw}. For a binary, the orbital motion is therefore governed by a competition between the inter-particle attraction and the curvature of the surrounding spacetime. Unlike the usual Kepler problem, which has no preferred length scale, the AdS two-body problem contains an additional scale set by the curvature radius. Depending on the separation and interaction strength, the motion can be dominated by the mutual attraction, by the background confinement, or by both. This makes AdS a particularly useful setting for understanding how the familiar two-body problem is modified by curvature.

There is a further motivation for studying binaries in AdS, arising from the AdS/CFT correspondence. According to this correspondence, the energy spectrum of multi-particle states in AdS is identical to the spectrum of scaling dimensions of operators in a conformal field theory (CFT). Two interacting particles in AdS correspond to a family of double-trace operators in the CFT, whose anomalous dimensions then correspond to the  binding energy of the AdS binary states\cite{Fitzpatrick:2010zm,Cornalba:2007zb,Kulaxizi:2018dxo}. For binary state made of heavy particles, the AdS binding energy can be computed non-relativistically and then related to the spectrum of the dual quantum theory.

This connection was explored further by Maxfield and Zahraee \cite{Maxfield:2022hkd,Maxfield:2022nat,Zahraee:2024dsj}, who studied a non-relativistic limit of AdS in which the curvature-induced harmonic potential remains finite. In this limit, the two-body problem resembles a familiar problem in non-relativistic mechanics: two particles moving under the combined influence of their mutual interaction and an external harmonic potential. The centre-of-mass motion separates from the relative motion, allowing the interacting spectrum to be studied in terms of an effective one-body problem. Their analysis interpolates between hydrogen-like bound states, when the attraction dominates, and harmonic-oscillator states, when AdS confinement dominates. In the semi-classical regime, the energy levels can also be related to classical orbital motion, with circular orbits providing a particularly simple description of states with large angular momentum and low radial excitation. Their companion work connects these energy shifts to CFT calculations involving scalar exchange, providing an independent way to study the same physics.

The analysis of Maxfield and Zahraee is primarily non-relativistic. It is then natural to ask how this picture changes when relativistic corrections are included. In particular, one would like to know how the background curvature modifies the familiar PN corrections to orbital motion, whether the centre-of-mass and relative degrees of freedom can still be separated, and how the resulting binding energies are reflected in the dual CFT. These questions require more than simply adding a harmonic potential to an otherwise flat-space calculation. Curvature affects both the motion of the particles and the propagation of the field mediating their interaction. Moreover, the symmetries that organise the two-body problem are now those of AdS rather than the Poincare symmetries of flat spacetime. A consistent relativistic treatment must account for these effects together.

In asymptotically flat spacetime, the worldline effective-field-theory approach of Goldberger and Rothstein separates the orbital and radiation scales and provides systematic power counting for the PN expansion \cite{Goldberger:2004jt}. Its application to conservative binary dynamics has been demonstrated explicitly at 2PN and 3PN order \cite{Gilmore:2008gq,Foffa:2011ub}. Eliminating the mediating fields at the classical level gives a Fokker action for the particle trajectories, a construction used in both scalar--tensor and general-relativistic PN calculations \cite{Damour:1995kt,Bernard:2015njp}. Action-level multipole expansions for scalar, electromagnetic and gravitational fields provide a related framework for organising propagation and source couplings \cite{Ross:2012fc,Birnholtz:2013nta}.

Related methods have been developed on curved backgrounds by authors of \cite{Loganayagam:2023pfb,Loganayagam:2025jmw}, who obtained influence functionals for scalar and electromagnetic sources in dS using multipole expansions and the appropriate Green functions . Besides curvature corrections to radiation reaction, the scalar extended-source action contains potential interactions between different parts of the source. For a source consisting of two worldlines, these terms describe their mutual interaction. This provides a direct methodological connection to the binary problem considered here.

In this paper, we take a step towards this goal by studying two massive particles interacting through a free scalar field on a fixed AdS background. Scalar exchange provides a simple setting in which the effects of relativistic motion and curved-spacetime propagation can be studied without the additional complications of dynamical gravity. We use a worldline effective-action approach, integrating out the scalar field to obtain an effective interaction between the particles. We do this by a sequence of multipole and non-relativistic expansions and then a multipole re-summation.  This leads to a Fokker action, from which we derive the two-body dynamics through second post-Newtonian (2PN) order. The calculation extends the usual PN treatment by retaining the effects of AdS curvature, both in the particle motion and in the scalar-mediated interaction.

We obtain the equations of motion, the Hamiltonian and the conserved charges associated with the AdS isometries. These symmetries strongly constrain the form of the Hamiltonian and the conserved charges: their Poisson brackets must reproduce the AdS isometry algebra order by order in the PN expansion. This requirement relates coefficients in the Hamiltonian to those in the other generators and provides a non-trivial check on the interplay between relativistic kinematics, curvature and scalar exchange. This use of symmetry parallels the Hamiltonian treatment of compact binaries in the gravitational-wave literature \cite{Schafer:1986,Schafer:2018jfw}. In particular, global Poincare invariance was used to fix a momentum-dependent regularisation ambiguity in the 3PN ADM Hamiltonian \cite{Damour:2000kk}.

From the holographic perspective, the AdS isometry algebra is the conformal algebra of the boundary conformal field theory. The PN Hamiltonian encodes the classical contributions to anomalous dimensions across infinite families of double-trace primaries $[\Phi_1\Phi_2]_{n,j}$ of the CFT \cite{Fitzpatrick:2010zm,Maxfield:2022hkd,Maxfield:2022nat,Zahraee:2024dsj}. Bulk interactions shift the primary dimensions away from their generalised mean-field values and modify the generators acting on the two-particle states, while preserving the conformal algebra. The Hamiltonian and conserved charges thus supply the semi-classical data fora whole class of  non-trivial interacting representations of this algebra, which fixes the descendant structure above each shifted primary.

Our results provide a relativistic extension of the non-relativistic AdS two-body problem studied in \cite{Maxfield:2022hkd, Maxfield:2022nat,Zahraee:2024dsj}. In particular, they supply the ingredients needed to determine the relativistic corrections to the orbital energy spectrum, as will be described  in detail in a forthcoming work. More broadly, the methods developed here offer a starting point for studying binary dynamics in curved backgrounds with more general interactions, including electromagnetic and gravitational interactions, and ultimately for investigating how spacetime curvature modifies the dynamics and radiation of inspiralling systems.

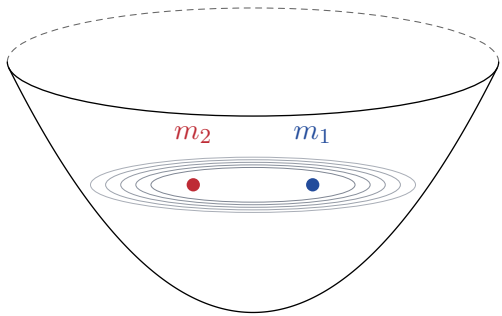
\begin{figure}
\centering
   \begin{tikzpicture}[
  x=1cm, y=1cm,
  line cap=round,
  line join=round,
  well boundary/.style={black, line width=0.5pt},
  well hidden/.style={
    black!60,
    line width=0.4pt,
    dash pattern=on 2.3pt off 2pt
  },
  mass label/.style={inner sep=2pt, outer sep=1pt}
]
  \definecolor{wellblue}{RGB}{35,76,155}
  \definecolor{wellred}{RGB}{190,43,53}
  \definecolor{wavegrey}{RGB}{104,113,128}

  \pgfmathsetmacro{\wellrim}{0.31*3.25*3.25-2.5}
  \pgfmathsetmacro{\wellside}{
    sqrt(3.25*3.25-(0.22/0.62)^2)
  }

  \draw[well hidden]
    (-3.25,\wellrim)
    arc[
      start angle=180, end angle=0,
      x radius=3.25, y radius=0.715
    ];

  \draw[well boundary]
    (-3.25,\wellrim)
    arc[
      start angle=180, end angle=360,
      x radius=3.25, y radius=0.715
    ];

  \draw[well boundary]
    plot[
      domain=-\wellside:\wellside,
      samples=150,
      variable=\q
    ]
    (\q,{0.31*\q*\q-2.5-0.22*0.22/(4*0.31)});

  \foreach \radius/\tone in {
    1.35/85,
    1.55/78,
    1.75/71,
    1.95/64,
    2.15/57
  } {
    \draw[wavegrey!\tone, line width=0.35pt]
      (0,-0.85)
      ellipse[
        x radius=\radius,
        y radius={0.17*\radius}
      ];
  }

  \fill[wellred]
    (-0.79,-0.85) circle[radius=2.5pt];
  \fill[wellblue]
    (0.79,-0.85) circle[radius=2.5pt];

  \node[mass label, text=wellred]
    at (-0.79,-0.19) {$m_2$};
  \node[mass label, text=wellblue]
    at (0.79,-0.19) {$m_1$};

\end{tikzpicture}
    \caption{Binary in AdS harmonic well}
    \label{fig:binary in AdS harmonic well}
\end{figure}

The paper is organised as follows-- we begin in \S\ref{sec:FPartAdS} by studying free particles in AdS, their conserved charges, the associated conformal algebra, and the non-relativistic limit.  In \S\ref{app:AdS} we define our conventions for various coordinates in AdS/dS as well as include a discussion of the corresponding isometries. In \S\ref{app:centroid}, we supplement the discussion by defining the AdS centroid and showing that it follows geodesic motion. In \S\ref{sec:Bulk-bulk Green's function in AdS}, we derive the non-relativistic expansion of the scalar bulk-bulk Green's function using a multipole expansion followed by resummation, with additional details provided in \S\ref{app:Scalar bulk-bulk Green's function in (A)dS}. In \S\ref{app:scalar-ads-ds} we supplement the discussion by examining the different field modes in AdS and dS, including near zone potential modes, radiative modes, and boundary/cosmic modes. We also explain how the dynamics can be restricted to the near-zone potential sector by omitting the radiative contributions. In \S\ref{sec:Effective Fokker action and equations of motion}, we use the PN-expanded Green's function to integrate out the scalar field and derive the effective Fokker action and the corresponding equations of motion, with supplementary expressions collected in \S\ref{app:2PN action and equations of motion}. In \S\ref{app:momInv} we show that the AdS Noether momentum retains its free particle form of the expression even after introducing interactions through 2PN order, under our coordinate choices. Finally, in \S\ref{sec:Hamiltonian and conformal algebra}, we formulate the dynamics in phase space and derive the Hamiltonian and symmetry generators through 2PN order. We demonstrate that these generators satisfy the conformal algebra, providing a non-trivial consistency check of our results. The supplementary expressions for this section are given in \S\ref{app:2PN Hamiltonian and conserved charges}.

\section*{Notations and conventions}

\begin{center}
\renewcommand{\arraystretch}{1.8}
\setlength{\tabcolsep}{10pt}

\begin{tabular}{@{}p{0.24\linewidth}p{0.70\linewidth}@{}}

\midrule

Integral shorthands &
$\displaystyle
\begin{aligned}
\int_{\vec{x}} &\equiv \int d^D\vec{x},
\qquad \int_t &\equiv \int dt,\qquad
\int_x &\equiv \int d^{D+1}x
= c\int_{t,\vec{x}}.
\end{aligned}
$ \\

\addlinespace[8pt]

Vector dot product &
$\displaystyle
p^2 \equiv \vec{p}\cdot\vec{p}
$ \\

\addlinespace[8pt]

Delta function shorthands &
$\displaystyle
\begin{aligned}
\delta_{t_1t_2}
&\equiv \delta(t_1-t_2),\qquad
\delta_{\vec{x},\vec{r}}^{(D)}
&\equiv \delta^{(D)}(\vec{x}-\vec{r}).
\end{aligned}
$ \\

\addlinespace[8pt]

Index conventions &
$\displaystyle
\begin{aligned}
\mu &\equiv \{0,\ldots,D\},
& i &\equiv \{1,\ldots,D\},\\
\bar{\mu} &\equiv \{0,\ldots,D-1\},
& \bar{i} &\equiv \{1,\ldots,D-1\}.
\end{aligned}
$ \\

\addlinespace[8pt]

Coordinate shorthands &
$\displaystyle
\begin{aligned}
\vec{r}_{\rm av}
&= \frac{1}{2}(\vec{r}_1+\vec{r}_2),\qquad
\vec{r}_{12}
&= \vec{r}_1-\vec{r}_2,\qquad
r_{12}
&= \left|\vec{r}_{12}\right|,
& \hat{n}_{12}
&= \frac{\vec{r}_{12}}{r_{12}}.
\end{aligned}
$ \\
\addlinespace[8pt]
\bottomrule
\end{tabular}
\end{center}
\pagebreak

\section{Free particles in (A)dS }\label{sec:FPartAdS}

In this work, we will find it convenient to convert to work with Cartesian-like global coordinates, where the AdS metric takes the form,
\begin{equation}
ds^2
=
-f(r)c^2dt^2
+
d\vec{x}^{\,2}
-
\frac{f(r)-1}{f(r)}
\frac{\left(\vec{r}\cdot d\vec{r}\right)^2}{r^2}.
\end{equation}
where $\vec{r}\equiv \left\{x^i\right\}$ with the spatial index $i$ varying from $1$ to $D$. We also have $r\equiv\sqrt{\sum_i(x^{i})^2}$ and $\cdot$ is the spatial dot product. The redshift factor\footnote{The factor $f(r)$ is a redshift factor in dS, but it is actually a blueshift factor in AdS: the observer at the centre of AdS sees far away objects blue shifted by this factor. Moving forward, we will just use the term redshift factor.} $f(r)$ appearing in the above metric is given  by
\begin{equation}
f(r)=\begin{cases}1+\frac{r^2\omA^2}{c^2}\;(\text{global anti-de Sitter})\ ,\\\\
1-\frac{r^2\hds^2}{c^2}\;(\text{de Sitter static patch})\ .
\end{cases}
\end{equation}
In this formula, we have also included the case of dS spacetime. In what follows, we will mainly quote the answers for AdS case : as we shall see, for the near zone dynamics, the corresponding statements for dS can always be obtained  via the analytic continuation $\omA\rightarrow \compi\,\hds$.

\subsection{Conserved quantities in AdS}
The point  particles are described by their worldline action,
\begin{equation}\label{eq:worldline action}
\mathcal{S}^{\text{free}}=\int_t\mathcal{L}^{\text{free}}=-\sum_a\,m_{a}c^{2}\intop d\tau_{a}
\end{equation}
where $m_a$ is the mass of particle with label ``$a$", and $\tau_a$ is the corresponding proper time. 
The proper time is defined via 
$\dot{\tau}_a\equiv \sqrt{-g_{\mu\nu}\dot{x}_a^\mu \dot{x}_a^\nu}$ with over-dot denoting derivative with respect to the coordinate time $t$.

The dynamics of a free-particle in AdS is completely integrable and has $\frac{1}{2}(D+2)(D+1)$ conserved quantities corresponding to that many continuous isometries of AdS$_{D+1}$.
We  relegate to appendix~\ref{app:AdS} a detailed discussion about AdS isometries, and present here only the corresponding conserved quantities, whose physics can be understood at an intuitive level without such details.  Given the manifest rotational/time-translation invariance of this system, conservation of the following Noether charges is self-evident:
\begin{itemize}
\item We have $\frac{1}{2}D(D-1)$ conserved angular momenta  
\begin{equation}
\mathbb{J}^{\text{free}}_{ij}\equiv \sum_a(x_{ai}p_{aj}-x_{aj}p_{ai})\ .
\end{equation}
 Here   $\vec{p}_a$ are the conjugate momenta defined via
\begin{equation}\label{eq:conjugate-momentum-free}\begin{split}
\vec{p}_{a}^{\;\text{free}}&\equiv \frac{\partial}{\partial\vec{v}_{a}}\mathcal{L}^{\,\text{free}}=\frac{m_{a}}{\sqrt{f\left(r_{a}\right)-\frac{v_{a}^{2}}{c^{2}}+\frac{\omA^{2}}{c^{4}}\frac{(\vec{r}_{a}.\vec{v}_{a})^{2}}{f(r_{a})}}}\left(\vec{v}_{a}-\frac{\omA^{2}}{c^{2}}\frac{(\vec{r}_{a}.\vec{v}_{a})\vec{r}_{a}}{f(r_{a})}\right).
\end{split}\end{equation}
\item We have a conserved Hamiltonian given by 
\begin{equation}\label{eq:Hamiltonian-free}\begin{split}
\mathbb{H}^{\,\text{free}}&\equiv \sum_a\mathbb H_a\equiv\;\sum_a\vec{p}_{a}^{\;\text{free}}.\,\vec{v}_{a}-\mathcal{L}^{\,\text{free}}\\
&=\sum_a\sqrt{f(r_{a})}\times\sqrt{m_{a}^{2}c^{4}+p_{a}^{2}c^{2}+\omA^{2}(\vec{r}_{a}.\vec{p}_{a})^{2}}\ .
\end{split}\end{equation}

\item The remaining 2D number of conserved quantities are more subtle.\footnote{They correspond to isometries in the second line of Eq.\eqref{eq:AdSKilling-globII}, which are not as immediately obvious in the coordinate system we are using. } They are encapsulated in the following conserved complex vector:
\begin{align}\label{eq:eitK-free}
e^{\compi\,\omA t}\ \vec{\mathbb{A}}^{\text{free}}
&= e^{\compi\,\omA t}\ 
\sum_a
\left[
\frac{\vec{r}_a}{c^2}
\sqrt{
m_a^2c^4
+p_a^2c^2
+\omA^2\left(\vec{r}_a\cdot\vec{p}_a\right)^2
}
+
\frac{\compi}{\omA}\,
\vec{p}_a
\sqrt{f(r_a)}
\right].
\end{align}
Conservation of this complex vector can be checked  directly by using either the Euler-Lagrange equations, or Hamilton equations.\footnote{The Euler Lagrange equation in this case is the AdS geodesic equation
\begin{equation}
\begin{split}
\vec{F}_{a}
\equiv m_a\frac{d^2\vec{r}_a}{dt^2}=&-m_{a}\omA^{2}f(r_{a})\,\vec{r}_{a}+\frac{m_{a}\omA^{2}}{c^{2}}\Biggl\{
\left[
v_{a}^{2}
-\frac{\omA^{2}}{c^{2}}
\frac{\left(\vec{r}_{a}\cdot\vec{v}_{a}\right)^{2}}{f(r_{a})}
\right]\vec{r}_{a}+\frac{2}{f(r_{a})}
\left(\vec{r}_{a}\cdot\vec{v}_{a}\right)\vec{v}_{a}\Biggr\}\ .
\end{split}
\end{equation}
The corresponding Hamilton's equations are: \begin{equation}
\dot{\vec{r}}_a
=
\frac{f(r_a)}{\mathbb H_a}
\left[
c^{2}\vec{p}_a
+\omA^{2}(\vec{r}_a\cdot\vec{p}_a)\vec{r}_a
\right],
\qquad
\dot{\vec{p}}_a
=
-\omA^{2}\left\{\frac{\mathbb H_a}{c^{2}f(r_a)}\,\vec{r}_a
+\frac{f(r_a)}{\mathbb H_a}
(\vec{r}_a\cdot\vec{p}_a)\vec{p}_a\right\}.
\end{equation}
As pointed out by Maxfield-Zaharee\cite{Maxfield:2022hkd,Maxfield:2022nat,Zahraee:2024dsj}, if we take a $c\to\infty$ limit keeping $\omA$ fixed, all these equations reduce to that of a harmonic oscillator with mass $m_a=\lim_{c\to\infty}\frac{\mathbb H_a}{c^{2}}$ and frequency $\omA$.}
\end{itemize}

To physically interpret this complex conserved vector, we consider the flat space (i.e., $\omA\to 0$) limit of the above vector. The leading order contribution 
\begin{align}
\frac{\compi}{\omA}\,\sum_a
\vec{p}_a\nn
\end{align}
is purely imaginary and is, up to a factor, just the total momentum ( which is known to be conserved in flat spacetime). The next to leading correction at $(\omA)^0$ order is 
\begin{align}
\sum_a
\left[
\frac{\vec{r}_a}{c^2}
\sqrt{
m_a^2c^4
+p_a^2c^2
}
-t\,
\vec{p}_a
\right].\nn
\end{align}
This purely real vector can be recognised as the Lorentz boost generator for the free particles in flat spacetime, whose conservation follows from Noether's theorem for boosts. Thus, the conserved complex vector in Eq.\eqref{eq:eitK-free} should be thought of as the AdS generalisation of boost/translation generators. 

\begin{figure}
    \centering
\begin{tikzpicture}[x=1cm,y=1cm,line cap=round,line join=round]
  \foreach \height in {-3.2,3.2} {
    \draw[fig boundary,fig hidden]
      (-1.7,\height) arc[start angle=180,end angle=0,
        x radius=1.7,y radius=0.34];
    \draw[fig boundary]
      (-1.7,\height) arc[start angle=180,end angle=360,
        x radius=1.7,y radius=0.34];
  }
  \draw[fig boundary] (-1.7,-3.2) -- (-1.7,3.2);
  \draw[fig boundary] ( 1.7,-3.2) -- ( 1.7,3.2);
  \draw[fig time arrow] (-2.14,-3.02) -- (-2.14,3.04);
  \node[fig label,above=6pt] at (-2.14,3.04) {$t$};

  \draw[fig worldline,figblue]
    plot[domain=-3.1:3.1,samples=260,variable=\q]
      ({0.45*cos(100*(\q-0.25))},
       {\q-0.24*0.45*sin(100*(\q-0.25))});
  \draw[fig worldline,figred]
    plot[domain=-3.1:3.1,samples=260,variable=\q]
      ({-0.65*cos(100*(\q-0.25))},
       {\q+0.24*0.65*sin(100*(\q-0.25))});
  \fill[figblue] ( 0.45,0.25) circle[radius=1.25pt];
  \fill[figred]  (-0.65,0.25) circle[radius=1.25pt];

  \node[fig label,text=figblue] (massone) at ( 1.03,0.87) {$m_1$};
  \node[fig label,text=figred]  (masstwo) at (-1.05,0.87) {$m_2$};
  \draw[figblue,line width=0.4pt,shorten <=4pt,shorten >=3pt]
    (massone.south west) -- ( 0.45,0.25);
  \draw[figred,line width=0.4pt,shorten <=4pt,shorten >=3pt]
    (masstwo.south east) -- (-0.65,0.25);
\end{tikzpicture}
    \caption{Binary worldlines in AdS}
    \label{fig:Binary worldlines in AdS}
\end{figure}
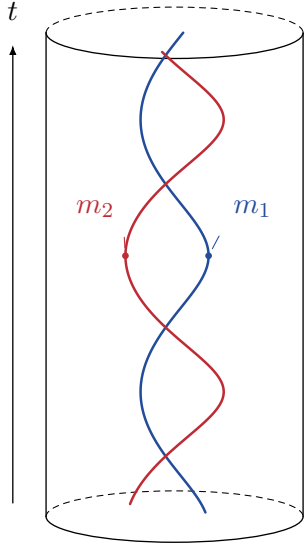

The same flat limit is obtained, if we focus on a spacetime region around origin, i.e., if we are interested in the dynamics of particles that keeps them within the region
\begin{equation}
t\ll \frac{1}{\omA}\ ,\quad r\ll \frac{c}{\omA}\ .
\end{equation}
In this regime, all statements we make about particle dynamics in AdS should reduce to the corresponding statements in flat spacetime. We will find it convenient to adopt definitions in such a way that this flat limit is always manifest. With this in mind, we define the vectors 
\begin{equation}\label{eq:GPfree}
\vec{\mathbb{G}}^{\;\text{free}}
\equiv \frac{1}{c^{2}}\sum_{a}\vec{r}_{a}
\sqrt{
m_{a}^{2}c^{4}
+p_{a}^{2} c^{2}
+\omA^{2}\left(\vec{r}_{a}.\vec{p}_{a}\right)^{2}
}\ ,\quad  \vec{\mathbb{P}}^{\,\text{free}}\equiv \sum_a\;\sqrt{f(r_a)}\, \vec{p}_{a}\ ,
\end{equation}
which can be thought of as AdS generalisation of the first moment of inertia (i.e., node of inertia) and the total momentum. In terms of these vectors,  we can  write 
\begin{equation}
\vec{\mathbb{A}}^{\,\text{free}}\equiv \vec{\mathbb{G}}^{\,\text{free}}+\frac{\compi}{\omA}\vec{\mathbb{P}}^{\,\text{free}}\ ,
\end{equation}
and the conservation of the  complex vector in Eq.\eqref{eq:eitK-free} then implies the following equations of motion,
\begin{equation}\label{eq:KdotFree}
\frac{d}{dt}\vec{\mathbb{A}}^{\,\text{free}}=-\compi\ \omA\ \vec{\mathbb{A}}^{\,\text{free}}
\end{equation}
or equivalently,
\begin{equation}\label{eq:GPdotFree}
\frac{d}{dt}\vec{\mathbb{G}}^{\,\text{free}}=\vec{\mathbb{P}}^{\,\text{free}}\ , \quad \frac{d}{dt}\vec{\mathbb{P}}^{\,\text{free}}=-\omA^{2}\vec{\mathbb{G}}^{\,\text{free}}\ .
\end{equation}
These equations can be interpreted as follows : \emph{the node of inertia $\vec{\mathbb{G}}^{\,\text{free}}$ of a collection of free particles in AdS moves like a harmonic oscillator with momentum $\vec{\mathbb{P}}^{\,\text{free}}$ and a natural frequency $\omA$}. This is the AdS generalisation of conservation of  momentum as well as centre of inertia theorem\footnote{We remind the reader that the centre of inertia theorem states that, for any isolated system in flat spacetime,  the node of inertia $\vec{\mathbb{G}}$ moves like a free particle with momentum $\vec{\mathbb{P}}$. } in flat spacetime. Upto a pre-factor, the complex combination $\vec{\mathbb{A}}^{\,\text{free}}$ then  acts like the annihilation operator for the oscillator (more on this later).

As we mentioned before,  the corresponding statements for dS can be obtained  via the analytic continuation $\omA\rightarrow \compi\,\hds$. The node of inertia in dS moves then like an \emph{inverted} oscillator of natural frequency $\hds$. In this case, we have two real  vector combinations
\begin{equation}
\vec{\mathbb{A}}^{\,\text{free}}\equiv \vec{\mathbb{G}}^{\,\text{free}}+\frac{1}{\hds}\vec{\mathbb{P}}^{\,\text{free}}\ ,\quad 
\vec{\widetilde{\mathbb{A}}}^{\,\text{free}}\equiv \vec{\mathbb{G}}^{\,\text{free}}-\frac{1}{\hds}\vec{\mathbb{P}}^{\,\text{free}}\ , 
\end{equation}
which are independent, and not related by complex conjugation like in AdS. Their time evolution is governed by 
\begin{equation}\label{eq:KdotFreedS}
\frac{d}{dt}\vec{\mathbb{A}}^{\,\text{free}}=\hds\ \vec{\mathbb{A}}^{\,\text{free}}\ ,\quad \frac{d}{dt}\vec{\widetilde{\mathbb{A}}}^{\,\text{free}}=-\hds\ \vec{\widetilde{\mathbb{A}}}^{\,\text{free}}\ ,
\end{equation}
i.e., the first one grows exponentially in time, whereas second vector decays exponentially. Thus, at very late times, we have
\begin{equation}
\vec{\mathbb{P}}^{\,\text{free}}\approx\hds\vec{\mathbb{G}}^{\,\text{free}}\ , 
\end{equation}
which is essentially a statement of Hubble's law in dS.

\subsection{The $\mathfrak{so}(D,2)$ 
  algebra and centroid dynamics}

We will now argue that the various features we described above generalise to interacting particles. In fact, all that is needed for this conclusion are the following: the asymptotic symmetries of AdS, the associated Noether charges and their algebra. We will show that, under these general assumptions, we still have an oscillator like dynamics for an interacting multi-particle system. The \emph{point} which oscillates is a centroid which generalises the notion of centre of inertia to relativistic particles in AdS. In this subsection, we will argue why an existence of such a centroid is intuitively plausible by examining the dynamics of charges, relegating the precise construction to appendix~\ref{app:centroid}.

The conserved  quantities described above generate a deformed Poincare action on the phase space of free particles. To see this, let us compute the following Poisson brackets : 
\begin{subequations}\label{eq:ads-algebraMain}
\begin{align}
\left[\mathbb{P}_i,\mathbb{P}_j\right]_{\rm PB}
=&\ 
-\frac{\omA^2}{c^2}\ \mathbb{J}_{ij},
\qquad
\left[\mathbb{H},\mathbb{P}_i\right]_{\rm PB}
=\  
\omA^2\,\mathbb{G}_i,
\\
\left[\mathbb{J}_{ij},\mathbb{J}_{kl}\right]_{\rm PB}
=&\ 
\delta_{ik}\mathbb{J}_{jl}
+\delta_{jl}\mathbb{J}_{ik}
-\delta_{il}\mathbb{J}_{jk}
-\delta_{jk}\mathbb{J}_{il},
\\[3pt]
\left[\mathbb{G}_i,\mathbb{G}_j\right]_{\rm PB}
=&
-\frac{1}{c^2}\mathbb{J}_{ij},
\qquad
\left[\mathbb{J}_{ij},\mathbb{G}_k\right]_{\rm PB}
=\ 
\delta_{ik}\mathbb{G}_j
-\delta_{jk}\mathbb{G}_i,
\\
\left[\mathbb{J}_{ij},\mathbb{P}_k\right]_{\rm PB}
=&\ 
\delta_{ik}\mathbb{P}_j
-\delta_{jk}\mathbb{P}_i,
\qquad
\left[\mathbb{J}_{ij},\mathbb{H}\right]_{\rm PB}
=\ 0,
\\
\left[\mathbb{G}_i,\mathbb{P}_j\right]_{\rm PB}
=&\ 
\frac{\delta_{ij}}{c^2}\mathbb{H},
\qquad
\left[\mathbb{G}_i,\mathbb{H}\right]_{\rm PB}
=\ 
\mathbb{P}_i.
\end{align}
\end{subequations}
Here $\left[\;,\;\right]_{\rm PB}$ denotes the Poisson bracket in phase space with the convention that $\left[x_{ai},p_{bj}\right]_{\rm PB} = \delta_{ab}\delta_{ij}$. We have also dropped all the `free' superscripts, since the above algebra is universal and will be valid even after interactions are taken into account (as we will show explicitly later).\footnote{The above Poisson algebra  holds also for interacting field theories in AdS.} We note that the first line of Eq.\eqref{eq:ads-algebraMain}  is  the translation algebra deformed by adding $\omA^2$ times the Lorentz generators in RHS. Apart from this first line, the rest of the algebra coincides with Poincare algebra $\mathfrak{iso}(D,1)$ to which it reduces in the flat (i.e., $\omA\to 0$) limit .

At the level of Lie algebras, the flat limit is the  Inonu-Wigner contraction of $\mathfrak{so}(D,2)$ algebra  to the Poincare algebra $\mathfrak{iso}(D,1)$.
The algebra of Poisson brackets above can be cast into standard $\mathfrak{so}(D,2)$ form with the identifications
\begin{equation}\label{eq:Jsod2Id}
\mathbb{J}_{0i}\equiv c\,\mathbb{G}_i\ ,\quad
\mathbb{J}_{0(D+1)}\equiv \frac{\mathbb{H}}{\omA}\ ,\quad \mathbb{J}_{(D+1)i}\equiv \frac{c}{\omA}\mathbb{P}_i\ .
\end{equation}
All the Poisson brackets can then be written succinctly as  
\begin{equation}
\left[\mathbb{J}_{AB},\mathbb{J}_{CD}\right]_{\rm PB}=\eta_{AC}\mathbb{J}_{BD}+\eta_{BD}\mathbb{J}_{AC}-\eta_{AD}\mathbb{J}_{BC}-\eta_{BC}\mathbb{J}_{AD}\ .
\end{equation}
Here  we take  $A,B$ indices to vary over $(D+2)$ values $\{0,1,2,\ldots, D+1\}$ and use $\eta_{AB}$ to denote the  $\mathfrak{so}(D,2)$ invariant metric on $\mathbb{R}^{D,2}$ with $\{0,(D+1)\}$ directions being time-like. With these identifications, the flat limit is obtained by scaling out the time-like $(D+1)$ direction. The corresponding statements for dS can be obtained by the analytic continuation 
\begin{equation}
\omA\rightarrow \compi\,\hds\ ,\quad X^{(D+1)}\rightarrow \compi\,X^{(D+1)}_{\rm dS}\ , 
\end{equation}
i.e., when we move to dS we also Wick-rotate in $(D+1)$ direction to make it into a space-like direction. We then have the identifications
\begin{equation}
\mathbb{J}_{0i}\equiv c\,\mathbb{G}_i\ ,\quad
\mathbb{J}_{0(D+1)}\equiv \frac{\mathbb{H}}{\hds}\ ,\quad \mathbb{J}_{(D+1)i}\equiv \frac{c}{\hds}\mathbb{P}_i\ , 
\end{equation}
and the algebra of Poisson brackets can now be cast into standard $\mathfrak{so}(D+1,1)$ form.  The flat limit is then the Inonu-Wigner contraction again  obtained by scaling out the $(D+1)$ direction, the only difference is that  $(D+1)$ direction is space-like in the dS case. 

\subsubsection*{Non-relativistic limit}

A more interesting Inonu-Wigner contraction from the point of view of this work is the non-relativistic limit. By this, we mean taking $c\to\infty$ keeping $\omA$ as well as $\dfrac{\mathbb{H}}{c^2}$ fixed at some non-zero values. As pointed out by Maxfield-Zahraee\cite{Maxfield:2022hkd,Maxfield:2022nat,Zahraee:2024dsj}, in this limit we land on an algebra of symmetries describing a non-relativistic harmonic oscillator. To see this, let us first rewrite the above algebra above to show how it can be thought of as a relativistic generalisation of an isotropic harmonic oscillator in $\mathbb{R}^d$.  We absorb the $\omA$ factors by defining
\begin{equation}
\mathbb{A}_i\equiv \mathbb{G}_i+\frac{\compi}{\omA}\mathbb{P}_i\ ,\quad
\widetilde{\mathbb{A}}_i\equiv \mathbb{G}_i-\frac{\compi}{\omA}\mathbb{P}_i\ ,\quad  \mathbb{D}\equiv \frac{1}{\omA}\mathbb{H}\ .  
\end{equation}
Up to overall factors, the first two definitions mimic annihilation/creation operators in an isotropic quantum oscillator, whereas the last definition mimics the definition of the total excitation number operator.\footnote{To get the corresponding definitions for dS, we take $\omA\rightarrow \compi\,\hds$ and replace $\mathbb{D}$ everywhere by $\dfrac{1}{\compi}\mathbb{D}$ so that we have 
\begin{equation}
\mathbb{A}_i\equiv \mathbb{G}_i+\frac{1}{\hds}\mathbb{P}_i\ ,\quad
\widetilde{\mathbb{A}}_i\equiv \mathbb{G}_i-\frac{1}{\hds}\mathbb{P}_i\ ,\quad  \mathbb{D}\equiv \frac{1}{\hds}\mathbb{H}\ .  
\end{equation}}  The Poisson bracket algebra  of rewritten in terms of the new variables is 
\begin{subequations}\label{eq:ads-algebraAlt}
\begin{align}
\left[\mathbb{J}_{ij},\mathbb{J}_{kl}\right]_{\rm PB}
=&\,
\delta_{ik}\mathbb{J}_{jl}
+\delta_{jl}\mathbb{J}_{ik}
-\delta_{il}\mathbb{J}_{jk}
-\delta_{jk}\mathbb{J}_{il}\ ,\quad
\left[\mathbb{J}_{ij},\mathbb{D}\right]_{\rm PB}
=0\ ,
\label{eq:ads-algebraAlt-a}
\\
\left[\mathbb{J}_{ij},\mathbb{A}_k\right]_{\rm PB}
=&\,
\delta_{ik}\mathbb{A}_j
-\delta_{jk}\mathbb{A}_i,\quad
\left[\mathbb{J}_{ij},\widetilde{\mathbb{A}}_k\right]_{\rm PB}
=
\delta_{ik}\widetilde{\mathbb{A}}_j
-\delta_{jk}\widetilde{\mathbb{A}}_i\ ,
\label{eq:ads-algebraAlt-b}\\
\left[\compi\mathbb{D},\mathbb{A}_i\right]_{\rm PB}
=&-\mathbb{A}_i,\quad
\left[\compi\mathbb{D},\widetilde{\mathbb{A}}_i\right]_{\rm PB}
=
\widetilde{\mathbb{A}}_i\ ,
\label{eq:ads-algebraAlt-c}
\\[2pt]
\left[\mathbb{A}_i,\mathbb{A}_j\right]_{\rm PB}
=&
\left[\widetilde{\mathbb{A}}_i,\widetilde{\mathbb{A}}_j\right]_{\rm PB}
=0,\quad
\left[\mathbb{A}_i,\widetilde{\mathbb{A}}_j\right]_{\rm PB}
=
-\frac{2}{c^2}
\left(\compi\mathbb{D}\,\delta_{ij}+\mathbb{J}_{ij}\right)
\ .
\label{eq:ads-algebraAlt-d}
\end{align}
\end{subequations}
To take the non-relativistic limit, we write 
\begin{equation}
\mathbb{D}\approx \frac{\mathbb{M}\,c^2}{\omA} + \mathbb{D}_{\rm NR}\ ,  
\end{equation}
and then take $c\to\infty $ keeping $\omA,\mathbb{M},\mathbb{D}_{\rm NR}$ and all other charges fixed. As in the usual non-relativistic limit, we will take $\mathbb{M}$ to be a constant (or more precisely, as a central element that Poisson commutes with the entire algebra). We see that  all Poisson brackets except the last one remain the same with the replacement $\mathbb{D}\mapsto \mathbb{D}_{\rm NR}$. The last Poisson bracket becomes the algebra of isotropic oscillator, i.e.,
\begin{equation}
\left[\mathbb{A}_i,\widetilde{\mathbb{A}}_j\right]_{\rm PB}
=
-\compi\frac{2\mathbb{M}}{\omA}
\delta_{ij}\ .  
\end{equation}

From Eq.\eqref{eq:Jsod2Id}, we can see that taking $c\to \infty$ limit this way corresponds to scaling out \emph{both} the time-like directions of $\mathfrak{so}(D,2)$ equally. This can also be more directly seen by introducing two mutually complex conjugate null directions $X^\pm\equiv X^0\mp\compi\,X^{(D+1)}$ on the time-like plane of the embedding space. We can then write\footnote{The same equation holds in dS, provided we remove the $\compi$ from the last equation. }
\begin{equation}\begin{split}
\mathbb{A}_i&=\frac{2}{c}\,\mathbb{J}_{+i}\ ,\quad \widetilde{\mathbb{A}}_i=\frac{2}{c}\,\mathbb{J}_{-i}\ ,\quad \dfrac{\compi\,\mathbb{D}}{c^2}=\dfrac{2}{c^2}\,\mathbb{J}_{-+}\ .
\end{split}\end{equation}
These forms then make it evident that $c\to\infty $ limit corresponds to equally scaling out both the $\pm$ directions. All this can also be cast into the language of embedding space and linked to appropriate AdS Killing vectors, as we show in appendix~\ref{app:AdS}. 

Going back to the original algebra, we can expand all quantities about this non-relativistic oscillator limit in a $1/c$ expansion. In this work, we will show explicitly how this works for a 2-particle system interacting via a light scalar field and derive the Hamiltonian till $1/c^4$ order. The reader can look at  Eqs.\eqref{eq:1PN H} and  \eqref{eq:1PN G} to see how the $1/c^2$ corrections look like for the Hamiltonian $\mathbb{H}$ and node of inertia $\vec{\mathbb{G}}$.

\subsubsection*{Consequences of $\mathfrak{so}(D,2)$}
Let us now see how the conserved combinations stated in the beginning of this subsection directly follow from the above Poisson algebra :
\begin{itemize}
\item Conservation of $\mathbb{J}_{ij}$ and $\mathbb{H}$ follows from the fact that they Poisson-commute with   $\mathbb{H}$.
\item To get the remaining conservation, we start with  the Poisson bracket $\left[\vec{\mathbb{A}},\mathbb{H}\right]_{\rm PB}
=-
\compi\,\omA\, \vec{\mathbb{A}}$. The conservation of $e^{\compi\,\omA t}\ \vec{\mathbb{A}}$ then follows from the Hamiltonian evolution
\begin{equation}
\frac{d}{dt} \vec{\mathbb{A}}=\left[\vec{\mathbb{A}},\mathbb{H}\right]_{\rm PB}\ .   \end{equation}
\end{itemize}
This Hamiltonian re-statement of the conservation laws directly from AdS symmetry generators makes clear that they will hold generally even in interacting systems, as long as interactions preserve this algebra. We can also consider dynamical gravity, and general space times that approach AdS  asymptotically as $r\to\infty$, and these conclusions would still hold under appropriate boundary conditions.\footnote{Similar statements are true for spacetimes that asymptote to dS in future.} All this is, of course, completely analogous to the role played by Poincare invariance in asymptotically flat space times.

For the purposes of this work, we will note a  well-known statement  within post-Newtonian literature about flat spacetime :  the terms that occur in these PN expansions are strongly constrained by Poincare algebra (See \cite{Schafer:2018jfw} for a review). In fact, in one notable instance, Poincare invariance was used to fix a regularisation ambiguity in post-Newtonian expansion\cite{Damour:2000kk}. The invariance is also useful in constraining spin-dependent terms\cite{Hergt:2008jn}. We anticipate that the  symmetries of AdS would be similarly useful for AdS post-Newtonian expansions. In due course, we will support this expectation by showing how the AdS algebra  serves as a non-trivial check for our calculations.

Another important use of these symmetries is to decouple the `centre of mass'/centroid motion from the relative motion between particles. Given an interacting multi-particle (or field) configuration in AdS, we can always move to the `centre of mass/centroid frame', i.e., we can always bring to rest the oscillator corresponding to centroid to rest. Classically, this corresponds to setting the complex vector $\vec{\mathbb{A}}$ to zero, i.e., to take $\mathbb{P}_i=0=\mathbb{G}_i$. More precisely, we should first identify a centroid for the given  multi-particle configuration and then figure out how to bring it to rest at the origin. We will not say much more here except to say that it can be done: the reader interested in details can look at the Appendix~\ref{app:centroid}. There, under some mild assumptions spelt out carefully in that appendix, we prove that, corresponding to a set of $\mathfrak{so}(D,2)$ charges $\mathbb{J}^{AB}$, we can always assign a centroid which moves along a time-like geodesic. The construction is especially clean in the embedding space, where the worldline of the centroid is simply the intersection of time-like eigen-plane of the matrix $\mathbb{J}^A{}_{B}$ with the AdS hyperboloid. We  also compute there the finite $SO(D,2)$ transformations which bring this centroid to rest at the origin.

The multi-particle configurations in AdS where the centroid is not oscillating are said to be \emph{primary states}. All other states can be obtained from primary states by translation/boosts that set the centre of mass into oscillation : such states are then referred to as \emph{descendant states}. Since descendant states have additional centre of mass kinetic energy, they are always higher in energy than their mother primary state. In other words, the descendants are blue-shifted relative to the primary.  Thus the primary state can also be defined as the least energetic/least blue-shifted state among the family of states related by AdS isometries. In this language, our construction in Appendix~\ref{app:centroid} can then be thought of as an algorithm to bring an arbitrary descendant state back to its mother primary. Similar statements hold in the quantum theory, if we replace the phase space with the corresponding coherent states.

The corresponding statements in dS are not so clear cut. We can still define the multi-particle configurations in dS where the centroid is resting at origin, i.e., states with $\mathbb{P}_i=0=\mathbb{G}_i$ as \emph{primary states}. But such states are neither the least energetic nor the most energetic. This can be seen by the following argument: by translating a primary configuration away from the origin, we can use the redshift in dS to arbitrarily decrease the energy. Similarly, by boosting a primary configuration, we can arbitrarily increase its energy. Thus a dS primary state sits in the middle of a continuos energy spectrum where we can move up or down by boosts/translations, and has no special energy-based interpretation.

\section{Bulk-bulk Green functions : Non-relativistic  expansion}\label{sec:Bulk-bulk Green's function in AdS}
We now turn to inter-particle scalar interactions. Our goal is to solve for and eliminate the scalar field, retaining only the particle degrees of freedom in the post-Newtonian (PN) equations of motion\footnote{This is similar in spirit to solving Maxwell's equations for the electromagnetic field and then substituting the solution into the Lorentz force.}. In asymptotically flat spacetimes, this calculation is commonly organised in momentum space using worldline Feynman-like diagrams\cite{Porto:2016pyg,Levi:2018nxp,Goldberger:2004jt,Goldberger:2006bd}. Since there is no manifest coordinate translation invariance in AdS global coordinates, it is hard to formulate momentum space worldline diagrammatics in AdS.\footnote{See however \cite{Gadde:2022ghy,Banerjee:2024yiq} which use a momentum space inherited from embedding space. It is not a priori clear how non-relativistic expansion can be formulated within such a framework. } Therefore we use a multipole summation approach. We therefore begin with the radial Green's function $G_{l\om}$ in each multipole, expand it in powers of $1/c$, and then re-sum the multipoles at each PN order to obtain the full position-space Green's function $\mathcal G$. In this section, we describe this construction and specify the part of the Green's function retained in our calculation.

\subsection{Scalar field modes and AdS boundary conditions}
In this section, we derive the PN expansion of the bulk-bulk Green's function of the \emph{designer} scalar field given by \cite{Ghosh:2020lel,Loganayagam:2023pfb},
\begin{equation}\label{eq:scalar field action}\begin{split}
\mathcal{S}^{\phi}=&-\frac{1}{2}\int_{x}\,\sqrt{-g\left(x\right)}\ r^{\,\mathcal{N}+1-D}\\&\quad\times\left[\left(\partial\phi\right)^{2}+\left(D+\mathcal{N}-3\right)\left(D-\mathcal{N}-1\right)\frac{\phi^{2}}{4r^{2}}-\frac{\omA^{2}}{c^{2}}\left(\mu^{2}-\frac{\left(\mathcal{N}+1\right)^{2}}{4}\right)\phi^{2}\right].
\end{split}\end{equation}
Here, $\mathcal{N}$ and $\mu$ are parameters in the action. The above scalar field obeys the wave equation,
\begin{equation}\label{eq:scalar-wave-equation-general}
\begin{split}
&\left[
-\frac{1}{c^{2}f(r)}\frac{\partial^{2}}{\partial t^{2}}
+\frac{1}{r^{\mathcal{N}}}\frac{\partial}{\partial r}
\left(r^{\mathcal{N}}f(r)\frac{\partial}{\partial r}\right)
\right.\\
&\left.\qquad
+\frac{1}{r^{2}}\nabla^{2}_{S^{D-1}}
-\frac{(D+\mathcal{N}-3)(D-\mathcal{N}-1)}{4r^{2}}
-\frac{\omA^2}{c^{2}}
\left(\mu^{2}-\frac{(\mathcal{N}+1)^{2}}{4}\right)
\right]\phi=0.
\end{split}
\end{equation}
The above wave equation can be interpreted in two ways. First, for appropriately chosen values of $\mathcal{N}$ and $\mu$, it is the radial wave equation corresponding to various \emph{massless} fields such as massless Klein-Gordon scalar, electromagnetic and gravitational fields. The corresponding values of $\mathcal{N}$ and $\mu$ are collected in Table~\ref{table:N mu values for massless fields}.

\begin{table}
\centering
\renewcommand{\arraystretch}{1.6}
\setlength{\tabcolsep}{14pt}

\begin{tabular}{@{}lcccccc@{}}
\toprule
& Massless Klein-Gordon & \multicolumn{2}{c}{Electromagnetism}
     & \multicolumn{3}{c}{Gravity} \\
\cmidrule(lr){2-2}
\cmidrule(lr){3-4}
\cmidrule(l){5-7}
& Scalar & Vector & Scalar & Tensor & Vector & Scalar \\
\midrule
$\mathcal{N}$
& $D-1$ & $D-3$ & $3-D$
& $D-1$ & $1-D$ & $3-D$ \\

$\mu$
& $\frac{D}{2}$ & $\frac{D}{2}-1$ & $\frac{D}{2}-2$
& $\frac{D}{2}$ & $\frac{D}{2}-1$ & $\frac{D}{2}-2$ \\
\bottomrule
\end{tabular}

\caption{$\mathcal{N}$ and $\mu$ for different massless fields.}
\label{table:N mu values for massless fields}
\end{table}

Alternatively, if we set $\mathcal{N}=D-1$, we can also interpret
this as the \emph{massive} Klein-Gordon equation,
\begin{equation}\label{eq:KG eqn}
\left[\partial^2+m_\phi^2\,c^2\right]\phi=0
\end{equation}
with a cosmologically small mass,
\begin{equation}\label{eq:scalar-mass}
m_\phi^2=\frac{\omA^{2}}{c^{4}}\left(\mu^{2}-\frac{D^{2}}{4}\right)
\end{equation}
with the scalar field action Eq.\eqref{eq:scalar field action} reducing to\footnote{Another advantage of parametrising the scalar field mass in this way is that the scaling dimension of the dual CFT operator takes the simple form
\begin{equation}
\Delta_{\pm}=\frac{D}{2}\pm\mu,\qquad
\Delta_{\pm}\left(\Delta_{\pm}-D\right)=\frac{m_\phi^2c^4}{\omA^2}.
\end{equation}},
\begin{equation}\label{eq:scalar field action 3}\begin{split}
\mathcal{S}^{\phi}=&-\frac{1}{2}\int_{x}\,\sqrt{-g\left(x\right)}\left[\left(\partial\phi\right)^{2}+\frac{\omA^{2}}{c^{2}}\left(\mu^{2}-\frac{D^{2}}{4}\right)\phi^{2}\right].
\end{split}\end{equation}
Since the scalar field mass is cosmologically small (i.e. suppressed by a factor of $\omA^2/c^4$), the leading order interaction potential is still Newtonian as opposed to Yukawa.

We want to emphasize  that the analysis and results for the Green's function in this section are valid for both interpretations of the scalar wave equation Eq.\eqref{eq:scalar-wave-equation-general} (i.e. as radial equation for massless higher spin fields as well as massive Klein-Gordon scalar with cosmologically small mass) but the statements in later sections involving that scalar field coupling to the worldline and then onwards only allow for the massive scalar interpretation.

We can solve these free field equations by variable separation 
\begin{equation}
\begin{split}
 \phi(t,r,\Omega)
 &=e^{-\compi\omega t}Y_{ l \bar m}\left(D;\hat{\Omega}\right)\,\Xi_{ l \omega}(r),\\
 \nabla_{S^{D-1}}^{2}Y_{ l \bar m}
 &=- l ( l +D-2)Y_{ l \bar m}\ ,
\end{split}
\label{eq:separation-main}
\end{equation}
where $Y_{l\bar{m}}\left(D;\hat{\Omega}\right)$ are the orthonormal scalar spherical harmonics on a $(D-1)$ sphere. the radial ODE for $\Xi_{ l \omega}(r)$ is given in Eq.\eqref{eq:ads-radial-equation}, and  can be solved in terms of hypergeometric functions. In appendix~\ref{app:scalar-ads-ds}, we review in detail the various ways to write down the solutions of the radial equation and its physical import. For now, we will state only the salient features of that analysis, leaving out details. 

A basis of solutions valid at odd $D$, which we will use  a lot later, 
is the pair of solutions regular/singular   at the origin. In terms of the variables
\begin{equation}
\begin{split}
 u_{\rm AdS}=-\frac{\omA ^{2}r^{2}}{c^{2}},
 \qquad
 \widehat\omega=\frac{\omega}{\omA }\ ,\ \nu\equiv l+\frac{D}{2}-1\ ,
\end{split}
\end{equation}
we define
\begin{equation}\label{eq:ads-reg-sing-alt}
\begin{split}
 \Xi_{\rm Reg}
 &\equiv {\frac{r^l}{2\nu}}f(r)^{-\widehat\omega/2}
 {}_2F_1\!\left(
 \frac{1+\nu-\mu-\widehat\omega}{2},
 \frac{1+\nu+\mu-\widehat\omega}{2};
 1+\nu;u_{\rm AdS}\right),\\
 \Xi_{\rm Sing}
 &\equiv r^{-( l +D-2)}f(r)^{-\widehat\omega/2}
 {}_2F_1\!\left(
 \frac{1-\nu+\mu-\widehat\omega}{2},
 \frac{1-\nu-\mu-\widehat\omega}{2};
 1-\nu;u_{\rm AdS}\right).
\end{split}
\end{equation}
The interaction between two sources of the field are given by a bulk-bulk Green function constructed using the above solutions. The detailed form of this Green function of course depends on what boundary conditions we would like to impose far away from the sources. AdS is like a cavity and a natural boundary condition in this regard is to insist that there be no energy flux escaping the AdS cavity. In the standard AdS/CFT parlance, this is termed as normalisable boundary conditions, and it amounts to choosing a sub-dominant fall-off at large $r$. The linear combination of the above two functions that has such a subdominant falloff can be written down as
\begin{equation}
\begin{split}
 \widehat\Xi_{\rm Sub}
 =\Xi_{\rm Sing}-\frac{K_{+}
   +\mathfrak R_{\rm Sub}K_{-}
   }{
   1+\mathfrak R_{\rm Sub}}\ \Xi_{\rm Reg}\ ,
\end{split}
\label{eq:hatXi-sigma-origin-alt}
\end{equation}
where we have normalised the coefficient of the singular solution to be unity.\footnote{Such a normalisation based on the near origin behaviour is somewhat unconventional in usual AdS/CFT where the emphasis is often on large $r$ behaviour. But, as we shall see, this is the natural normalisation to study interaction between bulk sources.} The functions appearing in front of the regular solution have the explicit forms
\begin{equation}\label{eq:Kpm-main}
\begin{split}
 K_\pm
 &\equiv 2e^{\mp\compi\pi\nu}\left(\frac{\omA }{ c}\right)^{2\nu}
 \frac{\Gamma(1-\nu)}{\Gamma(\nu)}
 \frac{\Gamma\!\left(\frac{1+\nu+\mu+\widehat\omega}{2}\right)
  \Gamma\!\left(\frac{1+\nu-\mu+\widehat\omega}{2}\right)
  }{
  \Gamma\!\left(\frac{1-\nu+\mu+\widehat\omega}{2}\right)
  \Gamma\!\left(\frac{1-\nu-\mu+\widehat\omega}{2}\right)},\\
  \mathfrak R_{\rm Sub}
 &\equiv\exp\!\left[\compi\pi
 \left(\widehat\omega-\nu-\mu\right)\right].
\end{split}
\end{equation}
We have written the subdominant solution in a suggestive `repeated reflection/multiple bounce' form with a reflection coefficient\footnote{The familiar normal modes of AdS shows up as places where $\mathfrak R_{\rm Sub}=-1$, i.e., when the $\widehat\Xi_{\rm Sub}$ has a pole. } $\mathfrak R_{\rm Sub}$. The functions $K_\pm$ describe a sort of renormalised self-interaction of the source that is ultimately responsible for radiation reaction.\footnote{See \cite{Loganayagam:2023pfb,Loganayagam:2025jmw} for a detailed discussion of this viewpoint in the context of dS static patch. As we show in appendix~\ref{app:scalar-ads-ds}, the functions appearing here follow directly by analytically continuing the radiation reaction kernel for a dS source.} We can formally rewrite the coefficient of the regular solution as\footnote{This kind of an expansion is justified after we give a slight positive imaginary part to the frequency so that $|\mathfrak R_{\rm Sub}|$ is slightly less than unity. This corresponds to working with a retarded boundary condition for the radiation. } 
\begin{equation}
\begin{split}
\frac{K_{+}
   +\mathfrak R_{\rm Sub}K_{-}
   }{
   1+\mathfrak R_{\rm Sub}}
=K_{+}+\left(\mathfrak R_{\rm Sub}-\mathfrak R_{\rm Sub}^2+\mathfrak R_{\rm Sub}^3\ldots\right)(K_{-}-K_{+})\ ,
\end{split}
\end{equation}
and then physically interpret the terms in $\widehat\Xi_{\rm Sub}$ as follows:
\begin{itemize}
\item The first term $\Xi_{\rm Sing}$ describes the near-zone quasi-static field of a point multipole source sitting at the origin. Since we are interested in the potential contributions in the near zone, we will be entirely focused on this term for the rest of this work.
\item The $K_{+}$ term without any $\mathfrak R_{\rm Sub}$ describes a purely  outgoing radiative correction to the field. For a proper justification of this viewpoint, see \S\S\ref{ssec:RadZone}. As discussed in detail in  the context of dS by authors of\cite{Loganayagam:2023pfb}, this term has a smooth flat limit and can be used to generalise the Abraham-Lorentz-Dirac force to cosmological spacetimes. All that analysis with appropriate sign changes applies to AdS.
\item After that, the term with $n$ powers of $\mathfrak R_{\rm Sub}$ computes the field after $n$ reflections at the boundary of AdS. This is a novel feature peculiar to AdS, not present in dS or flat spacetimes. Given that $\mathfrak R_{\rm Sub}$ has $\omA$ in the denominator of its exponent, these  represent  `cosmological memory' type radiative corrections which are non-perturbative in small $\omA$ expansion.
\end{itemize}
Thus, while the radiative contributions  are essential for the sub-dominant fall-off at large $r$, they could be ignored if we are interested in only potential contributions in the near zone. 

The normalisable bulk-bulk  Green function for the radial ODE
can be constructed by stitching together the homogeneous solutions such that we have regularity at the origin, sub-dominant fall-off at large $r$ and a unit jump. With our normalisations, the answer turns out to be
\begin{equation}
\begin{split}
 G_{ l \omega}(r_1,r_2)
 &=
 \Xi_{\rm Reg}(r_<)\,\widehat\Xi_{\rm Sub}(r_>)\\
 &=
 \Xi_{\rm Reg}(r_<)\Xi_{\rm Sing}(r_>)
 -\frac{K_{+}
   +\mathfrak R_{\rm Sub}K_{-}
   }{
   1+\mathfrak R_{\rm Sub}}\
  \Xi_{\rm Reg}(r_1)\Xi_{\rm Reg}(r_2)\ ,
\end{split}
\end{equation}
where we have defined $r_>\equiv \text{Max}(r_1,r_2)$ and  $r_<\equiv \text{Min}(r_1,r_2)$.

We retain the near-zone potential kernel (i.e., only the first term in the above equation) and omit the radiative contribution. The PN counting of the omitted term is discussed at the beginning of the next subsection. With this restriction, the radial Green's function used below is
\begin{equation}\begin{split}
G_{l\omega}\left(r_{1},r_{2}\right)=&\,\Xi_{_\mathrm{Reg}}(r_<)\Xi_{_\mathrm{Sing}}(r_>)
\\=&\, \frac{1}{2\nu}\frac{r_{<}^{l}}{r_{>}^{l+D-2}}f(r_<)^{\frac{\hat\omega}{2}}f(r_>)^{\frac{\hat\omega}{2}} \\
 & \times  \,_{2}F_{1}\left(\frac{1+\nu-\mu+\hat{\omega}}{2},\frac{1+\nu+\mu+\hat{\omega}}{2},1+\nu,-\frac{r_{<}^{2}\omA^{2}}{c^{2}}\right) \\
 & \times\,_{2}F_{1}\left(\frac{1-\nu+\mu+\hat{\omega}}{2},\frac{1-\nu-\mu+\hat{\omega}}{2},1-\nu,-\frac{r_{>}^{2}\omA^{2}}{c^{2}}\right).
\end{split}\end{equation}
The overall normalisation factor of $1$ is fixed by the jump condition,
\begin{equation}
r_2^{D-1}f(r_2)\left[
\left.\partial_{r_1}G_{l\omega}(r_1,r_2)\right|_{r_1\rightarrow r_2^{-}}
-\left.\partial_{r_1}G_{l\omega}(r_1,r_2)\right|_{r_1\rightarrow r_2^{+}}
\right]=1.
\end{equation}
Finally, we sum over the radial Green's functions to obtain the
position-space potential kernel,
\begin{align}\label{eq:multipole sum formula}
\mathcal{G}\left(x_{1},x_{2}\right) & =\sum_{l,\bar m}\int\frac{d\omega}{2\pi}e^{i\omega\left(t_{1}-t_{2}\right)}Y_{l \bar{m}}\left(D;\hat{\Omega}_{1}\right)Y_{l\bar{m}}^{*}\left(D;\hat{\Omega}_{2}\right)G_{l\omega}\left(r_{1},r_{2}\right)\nn \\
 & =\sum_{l}\int\frac{d\omega}{2\pi}e^{i\omega\left(t_{1}-t_{2}\right)}\frac{\mathcal{N}_{\text{HH}}\left(D,l\right)}{\left|\mathbb{S}^{D-1}\right|}\mathcal{P}_{l}\left(D;\hat{\Omega}_{1}.\hat{\Omega}_{2}\right)G_{l\omega}\left(r_{1},r_{2}\right)
\end{align}
where we have used the addition theorem for scalar spherical harmonics\cite{Loganayagam:2023pfb},
\begin{equation}
\sum_{\bar{m}}Y_{l\bar{m}}\left(D;\hat{\Omega}_{1}\right)Y_{l\bar{m}}^{*}\left(D;\hat{\Omega}_{2}\right)=\frac{\mathcal{N}_{\text{HH}}\left(D,l\right)}{\left|\mathbb{S}^{D-1}\right|}\mathcal{P}_{l}\left(D;\hat{\Omega}_{1}.\hat{\Omega}_{2}\right).
\end{equation}
Here $\mathcal{P}_{l}\left(D;\hat{\Omega}_{1}.\hat{\Omega}_{2}\right)$ are the generalised Legendre polynomials\footnote{The generalized Legendre polynomials $P_l(d,z)$ appearing
in our analysis are related to the Gegenbauer polynomials
and associated Legendre functions through the identities
\begin{equation}
C_l^{\frac{D}{2}-1}(z)
= (D-2)\frac{\mathcal{N}_{\rm HH}(D,l)}{2l+D-2}
P_l(D,z),
\qquad
P_\lambda^{-\mu}(z)
=
\frac{(1-z^2)^{\mu/2}}{2^\mu\Gamma(\mu+1)}
P_{\lambda-\mu}(2\mu+3,z).
\end{equation}}, $\left|\mathbb{S}^{D-1}\right|$ is the volume of $(D-1)$ sphere and $\mathcal{N}_{\text{HH}}\left(D,l\right)$ are the number of homogeneous harmonic polynomials of degree $l$ in $D$ dimensions. These are discussed in Appendix A. of \cite{Loganayagam:2023pfb}. Explicitly,
\begin{equation}
\mathcal{N}_{\text{HH}}\left(D,l\right)=(2l+D-2)\frac{(l+D-3)!}{l!(D-2)!},
\qquad
\left|\mathbb{S}^{D-1}\right|=\frac{2\pi^{D/2}}{\Gamma(D/2)}.
\end{equation}
Eq.\eqref{eq:multipole sum formula} gives the multipole expansion of the potential part of the scalar bulk-bulk Green's function in AdS. 

\subsection{PN expansion of bulk-bulk Green's function}
In this subsection, we perform a PN expansion of $G_{l\om}$ and re-sum the multipoles at each order. We take the PN limit with particle positions, frequencies $\om$ and $\omA$, and scalar parameter $\mu$ fixed.

Before expanding the potential kernel, let us comment on the scaling of the radiative contribution omitted above. As in flat spacetime, the radiative contributions of $l^{th}$ multipole come suppressed by a factor of $c^{-(2l+D-2)}$. If we are working in odd $D$ till 2PN, i.e., till $1/c^4$ accuracy, the only cases that matter are 
\begin{equation}
(D,l)=(3,0)_{0.5 {\rm PN}}\ ,\quad (3,1)_{1.5 {\rm PN}}\ ,\quad (5,0)_{1.5 {\rm PN}}\ .
\end{equation}
Explicit forms for these terms can be obtained from \cite{Loganayagam:2023pfb}.
The appropriate memory contributions from these radiative terms should be accounted for separately, in addition to terms that we compute in this work.

We now derive the PN expansion of the potential kernel $\mathcal{G}\left(x_1,x_2\right)$. It is convenient to expand the radial Green's function in a sequence of Jacobi polynomials
\footnote{The sequence of Jacobi polynomials appearing in this expansion can be defined via any one of the following representations valid when $n$ is a non-negative integer :
\begin{equation}\begin{split}
n! &\ \mathcal{P}_n^{\left(\nu, -1-2n\right)}\left(1-2 \tau^2\right)=n!\ (1-\tau^2)^n \mathcal{P}_n^{\left(\nu, -\nu\right)}\left(\frac{1+ \tau^2}{1-\tau^2}\right)
= \frac{\Gamma (\nu +n+1) }{\Gamma (\nu +1)}\, _2F_1\left(-n,\nu -n;\nu +1;\tau ^2\right).
\end{split}\end{equation}} as follows 
\begin{equation}\label{eq:radial-green-fn-exp}\begin{split}
G_{l\omega}\left(r_{1},r_{2}\right) =& \frac{r_{<}^{l}}{r_{>}^{l+D-2}}\sum_{n=0}^\infty n!\ r_{>}^{2n}\   \mathcal{P}_n^{(\nu,-2n-1)}\left(1-2\frac{r_{<}^2}{r_{>}^2}\right)\\
&\times \left\{ \frac{\Gamma(\nu-n)}{\Gamma(\nu+n+1)}\ \mathbb{A}_n(r_1,r_2) +\frac{\omA^2}{c^2} (r_{>}^2-r_{<}^2)\ \mathbb{B}_n(r_1,r_2)\right\}\ .
\end{split}\end{equation}
Here $\mathbb{A}_n(r_1,r_2)$ and $\mathbb{B}_n(r_1,r_2)$ 
are \emph{symmetric} functions of the two radial coordinates independent of multi-pole label $l$ as well as dimension $D$. 

They are especially suited for non-relativistic or post-Newtonian expansion since, we shall see that they scale at large $c$ as 
\begin{equation}\begin{split}
\mathbb{A}_n\ ,\ \mathbb{B}_n\ \sim \frac{1}{c^{2n}}\ .
\end{split}\end{equation}
We give explicit non-relativistic expansions of the $(\mathbb{A},\mathbb{B})$ functions up to 3PN order in equations Eq.\eqref{eq:AB0PN}-\eqref{eq:AB3PN}. These expressions show that, at k-PN order, the contribution to $(\mathbb{A}_n,\mathbb{B}_n)$ is a symmetric homogeneous polynomial of degree $(k-n)$ in variables $\{r_1^2,r_2^2\}$.

Another advantage of this expansion is that, since $(\mathbb{A}_n,\mathbb{B}_n)$ are independent of $l$, we can re-sum the multipole expansion. This can be done using the following two identities valid for $0\leq\tau <1$, $n$ a non-negative integer, and odd $D$: 
\begin{equation}\label{eq:MultFunI}\begin{split}
&\frac{\Gamma\left(\frac{D}{2}-1-n\right)}{\Gamma\left(\frac{D}{2}\right)\left(1+\tau^{2}-2z\tau\right)^{\frac{D}{2}-1-n}}=  \sum_{l}\mathcal{N}_{\text{HH}}\left(D,l\right)\mathcal{P}_{l}\left(D;z\right)\tau^l \frac{\Gamma (\nu -n) }{\Gamma (\nu +n+1)}\, n!\ \mathcal{P}^{(\nu,-2n-1)}_n(1-2\tau^2)
\ ,
\end{split}\end{equation}
as well as
\begin{equation}\label{eq:MultFunII}\begin{split}
\frac{\Gamma\left(\frac{D}{2}+n\right)(1-\tau^2)^{2n+1}}{\Gamma\left(\frac{D}{2}\right)\left(1+\tau^{2}-2z\tau\right)^{\frac{D}{2}+n}}
=&  \sum_{l}\mathcal{N}_{\text{HH}}\left(D,l\right)\mathcal{P}_{l}\left(D;z\right) \tau^{l} \ n!\ \mathcal{P}^{(\nu,-2n-1)}_{n}(1-2\tau^2)\ .
\end{split}\end{equation}
We will refer the reader to Appendix \S\ref{app:multProof} for a proof of these identities.
Equation (4.15) of \cite{Blanchet:2005ft} gives a similar identity for $D=3$. Our first identity corresponds to
$s=2n-1$ in that paper, up to normalization. Our second  identity corresponds to $s=-D-2n$, whose Jacobi-polynomial form follows
from Euler's hypergeometric transformation.

The multipole sum then gives the following PN expansion of the frequency-domain potential kernel:
\begin{equation}\label{eq:multipole-sum}\begin{split}
&\sum_{l}\mathcal{N}_{\text{HH}}\left(D,l\right)\mathcal{P}_{l}\left(D;\hat{\Omega}_{1}.\hat{\Omega}_{2}\right) G_{l\omega}\left(D;r_{1},r_{2}\right)\\
&\quad=\sum_{n=0}^\infty  \left\{ \mathbb{A}_n(r_1,r_2)\ \frac{\Gamma\left(\frac{D}{2}-1-n\right)}{\Gamma\left(\frac{D}{2}\right)r_{12}^{D-2n-2}}\  +\frac{\omA^2}{c^2}(r_1^2-r_2^2)^{2n+2} \ \mathbb{B}_n(r_1,r_2)\frac{\Gamma\left(\frac{D}{2}+n\right)}{\Gamma\left(\frac{D}{2}\right)r_{12}^{D+2n}} \right\}\ .
\end{split}\end{equation}

Hence, using Eq.\eqref{eq:multipole sum formula}, the PN expansion of the potential part of the bulk-bulk Green's function is
\begin{align}\label{eq:pn expansion of green function}
\mathcal{G}\left(x_{1},x_{2}\right)=&\int\frac{d\omega}{2\pi}\frac{e^{i\omega\left(t_{1}-t_{2}\right)}}{\left|\mathbb{S}^{D-1}\right|}\nn\\&\times\sum_{n=0}^{\infty}\left\{ \mathbb{A}_{n}(r_{1},r_{2})\ \frac{\Gamma\left(\frac{D}{2}-1-n\right)}{\Gamma\left(\frac{D}{2}\right)r_{12}^{D-2n-2}}\ +\frac{\omA^{2}}{c^{2}}(r_{1}^{2}-r_{2}^{2})^{2n+2}\ \mathbb{B}_{n}(r_{1},r_{2})\frac{\Gamma\left(\frac{D}{2}+n\right)}{\Gamma\left(\frac{D}{2}\right)r_{12}^{D+2n}}\right\} \ .
\end{align}

We now state the explicit Green's-function contributions through 2PN order. We label the individual terms by $\mathcal{G}_{p,q}$, where $p$ denotes the order in $1/c^2$ and $q$ counts powers of $\omA^2$. Thus,
\begin{equation}
\mathcal{G}^{(p)}(x_1,x_2)
=\sum_{q=0}^{p}\mathcal{G}_{p,q}(x_1,x_2).
\end{equation}
In frequency space, $\mathcal{G}_{p,q}$ carries the factor
$\omega^{2(p-q)}\omA^{2q}/c^{2p}$, which is included in its definition. The term $\mathcal{G}_{p,0}$ is the flat-space contribution at order $p$, while the terms with $q\geq1$ describe curvature corrections.

At Newtonian order, we recover the Kepler/Coulomb Green's function in $D$ spatial dimensions:
\begin{equation}
\begin{split}
\mathcal{G}^{(0)}(x_1,x_2)
&=\mathcal{G}_{0,0}(x_1,x_2)\\
&=\int\frac{d\omega}{2\pi}
\frac{e^{i\omega(t_1-t_2)}}{\left|\mathbb{S}^{D-1}\right|}
\frac{1}{(D-2)r_{12}^{D-2}}\\
&=\frac{\delta_{t_1t_2}}{(D-2)\left|\mathbb{S}^{D-1}\right|}
\frac{1}{r_{12}^{D-2}}.
\end{split}
\end{equation}

At first post-Newtonian order, the Green's function consists of a flat-space contribution $\mathcal{G}_{1,0}$ and a curvature correction $\mathcal{G}_{1,1}$:
\begin{equation}
\mathcal{G}^{(1)}(x_1,x_2)
=\mathcal{G}_{1,0}(x_1,x_2)+\mathcal{G}_{1,1}(x_1,x_2).
\end{equation}
These are given by
\begin{equation}
\begin{split}
\mathcal{G}_{1,0}(x_1,x_2)
&=\int\frac{d\omega}{2\pi}
\frac{e^{i\omega(t_1-t_2)}}{\left|\mathbb{S}^{D-1}\right|}
\frac{\omega^2}{c^2}
\frac{1}{2(D-2)(D-4)r_{12}^{D-4}},
\\
\mathcal{G}_{1,1}(x_1,x_2)
&=-\int\frac{d\omega}{2\pi}
\frac{e^{i\omega(t_1-t_2)}}{\left|\mathbb{S}^{D-1}\right|}
\frac{\omA^2}{c^2}\\
&\quad\times\Bigg[
\frac{\mu^2}{2(D-2)(D-4)r_{12}^{D-4}}
+\frac{r_1^2+r_2^2-\frac{D-2}{2}(\vec{r}_1+\vec{r}_2)^2}
{4(D-2)r_{12}^{D-2}}
+\frac{r_1^2r_2^2-(\vec{r}_1.\vec{r}_2)^2}
{2r_{12}^{D}}
\Bigg].
\end{split}
\end{equation}

At second post-Newtonian order, there is a flat-space term $\mathcal{G}_{2,0}$, a curvature contribution $\mathcal{G}_{2,1}$ proportional to $\omA^2$, and a curvature contribution $\mathcal{G}_{2,2}$ proportional to $\omA^4$:
\begin{equation}
\mathcal{G}^{(2)}(x_1,x_2)
=\mathcal{G}_{2,0}(x_1,x_2)
+\mathcal{G}_{2,1}(x_1,x_2)
+\mathcal{G}_{2,2}(x_1,x_2).
\end{equation}
Their explicit expressions are
\begin{subequations}
\begin{equation}
\begin{split}
\mathcal{G}_{2,0}(x_1,x_2)
&=\int\frac{d\omega}{16\pi}
\frac{e^{i\omega(t_1-t_2)}}{\left|\mathbb{S}^{D-1}\right|}
\frac{\omega^4}{c^4}
\frac{1}{(D-2)(D-4)(D-6)r_{12}^{D-6}},
\end{split}
\end{equation}
\begin{equation}
\begin{split}
\mathcal{G}_{2,1}(x_1,x_2)
&=\int\frac{d\omega}{2\pi}
\frac{e^{i\omega(t_1-t_2)}}{\left|\mathbb{S}^{D-1}\right|}
\frac{\omega^2\omA^2}{c^4}\\
&\quad\times\Bigg[
-\frac{2(\mu^2+1)}
{8(D-2)(D-4)(D-6)r_{12}^{D-6}}
-\frac{3(r_1^2+r_2^2)}
{8(D-2)(D-4)r_{12}^{D-4}}
+\frac{(r_1^2-r_2^2)^2}
{16(D-2)r_{12}^{D-2}}
\Bigg],
\end{split}
\end{equation}
\begin{equation}
\begin{split}
\mathcal{G}_{2,2}(x_1,x_2)
&=\int\frac{d\omega}{2\pi}
\frac{e^{i\omega(t_1-t_2)}}{\left|\mathbb{S}^{D-1}\right|}
\frac{\omA^4}{c^4}\\
&\quad\times\Bigg[
\frac{(\mu^2-1)^2}
{8(D-2)(D-4)(D-6)r_{12}^{D-6}}
+\frac{(2\mu^2+1)(r_1^2+r_2^2)}
{16(D-2)(D-4)r_{12}^{D-4}}\\
&\qquad\qquad
-\frac{(2\mu^2-5)r_1^4
-2(2\mu^2+1)r_1^2r_2^2
+(2\mu^2-5)r_2^4}
{32(D-2)r_{12}^{D-2}}\\
&\qquad\qquad
-\frac{3(r_1^2-r_2^2)^2(r_1^2+r_2^2)}
{32r_{12}^{D}}
+\frac{D(r_1^2-r_2^2)^4}
{128r_{12}^{D+2}}
\Bigg].
\end{split}
\end{equation}
\end{subequations}
In the flat-space limit $\omA\rightarrow0$ and for $D=3$, these expressions agree, after Fourier transformation to momentum space, with those obtained using traditional worldline diagrammatics \cite{Almeida:2024uph}.

\emph{Even though we explicitly work in AdS, every equation in this section \S\ref{sec:Bulk-bulk Green's function in AdS} can be analytically continued to dS by replacing $\omA\to\compi\hds$.}

\section{Effective Fokker action and equations of motion}\label{sec:Effective Fokker action and equations of motion}

Post-Newtonian (PN) methods are used to derive the dynamics of compact binary systems at successive orders in $1/c$. They are applicable during the early inspiral, when the orbital motion is non-relativistic and the gravitational interaction between the bodies is weak. The central idea is to solve for the gravitational field and eliminate it resulting in a description with only particle degrees of freedom. There are various methods within the PN framework which essentially implement this program in various ways-- {\bf The direct approach} \cite{Blanchet:2000nv,Blanchet:2000nu,Blanchet:2000ub,Blanchet:2000cw,Blanchet:2018yqa,deAndrade:2000gf,Blanchet:2002mb,Blanchet:2003gy} directly tackles the Einstein's equations to solve for the gravitational field order by order in PN expansion. The solution is then fed into the equations of motion of the particles.
{\bf The ADM Hamiltonian approach} \cite{Jaranowski:1997ky,Jaranowski:1999ye,Jaranowski:1999qd,Damour:2000kk,Damour:2001bu,Damour:2000ni} uses the ADM Hamiltonian and eliminates the conjugate field variables to obtain the point particle ADM Hamiltonian at each PN order.
{\bf The worldline effective field theory approach} \cite{Goldberger:2004jt,Gilmore:2008gq,Foffa:2011ub,Foffa:2013qca,Porto:2016pyg,Levi:2018nxp} employs worldline methods to solve for and eliminate the field degrees of freedom. The compact bodies are represented by particle worldlines, and the potential modes of the gravitational field are integrated out using diagrammatic methods from quantum field theory. This gives us an effective action with only particle degrees of freedom, called the \emph{Fokker action}.

The \emph{Fokker Lagrangian method} is an earlier variation of the worldline EFT method which falls somewhere in between full blown diagrammatic approach of worldline EFT and the direct approach mentioned earlier. Just like in wordline EFT, one begins with a field+particle worldline action,
\begin{equation*}
\mathcal{S}^{\rm field}[g]+\mathcal{S}^{\rm particle}[g,x_a]
\end{equation*}
with the particle action being the worldline action of the particle trajectory. One then varies the field action to obtain the field equations of motion. These can then be solved at each PN order to obtain the PN expanded field solution:
\begin{equation*}
\bar{g}_{\alpha\beta}=\bar{g}^{(0)}_{\alpha\beta}[t,x_a]+\frac{1}{c^2}\bar{g}^{(1)}_{\alpha\beta}[t,x_a]+\frac{1}{c^4}\bar{g}^{(2)}_{\alpha\beta}[t,x_a]+\ldots
\end{equation*}
This is then substituted back into the $\mathcal{S}^{\rm particle}[g,x_a]$ to obtain the Fokker action $S^{\rm eff}[x_a(t)]$ with only particle degrees of freedom. In this paper we use this technique but with the inter-particle forces mediated by a scalar field $\phi$ instead of a gravitational field $g$.

\subsection{Integrating out field degrees of freedom}
Consider the inter-particle interaction to be modelled by a free massive scalar field,
\begin{equation}\label{eq:scalar field action 1}\begin{split}
\mathcal{S}^{\phi}=&-\frac{1}{2}\int_{x}\,\sqrt{-g\left(x\right)}\left[\left(\partial\phi\right)^{2}+\frac{\omA^{2}}{c^{2}}\left(\mu^{2}-\frac{D^{2}}{4}\right)\phi^{2}\right].
\end{split}\end{equation}
In this case, the field equations are linear, and this elimination can be performed explicitly using the scalar bulk-to-bulk Green's function. We assume that the scalar field interacts with the particle worldline $x_a^\mu(t)\equiv\left\{ct,x^i_a(t)\right\}$ via the action,
\begin{equation}
\mathcal{S}^{\text{int}}=-\sum_am_{a}\lambda\intop d\tau_{a}\ \phi\left(x_{a}\right)\label{eq:worldline coupled to scalar}
\end{equation}
 where $\lambda$ is the wordline to scalar coupling. The total action describing the binary system is then given by,
\begin{equation}
\mathcal{S}=\mathcal{S}^{\text{free}}+\mathcal{S}^{\phi}+\mathcal{S}^{\text{int}}\label{eq:total action}
\end{equation}
where $\mathcal{S}^{\text{free}}$ was given in Eq.\eqref{eq:worldline action}.

We want to integrate out the field degrees of freedom to obtain an effective particle description in the form of a Fokker action. This can be achieved (in principle) by performing a path integral over the field degrees of freedom and retaining only terms at leading order in $\hbar$,
\begin{equation}
\mathcal{S}_{\text{eff}}=-\compi\log\int [d\phi]\,e^{\compi\,\mathcal{S}\left[x_{1},x_{2},\phi\right]}=\mathcal{S}^{\text{free}}[x]+W
\end{equation}
\begin{equation}
W=-\compi\,\log\int[d\phi]\,e^{\compi\,\left(\mathcal{S}^{\phi}+\mathcal{S}^{\text{int}}[x,\phi]\right)}
\end{equation}
Using standard Gaussian integration, we have,
\begin{equation}\label{eq:W}
W=\int_{x,y}\,J_1\left(x\right)\,\mathcal{G}\left(x,y\right)\,J_2\left(y\right)
\end{equation}
where we have defined the particle wordline current,
\begin{equation}
J_a=\frac{m_{a}\lambda}{c}\,\sqrt{-g_{00}-\frac{g_{ij}}{c^2}\frac{dx_{a}^{i}}{dt}\frac{dx_{a}^{j}}{dt}}\;\delta^{\left(D\right)}\left(\vec{x}-\vec{r}_{a}\left(t\right)\right)
\end{equation}

In Eq.\eqref{eq:W} we have also dropped the terms $\sim J_a.\mathcal{G}.J_a$. Due to time reversal symmetry of the Green's function, these terms can be removed after adding appropriate counterterms as discussed in \cite{Loganayagam:2023pfb}. We assume that such a procedure has been performed already and hence we ignore these terms. In flat space worldline EFT, removing these terms amounts to removing the contribution from the disconnected self force diagrams that are proportional to scale-free integrals in dimensional regularisation.

Computing Eq.\eqref{eq:W} then gives us the effective inter-particle potential $V_{\text{eff}}$ as,
\begin{equation}
W=-\int_tV_{\text{eff}}
\end{equation}
We want to obtain this potential in a post-Newtonian expansion. This will further gives us the Fokker action $\mathcal{S}_{\text{eff}}$.

\subsection{Effective potential in PN expansion}

To obtain the effective Fokker action for the binary, we must expand both the Green's function and the particle worldline currents in powers of $1/c^2$. The worldline factor has the expansion
\begin{equation}\label{eq:gamma-factor-expansion}
\begin{split}
\sqrt{-g_{00}-g_{ij}\frac{dx^{i}}{cdt}\frac{dx^{j}}{cdt}}
&=\sqrt{f(r)-\frac{1}{c^{2}}f(r)^{-1}\dot{r}^{2}
-\frac{1}{c^{2}}r^{2}\dot{\Omega}_{D-1}^{2}}\\
&=\sum_{n,s,k}\binom{n}{k}
\frac{(2s+2n+2k-5)!!(2s+2k-5)!!}
{2^{n+s-1}(2s+4k-5)!!n!(s-1)!}
\left(-\frac{r^{2}\omA^{2}}{c^{2}}\right)^{n}
\left(\frac{\dot{r}^{2}}{c^{2}}\right)^{k}
\left(\frac{v^{2}}{c^{2}}\right)^{s-1} \ ,
\end{split}
\end{equation}
where the sum is over all non-negative integers. Together with the PN expansion of the Green's function, this allows us to evaluate Eq.\eqref{eq:W} to any desired PN order.

We denote the contribution of $\mathcal{G}_{p,q}$ to the interaction action at total PN order $n$ by $W^{(n)}_{p,q}$. The Green's function supplies order $p$, while the expansion of the two worldline currents supplies the remaining order $n-p$. The corresponding potential contribution is defined by
\begin{equation}
W^{(n)}_{p,q}=-\int_t V^{(n)}_{p,q}.
\end{equation}
The complete contributions at this order are therefore
\begin{equation}
W^{(n)}=\sum_{p=0}^{n}\sum_{q=0}^{p}W^{(n)}_{p,q},
\qquad
V_{\text{eff}}^{(n)}
=\sum_{p=0}^{n}\sum_{q=0}^{p}V^{(n)}_{p,q}.
\end{equation}
All powers of $1/c$ and $\omA$ are included in these quantities. The index $q$ continues to label the curvature dependence of the Green's-function term, rather than that of the complete action or potential contribution, since additional powers of $\omA^2$ can arise from the worldline currents.

At Newtonian order, $\mathcal{O}(c^0)$, only $\mathcal{G}_{0,0}$ contributes:
\begin{equation}
W^{(0)}=W^{(0)}_{0,0}
=m_1m_2\lambda^2\int_{t_1,t_2,x_1,x_2}
\delta_{\vec{x}_1,\vec{r}_1}^{(D)}
\mathcal{G}_{0,0}(x_1,x_2)
\delta_{\vec{x}_2,\vec{r}_2}^{(D)}.
\end{equation}
The resulting Newtonian potential is
\begin{equation}
V_{\text{eff}}^{(0)}=V^{(0)}_{0,0}
=-\gsc\frac{m_1m_2}{r_{12}^{D-2}},
\end{equation}
where we have introduced the effective Newtonian coupling
\begin{equation}
\gsc=\frac{\lambda^2}{(D-2)\left|\mathbb{S}^{D-1}\right|}.
\end{equation}
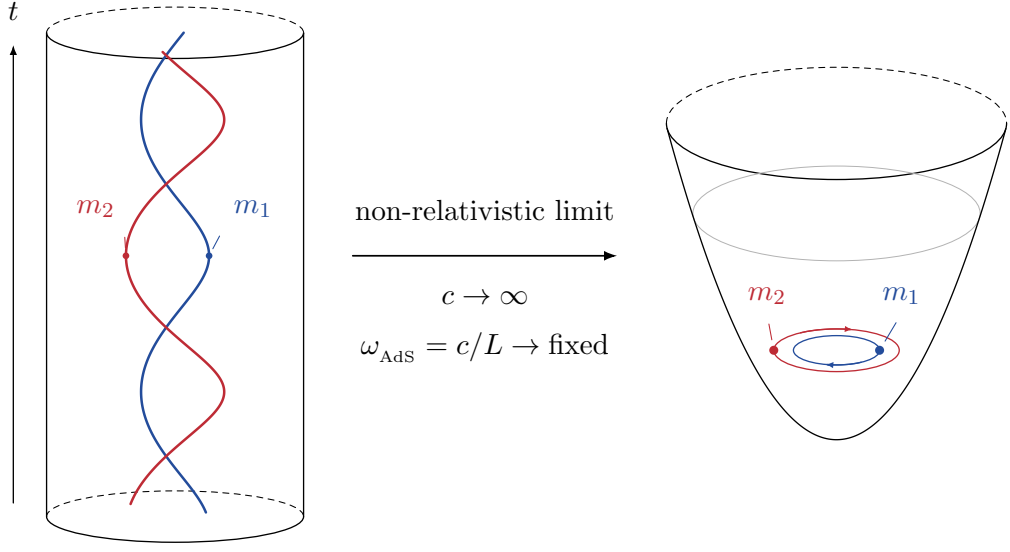
\begin{figure}
    \centering
    \begin{tikzpicture}[x=1cm,y=1cm,line cap=round,line join=round]
  \foreach \height in {-3.2,3.2} {
    \draw[fig boundary,fig hidden]
      (-1.7,\height) arc[start angle=180,end angle=0,
        x radius=1.7,y radius=0.34];
    \draw[fig boundary]
      (-1.7,\height) arc[start angle=180,end angle=360,
        x radius=1.7,y radius=0.34];
  }
  \draw[fig boundary] (-1.7,-3.2) -- (-1.7,3.2);
  \draw[fig boundary] ( 1.7,-3.2) -- ( 1.7,3.2);
  \draw[fig time arrow] (-2.14,-3.02) -- (-2.14,3.04);
  \node[fig label,above=6pt] at (-2.14,3.04) {$t$};

  \draw[fig worldline,figblue]
    plot[domain=-3.1:3.1,samples=260,variable=\q]
      ({0.45*cos(100*(\q-0.25))},
       {\q-0.24*0.45*sin(100*(\q-0.25))});
  \draw[fig worldline,figred]
    plot[domain=-3.1:3.1,samples=260,variable=\q]
      ({-0.65*cos(100*(\q-0.25))},
       {\q+0.24*0.65*sin(100*(\q-0.25))});
  \fill[figblue] ( 0.45,0.25) circle[radius=1.25pt];
  \fill[figred]  (-0.65,0.25) circle[radius=1.25pt];

  \node[fig label,text=figblue] (massone) at ( 1.03,0.87) {$m_1$};
  \node[fig label,text=figred]  (masstwo) at (-1.05,0.87) {$m_2$};
  \draw[figblue,line width=0.4pt,shorten <=4pt,shorten >=3pt]
    (massone.south west) -- ( 0.45,0.25);
  \draw[figred,line width=0.4pt,shorten <=4pt,shorten >=3pt]
    (masstwo.south east) -- (-0.65,0.25);

  \draw[black,line width=0.65pt,-{Latex[length=5pt]}]
    (2.35,0.25) -- (5.85,0.25);
  \node[fig label,above=10pt] at (4.1,0.25)
    {non-relativistic limit};
  \node[fig label,align=center,below=10pt] at (4.1,0.25)
    {$c\to\infty$\\[5pt]$\omA=c/L\to\text{fixed}$};

  \begin{scope}[xshift=8.75cm]
    \pgfmathsetmacro{\wellrim}{0.82*2.25*2.25-2.15}
    \pgfmathsetmacro{\wellside}{sqrt(2.25*2.25-(0.33/1.64)^2)}

    \draw[fig boundary,fig hidden]
      (-2.25,\wellrim) arc[start angle=180,end angle=0,
        x radius=2.25,y radius=0.7425];
    \draw[fig boundary]
      (-2.25,\wellrim) arc[start angle=180,end angle=360,
        x radius=2.25,y radius=0.7425];

    \draw[black!30,line width=0.35pt]
      (0,{0.82*1.9*1.9-2.15}) ellipse[x radius=1.9,y radius=0.627];

    \draw[fig boundary]
      plot[domain=-\wellside:\wellside,samples=150,variable=\q]
        (\q,{0.82*\q*\q-2.15-0.33*0.33/(4*0.82)});

    \draw[figblue,line width=0.5pt]
      (0,-1) ellipse[x radius=0.57,y radius=0.1938];
    \draw[figred,line width=0.5pt]
      (0,-1) ellipse[x radius=0.83,y radius=0.2822];
    \draw[figblue,line width=0.5pt,-{Latex[length=3pt,width=2.4pt]}]
      plot[domain=55:105,samples=25,variable=\q]
        ({0.57*cos(\q)},{-1-0.1938*sin(\q)});
    \draw[figred,line width=0.5pt,-{Latex[length=3pt,width=2.4pt]}]
      plot[domain=235:285,samples=25,variable=\q]
        ({0.83*cos(\q)},{-1-0.2822*sin(\q)});
    \fill[figblue] ( 0.57,-1) circle[radius=1.7pt];
    \fill[figred]  (-0.83,-1) circle[radius=1.7pt];

    \node[fig label,text=figblue] (wellmassone) at ( 0.86,-0.28) {$m_1$};
    \node[fig label,text=figred]  (wellmasstwo) at (-0.92,-0.28) {$m_2$};
    \draw[figblue,line width=0.4pt,shorten <=4pt,shorten >=4pt]
      (wellmassone.south) -- ( 0.57,-1);
    \draw[figred,line width=0.4pt,shorten <=4pt,shorten >=4pt]
      (wellmasstwo.south) -- (-0.83,-1);
  \end{scope}
\end{tikzpicture}
\caption{In the non-relativistic limit, the binary behaves as if it is placed in a harmonic well with harmonic frequency $\omA$.}
\label{fig:nr limit}
\end{figure}
Combining this interaction with the Newtonian-order terms in $\mathcal{S}^{\text{free}}$, and retaining the rest-mass term, gives the Newtonian order action Eq.\eqref{eq:0PN action}:
\begin{equation}
\begin{split}
\mathcal{S}_{\text{eff}}^{\text{N}}
&=\int_t\left[-Mc^{2}
+\frac{1}{2}m_1v_1^2+\frac{1}{2}m_2v_2^2+\gsc\frac{m_1m_2}{r_{12}^{D-2}}-
\frac{1}{2}m_1r_1^2\omA^2
-\frac{1}{2}m_2r_2^2\omA^2
\right].
\end{split}
\end{equation}
At this order, AdS curvature therefore contributes a harmonic potential for each particle. Under the scalar-mass scaling used here, mass-dependent corrections enter only at higher PN orders, so the leading interparticle potential is Newtonian rather than Yukawa.

At first post-Newtonian order, $\mathcal{O}(1/c^2)$, the interaction receives contributions from the leading Green's function combined with the 1PN current corrections, and from the 1PN Green's function combined with the leading currents:
\begin{equation}
W^{(1)}
=W^{(1)}_{0,0}+W^{(1)}_{1,0}+W^{(1)}_{1,1}.
\end{equation}
The contribution from $\mathcal{G}_{0,0}$ is
\begin{equation}
W^{(1)}_{0,0}
=m_1m_2\lambda^2\int_{t_1,t_2,x_1,x_2}
\left[
\frac{r_1^2\omA^2-v_1^2}{2c^2}
+\frac{r_2^2\omA^2-v_2^2}{2c^2}
\right]
\delta_{\vec{x}_1,\vec{r}_1}^{(D)}
\mathcal{G}_{0,0}(x_1,x_2)
\delta_{\vec{x}_2,\vec{r}_2}^{(D)}
\end{equation}
The two components of $\mathcal{G}^{(1)}$ are the flat-space term $\mathcal{G}_{1,0}$ and the curvature-dependent term $\mathcal{G}_{1,1}$. Their contributions are
\begin{equation}
\begin{split}
W^{(1)}_{1,0}
&=m_1m_2\lambda^2\int_{t_1,t_2,x_1,x_2}
\delta_{\vec{x}_1,\vec{r}_1}^{(D)}
\mathcal{G}_{1,0}(x_1,x_2)
\delta_{\vec{x}_2,\vec{r}_2}^{(D)}=\gsc\,m_1m_2\int_t
\partial_{t_1}\partial_{t_2}
\left\{\frac{1}{2c^2(D-4)r_{12}^{D-4}}\right\}_{t_{1,2}\rightarrow t},
\\
W^{(1)}_{1,1}
&=m_1m_2\lambda^2\int_{t_1,t_2,x_1,x_2}
\delta_{\vec{x}_1,\vec{r}_1}^{(D)}
\mathcal{G}_{1,1}(x_1,x_2)
\delta_{\vec{x}_2,\vec{r}_2}^{(D)}.
\end{split}
\end{equation}
Combining these contributions with the free-particle action  Eq.\eqref{eq:worldline action} gives the effective action through 1PN order:
\begin{equation}\begin{split}
\mathcal{S}_{\text{eff}}^{\text{1PN}}=&\ \mathcal{S}_{\text{eff}}^{\text{N}}+\frac{1}{8c^{2}}\int_t\;\left[m_{1}v_{1}^{4}+m_{2}v_{2}^{4}\right]\\&+\frac{\gsc}{c^{2}}\int_t\,\frac{m_{1}m_{2}}{2r_{12}^{D-2}}\,\left[\vec{v}_{1}.\vec{v}_{2}-\left(D-2\right)\left(\hat{n}_{12}.\vec{v}_{1}\right)\left(\hat{n}_{12}.\vec{v}_{2}\right)-v_{1}^{2}-v_{2}^{2}\right]\\&-\frac{\omA^{2}}{c^{2}}\int_t\;\left[\frac{1}{4}m_{1}r_{1}^{2}v_{1}^{2}+\frac{1}{4}m_{2}r_{2}^{2}v_{2}^{2}+\frac{1}{2}m_{1}\left(\vec{r}_{1}.\vec{v}_{1}\right)^{2}+\frac{1}{2}m_{2}\left(\vec{r}_{2}.\vec{v}_{2}\right)^{2}\right]\\&+\omA^{2}\frac{\gsc}{c^{2}}\int_t\,\frac{m_1 m_2}{2r_{12}^{D-2}}
\left[
r_{\mathrm{av}}^2
+ (D-2)\left(\hat{n}_{12}.\vec{r}_{\mathrm{av}}\right)^2
+ \left(\frac{1}{4}-\frac{\mu^2}{D-4}\right)r_{12}^2
\right]\\&+\frac{\omA^{4}}{c^{2}}\int_t\;\left[\frac{1}{8}m_{1}r_{1}^{4}+\frac{1}{8}m_{2}r_{2}^{4}\right].
\end{split}\end{equation}
In the flat-space limit, this agrees with the 1PN scalar-exchange action in $D=3$ without any coordinate transformation\cite{Porto:2016pyg,Levi:2018nxp}.

At second post-Newtonian order, $\mathcal{O}(1/c^4)$, the interaction receives contributions from $\mathcal{G}^{(0)}$, $\mathcal{G}^{(1)}$ and $\mathcal{G}^{(2)}$:
\begin{equation}
\begin{split}
W^{(2)}
&=W^{(2)}_{0,0}
+W^{(2)}_{1,0}+W^{(2)}_{1,1}+W^{(2)}_{2,0}+W^{(2)}_{2,1}+W^{(2)}_{2,2}.
\end{split}
\end{equation}
The leading Green's function $\mathcal{G}_{0,0}$ combines with the terms of total order $1/c^4$ in the product of the worldline currents:
\begin{equation}
\begin{split}
W^{(2)}_{0,0}
&=m_1m_2\lambda^2\int_{t_1,t_2,x_1,x_2}
\Bigg[
\frac{(r_1^2\omA^2-v_1^2)(r_2^2\omA^2-v_2^2)}{4c^4}
+\frac{(\vec{r}_1.\vec{v}_1)^2\omA^2}{2c^4}
+\frac{(\vec{r}_2.\vec{v}_2)^2\omA^2}{2c^4}\\
&\qquad
-\frac{(r_1^2\omA^2-v_1^2)^2}{8c^4}
-\frac{(r_2^2\omA^2-v_2^2)^2}{8c^4}
\Bigg]
\delta_{\vec{x}_1,\vec{r}_1}^{(D)}
\mathcal{G}_{0,0}(x_1,x_2)
\delta_{\vec{x}_2,\vec{r}_2}^{(D)}.
\end{split}
\end{equation}
The two components of $\mathcal{G}^{(1)}$ combine with the 1PN current corrections:
\begin{equation}
\begin{split}
W^{(2)}_{1,0}
&=m_1m_2\lambda^2\int_{t_1,t_2,x_1,x_2}
\left(
\frac{r_1^2\omA^2+r_2^2\omA^2-v_1^2-v_2^2}{2c^2}
\right)
\delta_{\vec{x}_1,\vec{r}_1}^{(D)}
\mathcal{G}_{1,0}(x_1,x_2)
\delta_{\vec{x}_2,\vec{r}_2}^{(D)}\\
&=\frac{\gsc}{c^2}\int_t
\partial_{t_1}\partial_{t_2}
\left\{
\frac{m_1m_2}{2(D-4)r_{12}^{D-4}}
\left(
\frac{r_1^2\omA^2+r_2^2\omA^2-v_1^2-v_2^2}{2c^2}
\right)
\right\}_{t_{1,2}\rightarrow t},
\\[3mm]
W^{(2)}_{1,1}
&=m_1m_2\lambda^2\int_{t_1,t_2,x_1,x_2}
\left(
\frac{r_1^2\omA^2+r_2^2\omA^2-v_1^2-v_2^2}{2c^2}
\right)
\delta_{\vec{x}_1,\vec{r}_1}^{(D)}
\mathcal{G}_{1,1}(x_1,x_2)
\delta_{\vec{x}_2,\vec{r}_2}^{(D)}.
\end{split}
\end{equation}
Finally, the three components of $\mathcal{G}^{(2)}$ combine with the leading worldline currents:
\begin{equation}
\begin{split}
W^{(2)}_{2,0}
&=m_1m_2\lambda^2\int_{t_1,t_2,x_1,x_2}
\delta_{\vec{x}_1,\vec{r}_1}^{(D)}
\mathcal{G}_{2,0}(x_1,x_2)
\delta_{\vec{x}_2,\vec{r}_2}^{(D)}=\frac{\gsc}{c^4}\int_t
\partial_{t_1}^2\partial_{t_2}^2
\left\{
\frac{m_1m_2}{8(D-4)(D-6)r_{12}^{D-6}}
\right\}_{t_{1,2}\rightarrow t},
\\[3mm]
W^{(2)}_{2,1}
&=m_1m_2\lambda^2\int_{t_1,t_2,x_1,x_2}
\delta_{\vec{x}_1,\vec{r}_1}^{(D)}
\mathcal{G}_{2,1}(x_1,x_2)
\delta_{\vec{x}_2,\vec{r}_2}^{(D)}\\
&=\gsc\,m_1m_2\frac{\omA^2}{c^4}\int_t
\partial_{t_1}\partial_{t_2}
\Bigg\{
-\frac{1+\mu^2}{4(D-6)(D-4)r_{12}^{D-6}}
-\frac{3(r_1^2+r_2^2)}{8(D-4)r_{12}^{D-4}}
+\frac{(r_1^2-r_2^2)^2}{16r_{12}^{D-2}}
\Bigg\}_{t_{1,2}\rightarrow t},
\\[3mm]
W^{(2)}_{2,2}
&=m_1m_2\lambda^2\int_{t_1,t_2,x_1,x_2}
\delta_{\vec{x}_1,\vec{r}_1}^{(D)}
\mathcal{G}_{2,2}(x_1,x_2)
\delta_{\vec{x}_2,\vec{r}_2}^{(D)}.
\end{split}
\end{equation}
The explicit expressions for the effective potential through 2PN order are given in \S\ref{app:2PN action and equations of motion}, from Eqns.\eqref{eq:0PN V} to \eqref{eq:2PN V}. In the flat-space limit $\omA\rightarrow0$, these expressions match term by term with \cite{Almeida:2024uph} for $D=3$, without any further contact transformation.

\subsection{Effective particle action and equations of motion}

The action obtained above depends on accelerations as well as positions and
velocities. Although its equations of motion can be derived directly, it is
convenient to obtain an ordinary Lagrangian that depends only on positions and
velocities. We do this in three steps: a double-zero term removes the quadratic
acceleration terms, a total time derivative simplifies the remaining linear
terms, and a change of particle variables removes the remaining linear order acceleration
dependence. These operations are standard in PN calculations
\cite{Barker:1980spx,Damour:1990jh}. Explicit applications of double-zero terms
and total derivatives can be found in general relativity
\cite{SCHAFER1984128,Damour:1985mt,Gilmore:2008gq} and in scalar-tensor theories
\cite{Almeida:2024uph,Bernard:2023eul}.

To explain the first step, recall that the Newtonian equations of motion are
\begin{equation}
\begin{split}
\vec a_1+\omA^2\vec r_1
+\frac{(D-2)\gsc m_2}{r_{12}^{D-1}}\hat n_{12}
&=0,
\\
\vec a_2+\omA^2\vec r_2
-\frac{(D-2)\gsc m_1}{r_{12}^{D-1}}\hat n_{12}
&=0.
\end{split}
\end{equation}
A double-zero term is quadratic in combinations that vanish by these
equations. Consequently, both the term and its first variation vanish when
the Newtonian equations are imposed: varying either factor leaves another
vanishing factor. Multiplying such a term by $1/c^4$ therefore does not
modify the dynamics through 2PN order.

We eliminate the quadratic acceleration terms in $W_{2,0}^{(2)}$ by adding
\begin{equation}
\begin{split}
\delta W^{(2)}_{\text{dz}}
={}&-\frac{\gsc}{c^4}\int_t
\frac{m_1m_2}{8(D-4)r_{12}^{D-4}}
\\
&
\Bigg[
\Bigg(
\vec a_1+\omA^2\vec r_1
+\frac{(D-2)\gsc m_2}{r_{12}^{D-1}}\hat n_{12}
\Bigg).
\Bigg(
\vec a_2+\omA^2\vec r_2
-\frac{(D-2)\gsc m_1}{r_{12}^{D-1}}\hat n_{12}
\Bigg)
\\
&
-(D-4)
\Bigg(
\hat n_{12}.\vec a_1
+\omA^2\hat n_{12}.\vec r_1
+\frac{(D-2)\gsc m_2}{r_{12}^{D-1}}
\Bigg)
\\
&\qquad\qquad\times\Bigg(
\hat n_{12}.\vec a_2
+\omA^2\hat n_{12}.\vec r_2
-\frac{(D-2)\gsc m_1}{r_{12}^{D-1}}
\Bigg)
\Bigg].
\end{split}
\end{equation}
Expanding this expression cancels the terms quadratic in accelerations
and replaces them with terms that are at most linear in accelerations.
The added term need not vanish away from a solution, but its contribution
to the equations of motion starts beyond the required PN accuracy. For $D=3$ and $\omA\rightarrow0$, the double-zero term agrees with the
pure-scalar specialization of the term used in
\cite{Almeida:2024uph}.

We next add the total time derivative
\begin{equation}
\begin{split}
\delta W_{\text{td}}
={}&\int_t\frac{d}{dt}\Bigg[
\frac{\gsc}{c^4}
\frac{m_1m_2}{8(D-4)r_{12}^{D-4}}
\Bigg\{
\omA^2\left(\vec r_2.\vec v_1+\vec r_1.\vec v_2\right)
\\
&+\frac{(D-2)(D-5)\gsc m_1}{r_{12}^{D-1}}
\left(\hat n_{12}.\vec v_1\right)
-\frac{(D-2)(D-5)\gsc m_2}{r_{12}^{D-1}}
\left(\hat n_{12}.\vec v_2\right)\\
&-(D-4)\omA^2
\left(
(\hat n_{12}.\vec r_2)(\hat n_{12}.\vec v_1)
+(\hat n_{12}.\vec r_1)(\hat n_{12}.\vec v_2)
\right)
\Bigg\}
\Bigg].
\end{split}
\end{equation}
A total time derivative changes the action only by an endpoint contribution
and hence leaves the bulk equations of motion unchanged. Its usefulness here
comes from the velocity dependence of the expression being differentiated:
taking its time derivative produces terms linear in the accelerations.
The above choice cancels the curvature-dependent acceleration terms and the
remaining position-dependent acceleration coefficients generated by
$\delta W^{(2)}_{\text{dz}}$. This simplifies the action without changing the
particle variables.

The remaining acceleration terms are linear and independent of $\omA$.
Removing them requires a change of particle variables. Since this change
depends on velocities, it is a contact transformation
\cite{Damour:1990jh,Julie:2017pkb}. Replacing accelerations by lower-order
equations of motion in the action is justified here as a consequence of
this transformation; the resulting equations describe the same dynamics
in the transformed variables. To keep track of the symmetry generators,
including the red-shifted momentum introduced in the next section, we
specify the transformation explicitly and use the corresponding variables
when constructing those generators.

Our sign convention is
\begin{equation}
\vec r_a^{\,\text{old}}
=\vec r_a+\delta\vec x_a+\mathcal O(c^{-6}),
\end{equation}
where the unlabelled positions and velocities are the reduced variables.
The required shifts are
\begin{equation}\label{eq:deltax}
\begin{split}
\delta\vec x_1
={}&-\frac{\gsc}{c^4}
\frac{m_2}{8(D-4)r_{12}^{D-4}}
\Bigg[
-\frac{2(D-4)(\hat n_{12}.\vec v_2)}{r_{12}}\vec v_2
+\frac{(D-4)(D-2)(\hat n_{12}.\vec v_2)^2}{r_{12}}
\hat n_{12}
\\
&\quad
-\frac{(D-4)v_2^2}{r_{12}}\hat n_{12}
+\frac{4(D-4)(\hat n_{12}.\vec v_2)}{r_{12}}\vec v_1
\Bigg],
\\
\delta\vec x_2
={}&-\frac{\gsc}{c^4}
\frac{m_1}{8(D-4)r_{12}^{D-4}}
\Bigg[
\frac{2(D-4)(\hat n_{12}.\vec v_1)}{r_{12}}\vec v_1
-\frac{(D-4)(D-2)(\hat n_{12}.\vec v_1)^2}{r_{12}}
\hat n_{12}
\\
&\quad
+\frac{(D-4)v_1^2}{r_{12}}\hat n_{12}
-\frac{4(D-4)(\hat n_{12}.\vec v_1)}{r_{12}}\vec v_2
\Bigg].
\end{split}
\end{equation}
The velocities transform by the time derivatives of these shifts.
These expressions specify the transformation through 2PN order; no
higher-order completion is needed for the present calculation.

Since $\delta\vec x_a=\mathcal O(c^{-4})$, only its variation of the
Newtonian action contributes through 2PN. After an integration by parts,
this is equivalent, up to a boundary term, to adding the single-zero term
\begin{equation}
\begin{split}
\delta W^{(2)}_{\text{sz}}
=\int_t\Bigg[
&\delta\vec x_1.
\left(
\frac{\partial \mathcal{L}^{\rm N}}{\partial\vec r_1}
-m_1\vec a_1
\right)
+\delta\vec x_2.
\left(
\frac{\partial \mathcal{L}^{\rm N}}{\partial\vec r_2}
-m_2\vec a_2
\right)
\Bigg].
\end{split}
\end{equation}
Unlike a double-zero term, a single-zero term is only linear in the
Newtonian equations of motion. Its variation does not vanish upon
imposing those equations, so it must be accompanied by the stated change
of variables. In \S\ref{app:momInv} we show that this coordinate transformation is carefully chosen so as to not modify the AdS Noether momentum expression \eqref{eq:GPfree} through 2PN order.

For the shifts above, the remaining acceleration-dependent part is cancelled by the acceleration terms in
$\delta W^{(2)}_{\text{sz}}$. The terms that remain are precisely those
obtained by replacing these accelerations with their Newtonian values,
\begin{equation}
\begin{split}
\vec a_1
&\longrightarrow
-\omA^2\vec r_1
-\frac{(D-2)\gsc m_2}{r_{12}^{D-1}}\hat n_{12},
\\
\vec a_2
&\longrightarrow
-\omA^2\vec r_2
+\frac{(D-2)\gsc m_1}{r_{12}^{D-1}}\hat n_{12}.
\end{split}
\end{equation}
These replacements are made only in the remaining 2PN acceleration terms.
Corrections to the Newtonian accelerations would contribute at higher PN
order.

The resulting ordinary action through 2PN order is
\begin{equation}
\begin{split}
\mathcal S_{\text{eff}}
=\int_t\mathcal L
=\sum_{\mathit{n}_1,\mathit{n}_2,\mathit{n}_3}
\left(\frac{1}{c^2}\right)^{\mathit{n}_1}
\left(\gsc m_1m_2\right)^{\mathit{n}_2}
\left(\omA^2\right)^{\mathit{n}_3}
\int_t L^{(\mathit{n}_1,\mathit{n}_2,\mathit{n}_3)}.
\end{split}
\end{equation}
The non-zero coefficients through 2PN are given in \S\ref{app:2PN action and equations of motion} from
Eqns.\eqref{eq:-1PN action} to \eqref{eq:2PN action}.
For $D=3$ and $\omA\rightarrow0$, the scalar-exchange terms agree with
those in \cite{Almeida:2024uph}, with the scalar coupling normalized to
$\gsc$. We can compare our results with the scalar-tensor results after
decoupling tensor exchange and retaining a free scalar with a linear
worldline coupling. Our expressions match with \cite{Mirshekari:2013vb,Julie:2017pkb} up to a boundary term in the action. This requires no further contact transformation. 

The equations of motion then follow from the ordinary Euler--Lagrange equations,
\begin{equation}
\frac{d}{dt}\frac{\partial\mathcal L}{\partial\vec v_a}
-\frac{\partial\mathcal L}{\partial\vec r_a}
=0.
\end{equation}
Solving these equations perturbatively for the accelerations through
2PN order gives
\begin{equation}
\vec{\mathcal F}_a=m_a\vec a_a,
\end{equation}
where
\begin{equation}
\vec{\mathcal F}_a
=\sum_{\mathit{n}_1,\mathit{n}_2,\mathit{n}_3}
\left(\frac{1}{c^2}\right)^{\mathit{n}_1}
\left(\gsc m_1m_2\right)^{\mathit{n}_2}
\left(\omA^2\right)^{\mathit{n}_3}
\vec F_a^{(\mathit{n}_1,\mathit{n}_2,\mathit{n}_3)}.
\end{equation}
The explicit force coefficients are given in \S\ref{app:2PN action and equations of motion} from
Eqns.\eqref{eq:0PN Force} to \eqref{eq:2PN Force}.

\emph{Even though we explicitly work in AdS, every equation in this section \S\ref{sec:Effective Fokker action and equations of motion} can be analytically continued to dS by replacing $\omA\to\compi\hds$.}

\section{Hamiltonian and conformal algebra}\label{sec:Hamiltonian and conformal algebra}

To verify that the Fokker action and equations of motion obtained above are consistent with the isometries of the background spacetime, we perform a Legendre transform of the Fokker Lagrangian to obtain the Hamiltonian. This allows us to construct a representation of the isometry algebra on the phase space of the PN Hamiltonian. In the flat-space PN literature, this route is known to reproduce the reduced ADM Hamiltonian of general relativity after the necessary contact transformations, as demonstrated explicitly at 2PN and 3PN orders~\cite{Damour:1985mt,deAndrade:2000gf}.

\subsection{2PN Hamiltonian}
We now compute the full 2PN Hamiltonian. We begin by computing the conjugate momentum,
\begin{equation}\label{eq:conjugate-momentumExp}
\vec{p}_a=\;\frac{\partial}{\partial\vec{v}_a}\mathcal{L}
=\sum_{\mathit{n}_1,\mathit{n}_2,\mathit{n}_3}\left(\frac{1}{c^2}\right)^{\mathit{n}_1}\left(\gsc m_1m_2\right)^{\mathit{n}_2}\left(\omA^2\right)^{\mathit{n}_3}\vec{p}^{\,(\mathit{n}_1,\mathit{n}_2,\mathit{n}_3)}
\end{equation}
where the explicit expressions for $\vec{p}^{\,(\mathit{n}_1,\mathit{n}_2,\mathit{n}_3)}$ up to 2PN order are given in \S\ref{app:2PN Hamiltonian and conserved charges} from Eqns.\eqref{eq:0PN conjugate momentum} to \eqref{eq:2PN conjugate momentum}. Inverting these relations gives the 2PN Hamiltonian,
\begin{equation}\label{eq:HamiltonianExp}
\mathbb{H}=\;\sum_a\vec{p}_a.\,\vec{v}_a-\mathcal{L}
=\sum_{\mathit{n}_1,\mathit{n}_2,\mathit{n}_3}\left(\frac{1}{c^2}\right)^{\mathit{n}_1}\left(\gsc m_1m_2\right)^{\mathit{n}_2}\left(\omA^2\right)^{\mathit{n}_3}\mathbb{H}^{\,(\mathit{n}_1,\mathit{n}_2,\mathit{n}_3)}
\end{equation}
where the explicit expressions for $\mathbb{H}^{\,(\mathit{n}_1,\mathit{n}_2,\mathit{n}_3)}$ up to 2PN are given in \S\ref{app:2PN Hamiltonian and conserved charges} from Eqns.\eqref{eq:-1PN Hamiltonian} to \eqref{eq:2PN Hamiltonian}. From Eqn.\eqref{eq:0PN Hamiltonian}, we see that at Newtonian order the total potential energy is simply the sum of the Newtonian potential energy and the harmonic oscillator energy, in agreement with \cite{Maxfield:2022hkd,Maxfield:2022nat,Zahraee:2024dsj},
\begin{equation}
\mathbb{H}^{\text{N}}=Mc^{2}+\frac{p_{1}^{2}}{2m_{1}}+\frac{p_{2}^{2}}{2m_{2}}+\frac{1}{2}m_{1}r_{1}^{2}\omA^{2}+\frac{1}{2}m_{2}r_{2}^{2}\omA^{2}-\gsc\frac{m_{1}m_{2}}{r_{12}^{D-2}}
\end{equation}
Eqns.\eqref{eq:1PN Hamiltonian} and \eqref{eq:2PN Hamiltonian} provide the relativistic corrections at 1PN and 2PN order. In the flat-space limit $\omA\to0$ and for $D=3$, our result agrees with the pure-scalar specialisation of the two-body Hamiltonian in Appendix~C of Ref.~\cite{Julie:2017pkb}, with the boundary-function choice $f_3=f_4=\tfrac14$ and all other $f_i=0$.\footnote{In the notation of that reference, the matching parameters are $\bar\gamma_{AB}=-2$, $\bar\beta_A=\bar\beta_B=-\tfrac12$, $\delta_A=\delta_B=0$, $\epsilon_A=\epsilon_B=2$, and $\zeta=1$, with $G_*G_{AB}=\gsc$ after restoring the bare gravitational constant $G_*$.} This requires no further contact transformation.

We display the Hamiltonian through 1PN order below, while the considerably longer 2PN terms are given in \S\ref{app:2PN Hamiltonian and conserved charges}.
\begin{equation}\label{eq:1PN H}
\begin{split}
\mathbb{H}^{\text{1PN}}
&=\;\mathbb{H}^{\text{N}}-\frac{1}{c^{2}}\left[\frac{p_{1}^{4}}{8m_{1}^{3}}+\frac{p_{2}^{4}}{8m_{2}^{3}}\right]\\
&\quad+\frac{\gsc}{c^{2}}\frac{m_{1}m_{2}}{r_{12}^{D-2}}\left[\frac{p_{1}^{2}}{2m_{1}^{2}}+\frac{p_{2}^{2}}{2m_{2}^{2}}-\frac{\vec{p}_{1}.\vec{p}_{2}-\left(D-2\right)\left(\vec{p}_{1}.\hat{n}_{12}\right)\left(\vec{p}_{2}.\hat{n}_{12}\right)}{2m_{1}m_{2}}\right]\\
&\quad+\frac{\omA^{2}}{c^{2}}\left[\frac{\left(\vec{p}_{1}.\vec{r}_{1}\right)^{2}}{2m_{1}}+\frac{\left(\vec{p}_{2}.\vec{r}_{2}\right)^{2}}{2m_{2}}+\frac{p_{1}^{2}r_{1}^{2}}{4m_{1}}+\frac{p_{2}^{2}r_{2}^{2}}{4m_{2}}\right]\\
&\quad-\omA^{2}\frac{\gsc}{c^{2}}\frac{m_{1}m_{2}}{2\,r_{12}^{D-2}}
\left[
r_{\mathrm{av}}^2
+ (D-2)\left(\hat{n}_{12}.\vec{r}_{\mathrm{av}}\right)^2
+ \left(\frac{1}{4}-\frac{\mu^2}{D-4}\right)r_{12}^2
\right]\\
&\quad-\frac{\omA^{4}}{c^{2}}\left[\frac{1}{8}m_{1}r_{1}^{4}+\frac{1}{8}m_{2}r_{2}^{4}\right].
\end{split}
\end{equation}
Beyond $\mathbb{H}^{\text{N}}$, the five lines give, respectively, the flat-space relativistic correction to the kinetic energy, the flat-space interaction correction proportional to $\gsc$, the pure curvature correction proportional to $\omA^2$, the mixed curvature--interaction correction proportional to $\gsc\omA^2$, and the pure curvature correction proportional to $\omA^4$.

\subsection{Obtaining the symmetry generators in phase space}

Using the particle Hamiltonian, we can derive the various symmetry generators in phase space. We begin by defining the (A)dS Noether total momentum,
\begin{equation}\label{eq:total momentum AdS 3}
\vec{\mathbb{P}}=\sum_a\;\sqrt{f(r_a)}\,\vec{p}_a.
\end{equation}
which has the same expression as the free particle case. In \S\ref{app:momInv} we show that this is consistent with our choice of coordinates.
Using the 2PN equations of motion derived in \S\ref{sec:Effective Fokker action and equations of motion} and \S\ref{app:2PN action and equations of motion}, we can verify that this obeys the harmonic equation of motion through the retained PN order:
\begin{equation}\label{eq:harmonic P}
\frac{d^2\vec{\mathbb{P}}}{dt^2}=-\,\omA^2\,\vec{\mathbb{P}}.
\end{equation}
Thus, the expression Eqn.\eqref{eq:total momentum AdS 3} requires no interaction-dependent corrections through 2PN order. Similarly, the manifest spatial rotational invariance of the Hamiltonian implies conservation of the total angular momentum,
\begin{equation}\label{eq:angular-momentum2}
\mathbb{J}_{ij}=\sum_a\,(x_{ai}p_{aj}-x_{aj}p_{ai}).
\end{equation}
We can then define the centre-of-energy moment using the relation
\begin{equation}\label{eq:ptotdot=-omasq G}
\vec{\mathbb{G}}=\frac{1}{\omA^{2}}\left[\mathbb{H},\vec{\mathbb{P}}\right]
=-\frac{1}{\omA^{2}}\frac{d}{dt}\vec{\mathbb{P}}.
\end{equation}

We can use Hamilton's equations of motion to obtain an expression for $\vec{\mathbb{G}}$ in terms of the Hamiltonian,
\begin{equation}\label{centre of mass generator 2}
\begin{split}
\vec{\mathbb{G}}
&=-\frac{1}{\omA^{2}}\frac{d}{dt}
\left\{\sum_a\;\vec{p}_a\sqrt{f(r_a)}\right\}\\
&=\frac{1}{\omA^{2}}\sum_a\;
\frac{\partial\mathbb{H}}{\partial\vec{r}_a}\sqrt{f(r_a)}
-\frac{1}{c^{2}}\sum_a\;
\left(\vec{r}_a.\frac{\partial\mathbb{H}}{\partial\vec{p}_a}\right)
\frac{\vec{p}_a}{\sqrt{f(r_a)}}.
\end{split}
\end{equation}
Thus, the AdS symmetry relation provides a direct construction of $\vec{\mathbb{G}}$ from the PN Hamiltonian. This expression applies at any PN order. Once the curvature dependence of the Hamiltonian is known, we can obtain $\vec{\mathbb{G}}$ and subsequently take the flat-space limit $\omA\to0$. In the flat-space PN literature, the corresponding generator has been constructed from Lorentz invariance of the Lagrangian in harmonic coordinates~\cite{deAndrade:2000gf,Blanchet:2002mb}, or by imposing the Poincare algebra directly on canonical phase-space generators~\cite{Damour:2000kk}. In AdS, we have an advantage since the Hamiltonian completely determines $\vec{\mathbb G}$. Such a relation is not available in the flat limit.

It follows from Eqns.\eqref{eq:ptotdot=-omasq G} and \eqref{eq:harmonic P} that
\begin{equation}\label{eq:rcmdot=ptot}
\frac{d}{dt}\vec{\mathbb{G}}=\vec{\mathbb{P}}.
\end{equation}
Therefore, $\vec{\mathbb{G}}$ also satisfies the harmonic oscillator equation\footnote{One may ask whether the definition of $\vec{\mathbb{P}}$ in Eqn.\eqref{eq:total momentum AdS 3} could be modified by adding a constant multiple of $\vec{\mathbb{G}}$, since it would still obey the harmonic equation of motion. We choose the definition Eqn.\eqref{eq:total momentum AdS 3} because, as in flat space, the total momentum is odd under time reversal, whereas $\vec{\mathbb{G}}$ is even. Adding such a term would spoil this property.}:
\begin{equation}\label{eq:harmonic G}
\frac{d^2\vec{\mathbb{G}}}{dt^2}=-\,\omA^2\,\vec{\mathbb{G}}.
\end{equation}

Using the 2PN Hamiltonian we obtain explicit an expression for the node of inertia at 2PN order,
\begin{equation}
\vec{\mathbb{G}}
=\sum_{\mathit{n}_1,\mathit{n}_2,\mathit{n}_3}
\left(\frac{1}{c^2}\right)^{\mathit{n}_1}
\left(\gsc m_1m_2\right)^{\mathit{n}_2}
\left(\omA^2\right)^{\mathit{n}_3}
\vec{\mathbb{G}}^{\,(\mathit{n}_1,\mathit{n}_2,\mathit{n}_3)},
\end{equation}
where the explicit coefficients through 2PN order are given in \S\ref{app:2PN Hamiltonian and conserved charges} from
Eqns.\eqref{eq:0PN boost generator} to \eqref{eq:2PN boost generator}.
An interesting feature of this result is that the node of inertia receives no explicit curvature corrections at 1PN order.
Indeed, we find
\begin{equation}\label{eq:1PN G}
\vec{\mathbb{G}}^{\mathrm{1PN}}
=m_{1}\vec{r}_{1}+m_{2}\vec{r}_{2}
+\frac{1}{c^{2}}\left[
\frac{p_{1}^{2}}{2m_{1}}\,\vec{r}_{1}
+\frac{p_{2}^{2}}{2m_{2}}\,\vec{r}_{2}
-\gsc\,\frac{m_{1}m_{2}}{r_{12}^{D-2}}\,
\vec{r}_{\mathrm{av}}
\right],
\end{equation}
which has precisely the same functional form as its flat-space
counterpart.\footnote{This remains true in configuration space,
since the relation between the canonical momenta and velocities
at Newtonian order is simply $\vec{p}_a^{\,N}=m_a\vec{v}_a$.}
Explicit dependence on the background curvature first appears
at 2PN order, where additional terms proportional to powers
of $\omA^2$ enter.

For $D=3$, the flat-space limit can be compared with the canonical centre-of-mass generator of general relativity~\cite{Damour:2000kk}. This provides a useful comparison of the allowed structures, although their interaction-dependent coefficients differ between scalar and gravitational interactions. The corresponding expressions in harmonic coordinates are given in~\cite{Blanchet:2002mb}.

A more direct comparison is provided by the scalar--tensor centre-of-mass integral computed through 3PN order in harmonic coordinates in~\cite{Bernard:2018ivi}. To recover the massless flat-space limit of our scalar model, we decouple tensor exchange while retaining a finite scalar interaction and a linear Einstein-frame worldline coupling. In the notation of~\cite{Bernard:2018ivi}, this gives $\alpha\widetilde G\to\gsc$, $\bar\gamma\to-2$ and $\bar\beta_1=\bar\beta_2\to-\tfrac12$. After a momentum--velocity conversion, the resulting generator reproduces all our curvature-independent coefficients through 2PN order at $D=3$, including the term quadratic in $\gsc$. This requires no further contact transformation.

\subsection{Symmetry check for the 2PN Hamiltonian}
Using the generators defined above, we can verify that the conformal algebra is realised on the phase space of the 2PN Hamiltonian,
\begin{subequations}\label{eq:ads-algebra}
\begin{align}
\left[\mathbb{P}_i,\mathbb{P}_j\right]_{\rm PB}
=&\ 
-\frac{\omA^2}{c^2}\ \mathbb{J}_{ij},
\qquad
\left[\mathbb{H},\mathbb{P}_i\right]_{\rm PB}
=\  
\omA^2\,\mathbb{G}_i,\label{eq:ads-algebra-a}
\\[3pt]
\left[\mathbb{J}_{ij},\mathbb{J}_{kl}\right]_{\rm PB}
=&\ 
\delta_{ik}\mathbb{J}_{jl}
+\delta_{jl}\mathbb{J}_{ik}
-\delta_{il}\mathbb{J}_{jk}
-\delta_{jk}\mathbb{J}_{il},\label{eq:ads-algebra-b}
\\[3pt]
\left[\mathbb{J}_{ij},\mathbb{P}_k\right]_{\rm PB}
=&\ 
\delta_{ik}\mathbb{P}_j
-\delta_{jk}\mathbb{P}_i,
\qquad
\left[\mathbb{J}_{ij},\mathbb{H}\right]_{\rm PB}
=\ 0,\label{eq:ads-algebra-c}
\\[3pt]
\left[\mathbb{G}_i,\mathbb{P}_j\right]_{\rm PB}
=&\ 
\frac{\delta_{ij}}{c^2}\mathbb{H},
\qquad
\left[\mathbb{G}_i,\mathbb{H}\right]_{\rm PB}
=\ 
\mathbb{P}_i,\label{eq:ads-algebra-d}
\\[3pt]
\left[\mathbb{G}_i,\mathbb{G}_j\right]_{\rm PB}
=&
-\frac{1}{c^2}\mathbb{J}_{ij},
\qquad
\left[\mathbb{J}_{ij},\mathbb{G}_k\right]_{\rm PB}
=\ 
\delta_{ik}\mathbb{G}_j
-\delta_{jk}\mathbb{G}_i.\label{eq:ads-algebra-e}
\end{align}
\end{subequations}
where $\left[\;,\;\right]_{\rm PB}$ denotes the Poisson Bracket in phase space. Eqns.\eqref{eq:ads-algebra-a} and \eqref{eq:ads-algebra-b} follow from the definitions in Eqns.\eqref{eq:total momentum AdS 3}, \eqref{eq:angular-momentum2} and \eqref{eq:ptotdot=-omasq G}. The right equation in \eqref{eq:ads-algebra-c} follows from the manifest rotational invariance of the PN Hamiltonian. The left equation in \eqref{eq:ads-algebra-c} and the right equation in \eqref{eq:ads-algebra-e} follow from the $\mathrm{SO}(D)$ vector nature of $\vec{\mathbb P}$ and $\vec{\mathbb G}$. The equations in \eqref{eq:ads-algebra-d} require verification using the explicit 2PN expressions. These provide a symmetry check on the PN Hamiltonian. We checked these in \emph{Mathematica} for the 2PN Hamiltonian using the expressions provided in \S\ref{app:2PN Hamiltonian and conserved charges}. Finally, the left equation in Eqn.\eqref{eq:ads-algebra-e} further follows from the Jacobi identity assuming the Poisson brackets checked already,
\begin{equation}\begin{split}
\left[\mathbb{G}_i,\mathbb{G}_j\right]_{\rm PB} &= \frac{1}{\omA^2} \left[\left[\mathbb{H},\mathbb{P}_i\right],\mathbb{G}_j\right]_{\rm PB} = \frac{1}{\omA^2} \left[\left[\mathbb{H},\mathbb{G}_j\right],\mathbb{P}_i\right]_{\rm PB} +\frac{1}{\omA^2} \left[\mathbb{H},\left[\mathbb{P}_i,\mathbb{G}_j\right]\right]_{\rm PB}\\
&= -\frac{1}{\omA^2} \left[\mathbb{P}_j,\mathbb{P}_i\right]_{\rm PB} -\frac{\delta_{ij}}{c^2\omA^2} \left[\mathbb{H},\mathbb{H}\right]_{\rm PB}= \frac{1}{\omA^2} \left[\mathbb{P}_i,\mathbb{P}_j\right]_{\rm PB}\\
&=
-\frac{1}{c^2}\ \mathbb{J}_{ij}\ .
\end{split}\end{equation}

This explicitly demonstrates that, within the PN expansion and after integrating out the field degrees of freedom, the inter-particle potential is consistent with the isometries of AdS. The isometry algebra, which is the conformal algebra, is therefore realised in phase space as a curvature-dependent deformation of the Poincare algebra. The Poincare algebra is recovered in the flat space limit $\omA\rightarrow 0$.

\emph{Even though we explicitly work in AdS, every equation in this section \S\ref{sec:Hamiltonian and conformal algebra} can be analytically continued to dS by replacing $\omA\to\compi\hds$.}

\section{Summary of results}

In this work, we have derived the effective action, equations of motion, Hamiltonian and other symmetry generators for two massive particles interacting through a free scalar field in a cosmological background, up to second post-Newtonian (2PN) order.

At the leading order, the Newtonian dynamics combines mutual attraction with the harmonic confinement of AdS. In dS, the background curvature instead produces a repulsive force that competes with the attraction between the particles. The 1PN and 2PN includes both the relativistic corrections to the kinetic terms as well as velocity-dependent corrections to the non-relativistic potential. We see that, apart from  the terms already present in flat limit, we get curvature dependent corrections. 

We carried out the calculation by computing the effective action for the particles (i.e., the Fokker effective action) by integrating out the field dynamics. We do this by a sequence of multipole expansion of the bulk Green's function, followed by a PN expansion and then resummation of the multipoles to obtain a PN expanded bulk-bulk Green's function, which we then use to determine the Fokker effective action. At 2PN, double-zero terms remove quadratic acceleration dependence, a total time derivative simplifies the linear terms, and a contact transformation removes the remaining accelerations. This gives then an ordinary Lagrangian with out any higher derivative terms which can be Legendre transformed to yield the Hamiltonian. We perform this reduction while preserving the required form of the redshifted total momentum $\vec{\mathbb P}$ in the final canonical variables. 

The $\mathfrak{so}(D,2)$ isometries of AdS strongly constrain the action and the Hamiltonian and provide non-trivial consistency checks on the curvature corrections. We construct all the $\mathfrak{so}(D,2)$ charges in the phase space  of the binary system and check that their Poisson brackets correctly reproduce the  $\mathfrak{so}(D,2)$ algebra till 2PN order.

\section{Future directions}

The framework developed here opens several directions for studying binary dynamics in curved spacetime and its relation to CFT data. This would supplement leading order results already available in literature  \cite{Fitzpatrick:2010zm,Cornalba:2007zb,Kulaxizi:2018dxo}(see \cite{Maxfield:2022nat,Zahraee:2024dsj} for an analysis of leading order non-relativistic limit). In forthcoming work, we will further explore the AdS two-body problem, using the Hamiltonian and symmetry generators to extend the connection between bulk dynamics and CFT observables.

The same Hamiltonian could be used to compute curvature corrections to flat-space scattering amplitudes in the non-relativistic regime and compare them with the AdS $S$-matrix constructions of Refs.~\cite{Gadde:2022ghy,Banerjee:2024yiq}. A first step would be to match the non-relativistic expansion of the scalar bulk-to-bulk propagators used in these works with our position-space PN Green's function. This would connect our results to momentum-space calculations and, through the dictionary of Ref.~\cite{Gadde:2022ghy}, to Mellin amplitudes.

Scalar self-interactions offer a starting point for treating bulk non-linearities, an essential ingredient of a gravitational extension. Including electromagnetic and gravitational forces would allow us to study curvature corrections to the binding and orbital motion of charged or gravitating binaries. The gravitational case would also provide a basis for addressing gravitational-wave observables within worldline effective field theory~\cite{Goldberger:2004jt,Porto:2016pyg}.

Including spin and finite-size effects would extend the point-particle description to more general compact objects~\cite{Porto:2005ac,Porto:2016pyg}. The massive-vector propagators and amplitudes of Ref.~\cite{Banerjee:2024yiq} offer useful comparisons for extensions to vector fields and spinning particles. In AdS, spinning binaries would also probe double-twist operators built from spinning primaries.

\section*{Acknowledgments}

This paper and its forthcoming companion are the culmination of three years of work, during which we have benefited from numerous discussions with colleagues. In particular, we would like to thank Saikat Bera, Simon Caron-Huot, Parashar Chatterjee, Nava Gaddam, Abhijit Gadde, Tamanna Jain, Apratim Kaviraj, Prayush Kumar, Alok Laddha, Godwin Martin, Rahul Metya, Shiraz Minwalla, Kaushik Paul, David Poland, Arnab Rudra, M. V. S. Saketh, Ashoke Sen, Omkar Shetye, Kaustubh Singhi, Jan Steinhoff, Farman Ullah, Spenta Wadia and many others for valuable discussions and insights. Various aspects of both works were presented at conferences, including CAPRA 28 (2025, online) and Strings 2026. We are grateful to the organisers of these conferences for providing opportunities to present our work, and to the participants for stimulating discussions and feedback. Finally, we acknowledge the support of the Department of Atomic Energy, Government of India, under Project No. RTI4019.

Wolfram Mathematica was used to compute and verify the post-Newtonian (PN) expressions presented in this work. Several supplementary calculations were also cross-checked with the assistance of Wolfram Mathematica and OpenAI's ChatGPT. ChatGPT was additionally used to assist with manuscript preparation, including writing, typesetting, and literature review. The authors have reviewed the resulting material and take full responsibility for the accuracy and content of the manuscript.

\appendix
\section{Coordinates and isometries of (A)dS :}\label{app:AdS}
In this appendix, we review various coordinates of (A)dS spacetime: much of what we do here is standard material appearing in one way or another in lecture notes on AdS/CFT \cite{Aharony:1999ti,DHoker:2002nbb,Kaplan:AdSCFT,Penedones:2016voo}. Our reason for reviewing such well-known material is twofold: first, we want to clearly set out the notations we use, given the mild variations of notation in the current literature. Second, and more importantly, we restore all factors of AdS frequency $\omA$ and speed of light $c$ so as to enable a clear discussion of flat/non-relativistic limit. We also wanted to emphasise certain aspects about isometries of AdS that are crucial to our work. We encourage the reader who is familiar with AdS/CFT to skim through this appendix mainly picking up the notation. For the reader who might be more familiar with flat space post-Newtonian literature, this section hopefully clarifies how  various structures familiar from flat space PN expansion, get deformed due to AdS curvature.

This section is arranged as follows: we start by reviewing AdS spacetime and its isometries, as described in three commonly used coordinate charts : global, embedding (\S\S\ref{ssec:AdSEmb}) and Poincare(\S\S\ref{ssec:AdSPoinc}). In each case, we construct the Killing vectors with special emphasis on their flat limits. We then repeat this exercise for dS static patch, discussing both embedding (\S\S\ref{ssec:dSEmb}) and Poincare(\S\S\ref{ssec:dSPoinc}) coordinates. 

Let us begin by recalling AdS$_{D+1}$ spacetime described by  the following metric in \emph{global coordinates}, 
\begin{equation}
ds^{2}=g_{\mu\nu}dx^{\mu}dx^{\nu}=-f\left(r\right)c^{2}dt^{2}+f\left(r\right)^{-1}dr^{2}+r^{2}d\Omega_{D-1}^{2}
\end{equation}
where $d\Omega_{D-1}^{2}$ is the metric on the unit sphere $\mathbb{S}^{D-1}$, and we have the redshift factor
\begin{equation}
f(r)=1+\frac{r^2\omA^2}{c^2}\ .
\end{equation}
The same  metric describes dS static patch under the analytic continuation $\omA\rightarrow \compi\,\hds$. Throughout this work, we use  cartesian-like  coordinates, where the above metric takes the form,
\begin{equation}
ds^2
=
-f(r)c^2dt^2
+
d\vec{x}^{\,2}
-
\frac{f(r)-1}{f(r)}
\frac{\left(\vec{r}\cdot d\vec{r}\right)^2}{r^2}.
\end{equation}
Here $\vec{r}\equiv \left\{x^i\right\}$ with the spatial index $i$ varying from $1$ to $D$. We also have $r\equiv\sqrt{\sum_i(x^{i})^2}$ and $\cdot$ is the spatial dot product.

\subsection{Embedding coordinates and AdS isometries}\label{ssec:AdSEmb}
The isometries of AdS are better discussed in another set of coordinates called 
\emph{embedding coordinates} $\{X^A\}$. They are defined by
\begin{equation}\label{eq:embXdef}
X^{\pm}\equiv \pm\compi\,\left[\dfrac{\omA}{c}\frac{e^{\pm\compi\,\omA t}}{\sqrt{f(r)}}\right]^{-1}\ ,\quad X^i\equiv x^i\ . 
\end{equation}
These are a set of $(D+2)$ coordinates for AdS$_{D+1}$ which obey the constraint
\begin{equation}\label{eq:Rd2Hyp}
X^+X^--\vec{X}^2=\frac{c^2}{\omA^2}\ , 
\end{equation}
as well as the reality conditions $(X^+)^\ast=X^-$ and $(X^i)^\ast=X^i$. Even after accounting for this constraint, given the time periodicity in the  definition above, the embedding coordinates only cover a single $\frac{2\pi}{\omA}$ time-window\footnote{This is a single time-window in \emph{coordinate} time, or alternately, time as seen by a comoving observer at the centre of AdS. But given that all far away physics appears blue-shifted or speeded up as seen from the centre, this corresponds to longer and longer  \emph{physical} time periods at larger and larger $r$.  } in AdS$_{D+1}$. Despite this limitation, the  embedding coordinates are still useful in exhibiting the isometries of AdS. The AdS$_{D+1}$ metric takes a simple form
\begin{equation}
ds^2
=
-dX^+dX^-
+
d\vec{X}^{\,2}\ ,
\end{equation}
which we recognise as the metric on $\mathbb{R}^{D,2}$, restricted to the hypersurface Eq.\eqref{eq:Rd2Hyp}. In other words, a single period of global AdS can be thought of as  a hyperboloid within $\mathbb{R}^{D,2}$ described by  Eq.\eqref{eq:Rd2Hyp}. 

Our index conventions for embedding coordinates are as follows: we will take  $A,B$ indices to vary over $(D+2)$ values $\{+,-,1,2,\ldots, D\}$ and use $\eta_{AB}$ to denote the  $\mathfrak{so}(D,2)$ invariant metric on $\mathbb{R}^{D,2}$. In our conventions, we have 
\begin{equation}\label{eq:sod2eta}
\eta_{+-}=-\frac{1}{2}\ ,\quad \eta_{ij}=\delta_{ij}\ , \quad\text{and}\quad \eta_{++}=\eta_{--}=\eta_{\pm i}=0\ . \end{equation}
We can use this metric to define the embedding coordinates with lower indices as
$X_{\pm}\equiv -\frac{1}{2}X^{\mp}$ and $X_i\equiv X^i$.
Here we are using a complex null basis to characterise the time-like plane in $\mathbb{R}^{D,2}$. Alternately, we can use two time-like directions $X^0,X^{(D+1)}$ defined via $X^\pm\equiv X^0\mp\compi\, X^{(D+1)}$, i.e., we have  \begin{equation}\label{eq:embXdefAlt}\begin{split}
X^{(D+1)}&\equiv-X_{(D+1)}\equiv -\cos(\omA t)\left[\dfrac{\omA}{c}\frac{1}{\sqrt{f(r)}}\right]^{-1}\ ,\\
X^0&\equiv -X_0\equiv \sin(\omA t)\left[\dfrac{\omA}{c}\frac{1}{\sqrt{f(r)}}\right]^{-1}\ . 
\end{split}\end{equation} 
The region near the origin $t,\frac{r}{c}\ll \frac{1}{\omA}$  corresponds then to the patch $X^0\sim ct$ and $X^{(D+1)}\sim -\frac{c}{\omA}$ of the hyperboloid in Eq.\eqref{eq:Rd2Hyp}. Using Eq.\eqref{eq:Rd2Hyp}, the function $f(r)$ can be rewritten in embedding coordinates as \begin{equation}\label{eq:fEmb}
f(r)=\frac{\omA^2}{c^2}X^+X^-=-2\frac{\omA^2}{c^2}X^+X_+=-2\frac{\omA^2}{c^2}X^-X_-\ . \end{equation}

We can
write  the $\mathfrak{so}(D,2)$ Killing vector fields on $\mathbb{R}^{D,2}$ in terms of embedding coordinates as
\begin{equation}\label{eq:AdSKilling-Emb}
\begin{split}
J_{AB}\equiv -\compi\left(X_A\frac{\partial}{\partial X^B}-X_B\frac{\partial}{\partial X^A}\right)\ .
\end{split}
\end{equation}
These Killing fields obey $\mathfrak{so}(D,2)$  algebra in the standard form, i.e., 
\begin{equation}
\left[J_{AB},J_{CD}\right]=\compi\left(\eta_{AC}J_{BD}+\eta_{BD}J_{AC}-\eta_{AD}J_{BC}-\eta_{BC}J_{AD}\right)\ ,
\end{equation}
where the LHS denotes the Lie-bracket between the Killing vector fields.
Since all these vector fields are tangent to the hyperboloid in Eq.\eqref{eq:Rd2Hyp}, they can be thought of as push-forwards of the corresponding Killing fields in AdS$_{D+1}$. 

To find these vector fields, we begin by writing down the push-forward 
of the coordinate basis vectors in global AdS as
 \begin{equation}
\begin{split}
&\frac{\partial}{\partial t} =2\omA J_{+-}\ ,\quad \frac{\partial}{\partial x^i} 
=2\compi\,\frac{\omA^2}{c^2f(r)}\left(X_{-}J_{+i}+X_{+}J_{-i}\right)\ .
\end{split}
\end{equation}
These push-forwards can be rewritten entirely in terms of embedding coordinates using Eq.\eqref{eq:fEmb}. 

The above push-forward formulae are not  unique : the vector fields $J_{+-},J_{+i}$ and $J_{-i}$ obey the relation $X_iJ_{+-}=X_-J_{+i}-X_+J_{-i}$ and are hence not linearly independent. The above form is hence unique up to this linear relation. With this in mind, we can solve for the Killing vector fields in global AdS by inverting the above equations : we get
\begin{equation}\label{eq:AdSKilling-glob}
\begin{split}
 J_{ij}
&\equiv -\compi\left(
x_i\frac{\partial}{\partial x^j}
-x_j\frac{\partial}{\partial x^i}
\right)\ ,\qquad
J_{+-}
\equiv \frac{1}{2\compi}J_{0(D+1)}\equiv \frac{1}{2\omA}\frac{\partial}{\partial t}\ ,\\ 
J_{\pm i}
&\equiv \frac{1}{2}J^{i\mp}\equiv \frac{1}{2}\left( J_{0i}\pm \compi\, J_{(D+1)i}\right)
\equiv  \frac{c}{2}\,\frac{e^{\pm\compi\,\omA t}}{\sqrt{f(r)}}
\left(\compi\,\frac{x_i}{c^2}\frac{\partial}{\partial t}\pm \frac{f(r)}{\omA} \frac{\partial}{\partial x^i}\right)\ .
\end{split}
\end{equation}
The Killing vector fields in the first line correspond to spatial rotations and 
time-translation, both of which are manifest isometries of the global AdS$_{D+1}$ metric. On the other hand, as we shall see later, the Killing vector fields in the second line represent the AdS analog of boosts/spatial translations. In fact, by looking at the flat limit, we can directly find the linear combinations of the above vectors that reduce to the translation/boost generators as $\omA\to 0$ : 
\begin{equation}\label{eq:AdSKilling-globII}
\begin{split}
H &\equiv -\omA J_{(D+1)0}\equiv\compi\frac{\partial}{\partial t}\ ,\\ 
P_{i}
&\equiv \frac{\omA}{c}J_{(D+1)i}\equiv-\frac{\compi}{\sqrt{f(r)}}
\left\{\cos(\omA\,t)\ f(r)\frac{\partial}{\partial x^i}-\omA\sin(\omA\,t)\ \frac{x_i}{c^2}\frac{\partial}{\partial t}\right\}\ ,\\
 G_{i}
&\equiv \frac{1}{c}J_{0i}\equiv \frac{\compi\,}{\sqrt{f(r)}}
\left\{\cos(\omA\,t)\ \frac{x_i}{c^2}\frac{\partial}{\partial t}+ f(r)\frac{\sin(\omA\,t)}{\omA} \frac{\partial}{\partial x^i}\right\}\ .
\end{split}
\end{equation}
These become the corresponding Killing fields of Minkowski spacetime  for $t,\frac{r}{c}\ll \frac{1}{\omA}$. The above equations show that the Lorentz transformations in $\mathbb{R}^{D,2}$ that keep $X^{(D+1)}$ invariant coincide with physical Lorentz transformations near the origin of AdS. The remaining $\mathfrak{so}(D,2)$ transformations which change $X^{(D+1)}$ act as physical spacetime translations near the origin of AdS. Further, we have $J^{i\mp}=c\,\Bigl\{G^i\pm \frac{\compi}{\omA} P^i\Bigr\}$ .

It is instructive to write out explicitly the Lie-Bracket of these Killing vector fields given in Eq.\eqref{eq:AdSKilling-Emb}, since it plays an important role when we discuss particle dynamics in AdS in \S\ref{sec:FPartAdS} . 

The translation algebra gets deformed in AdS due to curvature
\begin{equation}
\left[P_i,P_j\right]
=\ 
-\compi\frac{\omA^2}{c^2}\ J_{ij},
\qquad
\left[H,P_i\right]
=\compi\, 
\omA^2\,G_i,   
\end{equation}
whereas the algebra of Lorentz generators remains the same as in flat space,i.e.,
\begin{equation}\begin{split}
\left[J_{ij},J_{kl}\right]
=&\compi\left( 
\delta_{ik}J_{jl}
+\delta_{jl}J_{ik}
-\delta_{il}J_{jk}
-\delta_{jk}J_{il}\right),
\\
\left[G_i,G_j\right]
=&
-\frac{\compi}{c^2}J_{ij},
\qquad
\left[J_{ij},G_k\right]
=\compi\left( 
\delta_{ik}G_j
-\delta_{jk}G_i\right)\ .  
\end{split}\end{equation}
The Lie-brackets between translation and Lorentz generators is also undeformed from the flat space
\begin{equation}\begin{split}
\left[J_{ij},P_k\right]
=&\compi\left( 
\delta_{ik}P_j
-\delta_{jk}P_i\right),
\qquad
\left[J_{ij},H\right]
=\ 0,
\\
\left[G_i,P_j\right]
=&\compi 
\frac{\delta_{ij}}{c^2}H,
\qquad
\left[G_i,H\right]
=\compi\, 
P_i.
\end{split}\end{equation}

The significance of these Killing vector fields and their Lie-brackets lies in their relation to conserved charges. Let us recall how this works. Consider a physical system in AdS, whether it is a collection of particles or fields or a particle-field hybrid, that respects the isometries generated by the AdS Killing fields. Then corresponding to every Killing field $\xi$ there is a Noether charge $\mathbb{Q}_\xi$ in the phase space of the system which implements that isometry on the phase space. What this means is that on a phase space function/observeable  $\mathbb{O}$, the isometry acts as an infinitesimal canonical transformation implemented via a Poisson bracket
\begin{equation}
\mathbb{O}\mapsto \mathbb{O}+
[\mathbb{O},\mathbb{Q}_{\xi}]_{\rm PB}+\text{O}(\xi^2)\ .
\end{equation}
If the observable $\mathbb{O}$ is a tensor field, by definition, this change is related to Lie derivative with respect to $\xi$, i.e.,
\begin{equation}
[\mathbb{O},\mathbb{Q}_{\xi}]_{\rm PB}=\frac{1}{\compi} \mathscr{L}_\xi \mathbb{O}\ .
\end{equation}
This means that, in a field theory, we can write the Noether charge associated with an isometry in terms of the energy-momentum tensor $T^{\mu\nu}$ of the field theory
\begin{equation}
\begin{split}
\mathbb{Q}_{\xi}=\compi\,\int_{\Sigma_t}T^{0}{}_{\mu}\xi^{\mu}\ ,
\end{split}
\label{eq:noether-charge}
\end{equation}
where the integral is over a time-slice with standard spatial measure. Another implication is that the Poisson bracket of charges is itself related to Lie-bracket between the Killing vector fields via
\begin{equation}
[\mathbb{Q}_{\xi_1},\mathbb{Q}_{\xi_2}]_{\rm PB}=\frac{1}{\compi} \mathbb{Q}_{[\xi_1,\xi_2]}  
\end{equation}
We see an example of this in free particles described in \S\ref{sec:FPartAdS} of the main text. We will see another example in the free scalar field described in appendix~\ref{app:scalar-ads-ds}.

\subsection{Poincare coordinates for AdS}\label{ssec:AdSPoinc}
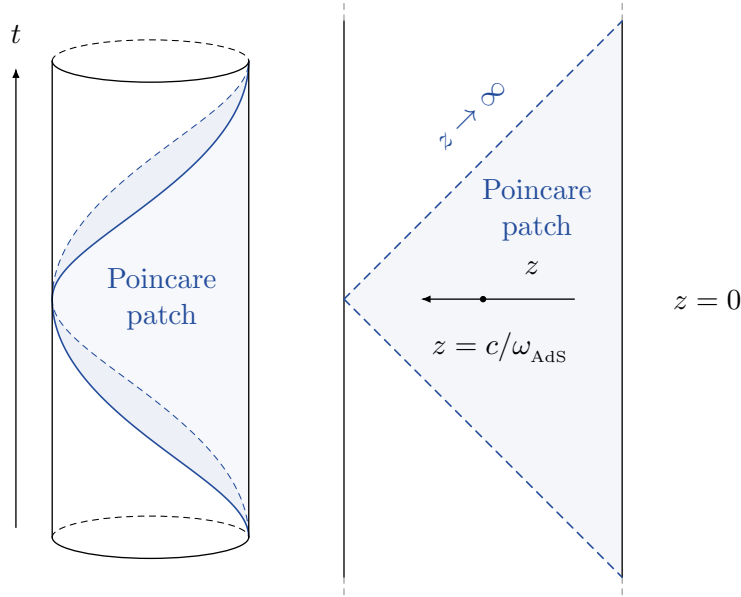
\begin{figure}
    \centering
    \begin{tikzpicture}[x=1cm,y=1cm,line cap=round,line join=round]
  \begin{scope}
    \fill[figblue!4]
      plot[domain=0:180,samples=120,variable=\q]
        ({-1.3*cos(\q)},{3.15*\q/180+0.28*sin(\q)})
      -- (1.3,-3.15)
      -- plot[domain=180:0,samples=120,variable=\q]
        ({-1.3*cos(\q)},{-3.15*\q/180-0.28*sin(\q)})
      -- cycle;
    \foreach \sign in {-1,1} {
      \fill[figblue!8]
        plot[domain=0:180,samples=120,variable=\q]
          ({-1.3*cos(\q)},
           {\sign*3.15*\q/180-0.28*sin(\q)})
        -- plot[domain=180:0,samples=120,variable=\q]
          ({-1.3*cos(\q)},
           {\sign*3.15*\q/180+0.28*sin(\q)})
        -- cycle;
      \draw[figblue,fig hidden]
        plot[domain=0:180,samples=120,variable=\q]
          ({-1.3*cos(\q)},
           {\sign*3.15*\q/180+0.28*sin(\q)});
      \draw[figblue,line width=0.6pt]
        plot[domain=0:180,samples=120,variable=\q]
          ({-1.3*cos(\q)},
           {\sign*3.15*\q/180-0.28*sin(\q)});
    }
    \foreach \height in {-3.15,3.15} {
      \draw[fig boundary,fig hidden]
        (-1.3,\height) arc[start angle=180,end angle=0,
          x radius=1.3,y radius=0.28];
      \draw[fig boundary]
        (-1.3,\height) arc[start angle=180,end angle=360,
          x radius=1.3,y radius=0.28];
    }
    \draw[fig boundary] (-1.3,-3.15) -- (-1.3,3.15);
    \draw[fig boundary] ( 1.3,-3.15) -- ( 1.3,3.15);
    \draw[fig time arrow] (-1.78,-3.02) -- (-1.78,3.06);
    \node[fig label,above=6pt] at (-1.78,3.06) {$t$};
    \node[fig label,align=center,text=figblue] at (0.15,0)
      {Poincare\\patch};
  \end{scope}

  \begin{scope}[xshift=4.4cm,scale=1.15]
    \fill[figblue!5] (-1.6,0) -- (1.6,3.2) -- (1.6,-3.2) -- cycle;
    \foreach \side in {-1.6,1.6} {
      \draw[fig boundary] (\side,-3.2) -- (\side,3.2);
      \draw[black!45,fig hidden] (\side,-3.2) -- (\side,-3.43);
      \draw[black!45,fig hidden] (\side, 3.2) -- (\side, 3.43);
    }
    \draw[figblue,fig horizon] (-1.6,0) -- (1.6, 3.2);
    \draw[figblue,fig horizon] (-1.6,0) -- (1.6,-3.2);
    \path (-1.6,0) -- node[fig label,pos=0.56,sloped,above=9pt,
      text=figblue] {$z\to\infty$} (1.6,3.2);
    \node[fig label,anchor=west] at (2.08,0) {$z=0$};
    \node[fig label,align=center,text=figblue] at (0.62,1.03)
      {Poincare\\patch};

    \draw[fig time arrow] (1.05,0) -- (-0.72,0);
    \node[fig label,above=6pt] at (0.56,0) {$z$};
    \fill (0,0) circle[radius=1.05pt];
    \node[fig label,below=9pt] at (0.2,0) {$z=c/\omA$};
  \end{scope}
\end{tikzpicture}
    \caption{AdS Poincare patch}
    \label{fig: AdS Poincare patch}
\end{figure}
A third set of planar coordinates called \emph{Poincare coordinates} is often useful to see how the isometries of AdS become conformal isometries of the boundary.\footnote{For this reason, it is a very useful way to describe the dual conformal field theory(CFT) in the context of AdS/CFT correspondence.} The Poincare coordinates, like embedding coordinates, only cover a single period in coordinate time. But, in addition, we zoom in near the north (i.e. positive $x^{(D)}$ direction) and adopt planar/stereographic coordinates with north pole as the centre.\footnote{i.e., we consider a stereographic projection through south pole of the sphere $\mathbb{S}^{D-1}$.} As we shall see, this choice of direction  breaks the manifest  $\mathfrak{so}(D)$ invariance, but makes a lower dimensional Poincare invariance $\mathfrak{iso}(D-1,1)$ manifest. Since, by AdS/CFT correspondence, this is indeed how the Poincare invariance of the dual CFT emerges, we will call this the `CFT Poincare invariance'.

Since we are breaking the full $SO(D)$ invariance by choosing south as a special direction, it is useful to introduce a transverse spatial index $\bar{i}$ that varies from $1$ to $(D-1)$, describing spatial directions perpendicular to the north-south axis (i.e.  $x^{(D)}$ axis). We will then introduce the coordinates
\begin{equation}\begin{split}
y^0&\equiv\frac{c}{\omA}\frac{\sqrt{f(r)}\,\sin(\omA t)}{\dfrac{\omA}{c}x^{(D)}+\sqrt{f(r)}\,\cos(\omA t)}=\frac{c}{\omA}\frac{X^0}{X^{(D)}-X^{(D+1)}}\ ,\\
y^{\bar{i}}&\equiv\frac{x^{\bar{i}}}{\dfrac{\omA}{c}x^{(D)}+\sqrt{f(r)}\,\cos(\omA t)}=\frac{c}{\omA}\frac{X^{\bar{i}}}{X^{(D)}-X^{(D+1)}}\ ,\\
z&\equiv \frac{c}{\omA}\frac{1}{\dfrac{\omA}{c}x^{(D)}+\sqrt{f(r)}\,\cos(\omA t)}=\left(\frac{c}{\omA}\right)^2\frac{1}{X^{(D)}-X^{(D+1)}}\ .
 \end{split}\end{equation}
A more succinct and convenient presentation is
 \begin{equation}\begin{split}
y^{\bar{\mu}}&=\frac{\omA}{c}\,z\,X^{\bar{\mu}}\ ,\quad \left(\frac{\omA}{c}\right)^2\,z=\frac{1}{X^{(D)}-X^{(D+1)}}\ ,
\end{split}\end{equation}
where the transverse spacetime index $\bar{\mu}$ varies from $0,1$ to $(D-1)$. We will call these the CFT indices.
We will also use $\eta_{\bar\mu\bar\nu}$ to denote the induced Minkowski metric over these indices : this is indeed the metric inherited from the embedding space.  

The region of validity of the Poincare coordinates is given by $z\geq 0$.  This condition can be thought of as giving for every spatial point, a finite coordinate time interval\footnote{This is a subset of the AdS period $\left(-\frac{\pi}{\omA},\frac{\pi}{\omA}\right)$} within which the Poincare coordinates are valid. The following limiting cases are useful to note:
\begin{itemize}
\item In the region $r,ct\ll\frac{c}{\omA}$ near the origin of global AdS, the above maps can be approximated as  $y^{\bar{\mu}}\approx x^{\bar{\mu}}$ and $z\approx \frac{c}{\omA}-x^{(D)}$. In other words, Poincare coordinates are effectively cartesian coordinates for the  near-origin region\footnote{With the caveat that $z$ has a huge shift of $\frac{c}{\omA}$ at the origin and is oriented oppositely to $x^{(D)}$}, and hence cover this region pretty well.
\item If we take $x^{(D)}\gg \frac{c}{\omA}$ to  go far north of origin, Poincare coordinates are valid essentially\footnote{This statement is more and more true as we go farther and farther north.} over a full AdS period $\left(-\frac{\pi}{\omA},\frac{\pi}{\omA}\right)$. In contrast, if we take $x^{(D)}\ll -\frac{c}{\omA}$ to go  far south, Poincare coordinates are valid for smaller and smaller time periods. Thus, once we exit the near-origin region, the Poincare coordinate system is mainly suited to study dynamics confined to the northern region.
\end{itemize}
Another interesting limiting case is to take $z\to 0^+$ at fixed $y^{\bar{\mu}}$ : this corresponds to taking $r\to\infty$ in global coordinates at a fixed $t$ within the AdS period, but also keeping 
$\cos\vartheta_{(D)}\equiv \dfrac{x^{(D)}}{r}$ fixed to an angle $\vartheta_{(D)}<\omA\,t$. This focuses onto a flat patch near the north pole.
 
Let us now discuss the isometries. The Lorentz transformations that transform CFT indices preserving $\eta_{\bar\mu\bar\nu}$ shall be dubbed `CFT Lorentz transformations'. Using the above presentation in embedding coordinates, we see that the 
CFT Lorentz transformations are isometries of AdS. The AdS metric in Poincare coordinates takes the form 
\begin{equation}
ds^{2}=\left(\frac{\omA}{c}\right)^{-2}\,\frac{dz^{2}+\eta_{\bar\mu\bar\nu}dy^{\bar\mu}dy^{\bar\nu}}{z^{2}}\ ,
\end{equation}
which makes it clear that translations of $y^{\bar\mu}$ (`CFT translations') are also isometries. Apart from the CFT Poincare transformations, the above metric is manifestly invariant under scaling $y^{\bar\mu}\mapsto \lambda\ y^{\bar\mu}$ and $z\to \lambda z$. The global AdS metric is also invariant under the north-south reflection $x^{(D)}\mapsto -x^{(D)}$, which, in Poincare coordinates, is   the conformal inversion
\begin{equation}
z\mapsto \frac{z}{z^2+\eta_{\bar\mu\bar\nu}y^{\bar\mu}y^{\bar\nu}}\ ,\quad  y^{\bar\lambda}\mapsto \frac{y^{\bar\lambda}}{z^2+\eta_{\bar\mu\bar\nu}y^{\bar\mu}y^{\bar\nu}}\ .
\end{equation}
This, in turn, implies that special conformal translations also extend to isometries of AdS. To conclude, all conformal transformations of CFT indices extend to full isometries in AdS. This can be made more explicit by writing down the Killing vector fields of AdS in the Poincare coordinates:
\begin{equation}\begin{split}
J_{\bar{\mu}\bar{\nu}}
&=-\compi\left(
y_{\bar{\mu}}\frac{\partial}{\partial y^{\bar{\nu}}}
-y_{\bar{\nu}}\frac{\partial}{\partial y^{\bar{\mu}}}
\right)\ ,\\
P_{(D)}
&\equiv \frac{\omA}{c}J_{(D+1)D}=\frac{\compi\,\omA}{c}\left(
y^{\bar\mu}\frac{\partial}{\partial y^{\bar\mu}}
+z\frac{\partial}{\partial z}
\right)\ ,\\
P_{\bar\mu}+\frac{\omA}{c}J_{D\bar\mu}&\equiv \frac{\omA}{c}\left(J_{(D+1)\bar\mu}+J_{D\bar\mu}\right)=-\compi\,\frac{\partial}{\partial y^{\bar{\mu}}}\ ,\\
P_{\bar\mu}-\frac{\omA}{c}J_{D\bar\mu}&\equiv \frac{\omA}{c}\left(J_{(D+1)\bar\mu}-J_{D\bar\mu}\right)\\
&=-\frac{\compi\,\omA^2}{c^2}\left[
\left(y^{\bar\nu} y_{\bar\nu}+z^2\right)
\frac{\partial}{\partial y^{\bar\mu}}-2y_{\bar\mu}\left(
y^{\bar\nu}\frac{\partial}{\partial y^{\bar\nu}}
+z\frac{\partial}{\partial z}
\right)
\right]\ .
\end{split}\end{equation}
Here we are using the same convention as before, of labelling Killing fields both by their flat limits near origin, as well as by their embedding space description. The last two equations can also be rewritten as
\begin{equation}\begin{split}
P_{\bar\mu}&= \frac{\omA}{c}J_{(D+1)\bar\mu}=-\frac{\compi}{2}\,\frac{\partial}{\partial y^{\bar{\mu}}}-\frac{\compi\,\omA^2}{2c^2}\left[
\left(y^{\bar\nu} y_{\bar\nu}+z^2\right)
\frac{\partial}{\partial y^{\bar\mu}}-2y_{\bar\mu}\left(
y^{\bar\nu}\frac{\partial}{\partial y^{\bar\nu}}
+z\frac{\partial}{\partial z}
\right)
\right]\ ,\\
J_{D\bar\mu}&= -\frac{\compi c}{2\omA}\,\frac{\partial}{\partial y^{\bar{\mu}}}+\frac{\compi\,\omA}{2c}\left[
\left(y^{\bar\nu} y_{\bar\nu}+z^2\right)
\frac{\partial}{\partial y^{\bar\mu}}-2y_{\bar\mu}\left(
y^{\bar\nu}\frac{\partial}{\partial y^{\bar\nu}}
+z\frac{\partial}{\partial z}
\right)
\right]\ .
\end{split}\end{equation}

The correctness of our labelling can be checked by taking the flat limit : this can be done by approximating $y^{\bar{\mu}}\approx x^{\bar{\mu}}$ and $z\approx \frac{c}{\omA}-x^{(D)}$ and then taking $\omA\to 0$ at fixed $x^{\bar{\mu}}$ and $x^{(D)}$. The reader should be able to check that this procedure does land us on the correct set of $\mathfrak{iso}(D,1)$ Killing fields for the Minkowski spacetime $\mathbb{R}^{D,1}$.

The above Killing vectors can also be studied near the northern boundary patch by taking $z\to 0^+$ at fixed $y^{\bar{\mu}}$. In this limit, we get the \emph{conformal} Killing fields of a lower dimensional Minkowski spacetime  $\mathbb{R}^{D-1,1}$ which satisfy $\mathfrak{so}(D,2)$ algebra . The above expressions at finite $z$ interpolate between these two limits. We see from the above formulae that
\begin{itemize}
\item CFT Lorentz transformations are realised as transverse\footnote{We are using the word `transverse' and `longitudinal' to distinguish between transformations that do not change $x^{(D)}$ and those that do.} Lorentz transformations near the centre of AdS.
\item Ordinary translations/special conformal translations of the CFT are realised as a mixture of transverse translations and `longitudinal' Lorentz transformations near the centre of AdS. This mixing is necessary since the CFT translations do not see the curvature whereas AdS translations do. In other words, we have the Lie brackets  $[P_{\bar{\mu}},P_{\bar{\nu}}]=-\compi\frac{\omA^2}{c^2}J_{\bar\mu\bar\nu}$ but as expected out of ordinary/conformal translations we have the Lie brackets
\begin{equation}\begin{split}
\left[P_{\bar\mu}+\frac{\omA}{c}J_{D\bar\mu}\ ,\ P_{\bar\nu}+\frac{\omA}{c}J_{D\bar\nu}\right]&=\left[P_{\bar\mu}-\frac{\omA}{c}J_{D\bar\mu}\ ,\ P_{\bar\nu}-\frac{\omA}{c}J_{D\bar\nu}\right]=0\ .
\end{split}\end{equation}
\item The scale transformations (i.e., dilatations) of the CFT are realised as longitudinal translations, i.e, translations in the $x^{(D)}$ direction. 
\end{itemize}
These statements form a crucial element of the AdS/CFT correspondence where the asymptotic data near the conformal boundary can be mapped to the corresponding CFT data.

\subsection{Embedding coordinates for dS}\label{ssec:dSEmb}
Next we move to dS. We will begin by discussing how embedding coordinates work in the dS case. All formulae go through if we perform the analytic continuation
\begin{equation}
\omA\rightarrow \compi\,\hds\ ,\quad X^{(D+1)}\rightarrow \compi\,X^{(D+1)}_{\rm dS}\ , 
\end{equation}
with the consequence that the $(D+1)$ direction becomes a space-like direction, and the isometries of dS$_{D+1}$ form a $\mathfrak{so}(D+1,1)$ algebra.
We then have the following definitions for the embedding coordinates
\begin{equation}\label{eq:embXdefdS}\begin{split}
X^{\pm}_{\rm dS}&\equiv \pm e^{\pm\hds t}\left[\dfrac{\hds}{c}\frac{1}{\sqrt{f(r)}}\right]^{-1}\ ,\quad X^i_{\rm dS}\equiv x^i\ ,\\ 
X^{(D+1)}_{\rm dS}&\equiv \cosh(\hds t)\left[\dfrac{\hds}{c}\frac{1}{\sqrt{f(r)}}\right]^{-1}\ ,\quad  X^0_{\rm dS}\equiv \sinh(\hds t)\left[\dfrac{\hds}{c}\frac{1}{\sqrt{f(r)}}\right]^{-1}\ ,
\end{split}\end{equation} 
with $X^\pm_{\rm dS}\equiv X^0_{\rm dS}\pm X^{(D+1)}_{\rm dS}$. These are a set of $(D+2)$ coordinates for dS$_{D+1}$ which embed it as a  hyperboloid within $\mathbb{R}^{D+1,1}$ via the constraint
\begin{equation}\label{eq:Rd2HypdS}
X^+_{\rm dS}X^-_{\rm dS}-\vec{X}^2_{\rm dS}=(X^0_{\rm dS})^2-(X^{D+1}_{\rm dS})^2-\vec{X}^2_{\rm dS}=-\frac{c^2}{\hds^2}\ .
\end{equation}
Thus, the embedding space metric is still Eq.\eqref{eq:sod2eta}, but with the crucial difference that $X^\pm_{\rm dS}$ are not complex conjugates of each other. Unlike the case of embedding coordinates in AdS, the embedding coordinates in dS cover more of dS (in fact all of dS) compared to static patch coordinates.

The  Killing vectors of dS are  given by analytically continuing our AdS answers:
\begin{equation}\label{eq:dSKilling-glob}
\begin{split}
J_{ij}
&\equiv -\compi\left(
x_i\frac{\partial}{\partial x^j}
-x_j\frac{\partial}{\partial x^i}
\right)\ ,\qquad
J_{-+}
\equiv \frac{1}{2}J_{0(D+1)}\equiv \frac{\compi}{2\hds}\frac{\partial}{\partial t}\ ,\\ 
J_{\pm i}
&\equiv \frac{1}{2}J^{i\mp}\equiv \frac{1}{2}\left( J_{0i}\pm J_{(D+1)i}\right)
\equiv  \compi\,\frac{c}{2}\,\frac{e^{\mp\hds t}}{\sqrt{f(r)}}
\left(\frac{x_i}{c^2}\frac{\partial}{\partial t}\mp\frac{f(r)}{\hds} \frac{\partial}{\partial x^i}\right)\ .
\end{split}
\end{equation}
The Killing vector combinations which reduce to standard translations/boosts are
\begin{equation}
\begin{split}
H &\equiv -\hds J_{(D+1)0}\equiv\compi\frac{\partial}{\partial t}\ ,\\
P_{i}
&\equiv \frac{\hds}{c}J_{(D+1)i}\equiv-\frac{\compi}{\sqrt{f(r)}}
\left\{\cosh(\hds\,t)\ f(r)\frac{\partial}{\partial x^i}+\hds\sinh(\hds\,t)\ \frac{x_i}{c^2}\frac{\partial}{\partial t}\right\}\ ,\\
 G_{i}
&\equiv \frac{1}{c}J_{0i}\equiv \frac{\compi\,}{\sqrt{f(r)}}
\left\{\cosh(\hds\,t)\ \frac{x_i}{c^2}\frac{\partial}{\partial t}+f(r)\frac{\sinh(\hds\,t)}{\hds} \frac{\partial}{\partial x^i}\right\}\ .
\end{split}
\end{equation}
We have $J^{i\mp}=c\,\Bigl\{G^i\pm \frac{1}{\hds} P^i\Bigr\}$.
The Lie brackets of these Killing vector fields can also be read off from those of AdS by a simple analytic continuation.

\subsection{Poincare coordinates for dS}\label{ssec:dSPoinc}
\begin{figure}
    \centering
    \begin{tikzpicture}[x=1cm,y=1cm,line cap=round,line join=round]
  \fill[figblue!5] (0,-3) -- (0,3) -- (6,3) -- cycle;
  \draw[fig boundary] (0,-3) rectangle (6,3);

  \draw[figred,fig horizon] (0,-3) -- (3,0);
  \draw[figred,fig horizon] (0, 3) -- (3,0);
  \draw[figblue,fig horizon] (3,0) -- (6,3);

  \node[fig label,anchor=east,align=right] at (-0.48,0)
    {observer\\[2pt]$r=0$};
  \node[fig label,align=center,text=figred] at (1.12,-0.08)
    {static\\patch};
  \path (0,3) -- node[fig label,pos=0.50,sloped,above=9pt,
    text=figred] {$r=\nicefrac{c}{\hds}$} (3,0);
  \node[fig label,align=center,text=figblue] at (3.5,2.3)
    {Poincare\\patch};
  \node[fig label,text=figblue] at (3.5,1.5) {$\eta<0$};
  \node[fig label,text=figblue,above=12pt] at (3,3) {$\eta=0$};
\end{tikzpicture}
    \caption{dS Poincare patch}
    \label{fig:dS Poincare patch}
\end{figure}
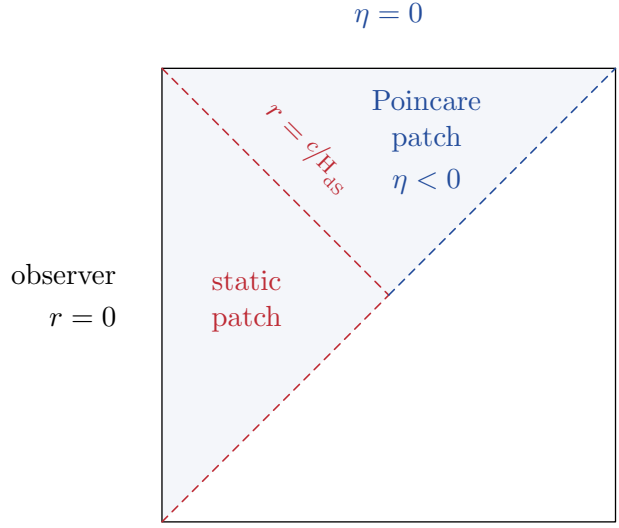
As we did for AdS, we can also introduce a planar set of coordinates for de-Sitter spacetime. There is, however, one crucial difference : in AdS, we singled out one of the spatial directions in embedding space, whereas in dS, we single out the time direction.\footnote{Alternately, the null combination $X^{(D)}-X^{(D+1)}$ in AdS is instead replaced by $X^0_{\rm dS}+X^{(D+1)}_{\rm dS}$.} The resulting constant-time slices are $D$-dimensional Euclidean spaces and the manifest symmetry subgroup is $\mathfrak{iso}(D)$. We thus define
a set of new coordinates
\begin{equation}
\begin{split}
y^i
&\equiv 
\frac{e^{-\hds t}}{\sqrt{f(r)}}\,x^i=\frac{c}{\hds}
\frac{X^i_{\rm dS}}
     {X^0_{\rm dS}+X^{(D+1)}_{\rm dS}}\ ,\\
\eta 
&\equiv -\frac{c}{\hds}
\frac{e^{-\hds t}}{\sqrt{f(r)}}=-\left(\frac{c}{\hds}\right)^2
\frac{1}
     {X^0_{\rm dS}+X^{(D+1)}_{\rm dS}}  ,
\end{split}
\label{eq:dSPoincareStatic}
\end{equation}
in which the dS metric takes the form 
\begin{equation}
\begin{split}
ds^2&=
\left(\frac{\hds}{c}\right)^{-2}
\frac{- d\eta ^2+\delta_{ij} dy^id y^j}{\eta ^2} .
\end{split}
\label{eq:dSPoincareMetric}
\end{equation}
The coordinate $\eta<0$ is usually termed \emph{conformal time}, and $\eta\to 0^-$ is the future time-like infinity, i.e., future space-like conformal boundary of dS. Poincare coordinates break down before we reach past time-like boundary of dS (somewhat analogous to the breakdown near southern regions in AdS Poincare), 

Our discussion about the flat limit
in AdS Poincare coordinates can be adapted to dS with some minimal changes. Given the similarities, we will necessarily be brief. The neighbourhood of the centre of the static patch, i.e., the region defined by $\hds t\ll1, \dfrac{\hds r}{c}\ll1 $ is mapped to
\begin{equation}
\begin{split}
y^i
&\simeq x^i,\quad
\eta 
\simeq-\frac{c}{\hds}+ct ,
\end{split}
\label{eq:dSPoincareFlatNearOrigin}
\end{equation}
i.e., apart from the large constant displacement $-c/\hds$ of the Poincare time coordinate, the variables $(\eta ,\vec y)$ reduce locally to ordinary Cartesian Minkowski coordinates. The flat limit is obtained by keeping
$\eta +\dfrac{c}{\hds}$ and $y^i
$ fixed as $\hds\to 0$. 

The Killing vector fields of dS are given by 
\begin{equation}
\begin{split}
J^{ij}
&=-\compi\left(
 y^i\frac{\partial}{\partial y^j}
-y^j\frac{\partial}{\partial y^i}
\right)\ ,\\
H
&\equiv -\hds J_{(D+1)0}\equiv\compi\frac{\partial}{\partial t}=-\compi\hds\left(
 y^i\frac{\partial}{\partial y^i}
+\eta \frac{\partial}{\partial \eta }
\right)\ ,\\
P_i-\hds G_i
&=-\frac{\hds}{c}J^{i+}=-\frac{\compi\, e^{\hds t}}{\sqrt{f(r)}}
\left[
 f(r)\frac{\partial}{\partial x^i}
 +\frac{\hds x_i}{c^2}\frac{\partial}{\partial t}
\right]=-\compi\frac{\partial}{\partial y^i},\\
P_i+\hds G_i
&=\frac{\hds}{c}J^{i-}=-\frac{\compi\, e^{-\hds t}}{\sqrt{f(r)}}
\left[
 f(r)\frac{\partial}{\partial x^i}
 -\frac{\hds x_i}{c^2}\frac{\partial}{\partial t}
\right]\\
&=\compi\frac{\hds^2}{c^2}
\bigg[
\left(y^2-\eta ^2\right)
\frac{\partial}{\partial y^i}
-2y_i\left(
 y^j\frac{\partial}{\partial y^j}
+\eta \frac{\partial}{\partial \eta }
\right)
\bigg].
\end{split}
\end{equation}
The last two equations can be inverted to yield 
\begin{equation}
\begin{split}
P_i
&=-\frac{\compi}{2}\frac{\partial}{\partial y^i}+\frac{\compi\hds^2}{2c^2}
\bigg[
\left(y^2-\eta ^2\right)
\frac{\partial}{\partial y^i}
-2y_i\left(
 y^j\frac{\partial}{\partial y^j}
+\eta \frac{\partial}{\partial \eta }
\right)
\bigg],\\
G_i
&=\frac{\compi}{2\hds }\frac{\partial}{\partial y^i}+\frac{\compi\hds}{2c^2}
\bigg[
\left(y^2-\eta ^2\right)
\frac{\partial}{\partial y^i}
-2y_i\left(
 y^j\frac{\partial}{\partial y^j}
+\eta \frac{\partial}{\partial \eta }
\right)
\bigg].
\end{split}
\end{equation}
Like in AdS, the dS Killing fields behave like Poincare Killing fields near the centre of the static patch (i.e., $\eta 
\simeq-\dfrac{c}{\hds}$), and they behave as conformal Killing fields of $\mathbb{R}^D$ near future time-like infinity ($\eta\to 0^-$). The time translations near the centre extrapolate to dilatations near the future boundary, whereas the ordinary/special-conformal translations of the future boundary are obtained by a mixture of translations and boosts around the centre.  In particular, $J_{ij}$ generates ordinary Euclidean rotations, $P_i-\hds G_i$ generates Euclidean translations, $H$ generates dilatations, and $P_i+\hds G_i$ generates special conformal transformations. Note that the two sets of generators separately commute,
\begin{equation}
\begin{split}
\left[
P_i-\hds G_i,
P_j-\hds G_j
\right]
&=0,
\quad
\left[
P_i+\hds G_i,
P_j+\hds G_j
\right]
=0.
\end{split}
\label{eq:dSPoincareCommuting}
\end{equation}
Together with $J_{ij}$ and $H$, these generators then give the conformal presentation of the full $\mathfrak{so}(D+1,1)$ isometry algebra.

\section{A centroid for AdS particles}\label{app:centroid}
In this appendix, we would like to prove that, given an isolated\footnote{By `isolated', we mean that particles are not flying in from or flying out to the boundary of AdS. We will also assume that there are no sourcs at the boundary giving external forcing, i.e., all non-normalisable modes are turned off.  } collection of heavy\footnote{We will assume that the total invariant mass of this configuration (which we define precisely using Casimir invariants later on) is large. On the dual CFT side, this is equivalent to assuming that this multi-particle state is a descendant of a generic heavy primary whose family has no extra null descendants.} interacting particles in AdS, there is a natural centroid which we can associate with that collection various desirable properties. The centroid only depends on the net $\mathfrak{so}(D,2)$ charges of the collection, and it moves along a time-like geodesic in AdS. As we shall argue,  the centroid we define moves like a massive spinning particle in AdS, and can be brought to rest at the origin of AdS by an AdS isometry.

This appendix is organised as follows : we begin in \S\S\ref{ssec:finIso} by constructing finite versions of AdS isometries described in the main text. This is followed by \S\S\ref{ssec:spinning} where we use these finite isometries to construct the space of states of a massive spinning particle, and the $\mathfrak{so}(D,2)$ charges associated with these states. Finally, in \S\S\ref{ssec:TDcen}, we describe how this logic can be flipped and we can instead start from the $\mathfrak{so}(D,2)$ charges and then construct a centroid from these charges which behaves like a massive spinning particle. While our discussion is focused on AdS, all our formulae can be adopted to dS static patch via the analytic continuation
\begin{equation}
\omA\rightarrow \compi\,\hds\ ,\quad X^{(D+1)}_{\rm dS}\rightarrow \compi\,X^{(D+1)}_{\rm dS}\ . 
\end{equation}
The construction described in here would then assign a centroid to any collection of particles in dS bound by interactions, and such a centroid would move along a dS geodesic.\footnote{This is especially easy to see from the point of view of global dS.} Some aspects of this is discussed in \S\S\ref{ssec:dSCent}.

The choice of centroid we describe here  is closely related to the usual
Tulczyjew--Dixon (TD) definition of the relativistic centre of mass.  In the
multipole description of collection of particles, a \emph{spin supplementary condition}
selects a representative centroid about which the multipole moments are defined; the TD choice is characterised by the orthogonality of the spacetime spin
tensor to the spacetime momentum.\footnote{The condition was introduced in the multipole description of relativistic
extended bodies by Tulczyjew \cite{Tulczyjew1959}, and was subsequently
developed systematically in Dixon's covariant multipole formalism
\cite{Dixon:1964cjb,Dixon:1970zza}. In this language, a spin supplementary
condition selects the representative worldline about which the total
angular momentum is decomposed into orbital and intrinsic parts; see
\cite{Costa:2014nta} for a detailed discussion of the different
choices of centroid and their physical interpretation.}  The condition
$\Pi_A S^{AB}=0$ appearing below is the corresponding condition in the
$SO(D,2)$ embedding space.  What is special here is the maximal symmetry of AdS: starting from the conserved $so(D,2)$ charges, the
TD condition can be used in reverse to reconstruct the centroid together
with its covariant momentum and spin.  In this sense, the construction
below can be thought of as an $SO(D,2)$-covariant implementation of the usual TD prescription.

\subsection{Finite form of AdS isometries}\label{ssec:finIso}
We will begin by describing the finite version of AdS isometries, obtained by exponentiating the Killing vectors described in the main text. We will find it convenient to describe these transformations in the language of embedding space.

The spatial translations in global AdS can be realised via an $SO(D,2)$ matrix of the form
\begin{equation}\label{eq:Gammx}
\Gamma_x\equiv (\Gamma_x)^{^A}{}_{_B}\equiv 
\begin{pmatrix}
1&0&0\\[1mm]
0&\delta^i{}_j+\frac{\omA^2}{c^2}\frac{x^i x_j}{1+\sqrt{f(r)}}&-\frac{\omA}{c}x^i\\[1mm]
0&-\frac{\omA}{c}x_j & \sqrt{f(r)}
\end{pmatrix}.    
\end{equation}
Here and in what follows, the row/column indices of $(D+2)\times (D+2)$ matrices are ordered as 
\[
A,B \in \{0,i,D+1\},\qquad \text{with}\quad i=1,\dots,D\ .
\]
The above matrix satisfies $\Gamma_x^{\rm T}\eta \Gamma_x =\eta$, where $\eta$ is the metric on $\mathbb{R}^{D,2}$. The explicit matrix form of $\eta$ is 
\begin{equation}
\eta\equiv \eta_{_{AB}}\equiv 
\begin{pmatrix}
-1&0&0\\[1mm]
0&\delta_{ij}&0\\[1mm]
0& 0 & -1
\end{pmatrix}.    
\end{equation}
When we say that $\Gamma_x$ realises spatial translations in global AdS, what we mean is the following: the matrix  $\Gamma_x$ acts on embedding coordinates corresponding to spatial origin  at $t=0$ and transports it to embedding coordinates corresponding to point ${x^i}$  at $t=0$, i.e.,
\begin{equation}
\Gamma_x \begin{pmatrix} 0\\ 0\\ -\frac{c}{\omA}\end{pmatrix}=\begin{pmatrix} 0\\ x^i\\ -\frac{c}{\omA}\sqrt{f(r)}\end{pmatrix}\ .    
\end{equation}
The matrix  $\Gamma_x$ is, in a sense, the most symmetric matrix with the above property. By this, we mean that  $\Gamma_x$ is the unique matrix that behaves covariantly under spatial translations/reflections etc., obeys $\Gamma_x^{-1}=\Gamma_{-x}$ (reciprocity), and that also has the property of not transforming $X^0$. The reciprocity property can be stated explicitly as
\begin{equation}
\Gamma_x^{-1}=\Gamma_{-x}=
\begin{pmatrix}
1&0&0\\[1mm]
0&\delta^i_j+\frac{\omA^2}{c^2}\frac{x^i x_j}{1+\sqrt{f(r)}}&\frac{\omA}{c}x^i\\[1mm]
0&\frac{\omA}{c}x_j & \sqrt{f(r)}
\end{pmatrix}.
\end{equation}
In the parlance of symmetric spaces, $\Gamma_x$ is the principal geodesic transvection\cite{Helgason} which parallel transports along a geodesic in a symmetric space.

The full set of isometries of the $t=0$ spatial slice combines the above translations with rotations. Note however that, due to the curvature of AdS, these translations and rotations do not complete to the standard Euclidean Lie algebra $\mathfrak{iso}(D)$. Rather the flat Euclidean isoetries are deformed by the AdS curvature  into a Lorentz like algebra $\mathfrak{so}(D,1)$ : it is made of  $\mathfrak{so}(D)$ rotations as well as the transvections described above that play the role of boosts. The most general spatial isometry is then of the form
\begin{equation}
\begin{pmatrix}
1&0&0\\[1mm]
0& R^i{}_j+\frac{\omA^2}{c^2}\frac{(Rx)^i x_j}{1+\sqrt{f(r)}}&-\frac{\omA}{c}(Rx)^i\\[1mm]
0&-\frac{\omA}{c}x_j & \sqrt{f(r)}
\end{pmatrix}\ ,  
\end{equation}
where $R$ is an $SO(D)$ rotation.

Let us now turn to AdS isometries that mix space and time. In a similar vein to the spatial transvection above, we can define a $SO(D,2)$ matrix $\Gamma_t$ which time translates along global AdS time as
\begin{equation}
\Gamma_t\equiv 
\begin{pmatrix}
\cos(\omA\,t)&0&-\sin(\omA\,t)\\[1mm]
0&\delta^i{}_j&0\\[1mm]
\sin(\omA\,t)&0&\cos(\omA\,t)
\end{pmatrix}\ ,
\end{equation}
such that it parallel transports to future an arbitrary point at $t=0$ along a geodesic, i.e., 
\begin{equation}
\Gamma_t \begin{pmatrix} 0\\ x^i\\ -\frac{c}{\omA}\sqrt{f(r)}\end{pmatrix}=\begin{pmatrix} \frac{c}{\omA}\sqrt{f(r)}\sin(\omA\,t)\\ x^i\\ -\frac{c}{\omA}\sqrt{f(r)}\cos(\omA\,t)\end{pmatrix}\ .    
\end{equation}
We can combine space and time translations to define a spacetime translation matrix $\Gamma_{t,x}\equiv \Gamma_t\Gamma_{x}$ which takes the point at the AdS origin at $t=0$ to any other spacetime point within an AdS period. This can, of course be extended to the full global AdS by passing to global coordinates.

Now we turn to finite $SO(D,2)$ matrices for Lorentz transformations at a point in AdS. A general Lorentz transformation is an element of $SO(D,1)\subset SO(D,2)$ that keeps $X^{(D+1)}$ invariant, and it is determined a rotation in  $SO(D)$, along with a vector in $\mathbb{R}^D$ specifying the boost. Let us first describe a pure boost :  we will also find it convenient to phrase the boost in terms of its action on a spinning particle of mass $m$ resting at the origin of AdS. If we want  the final momentum post the boost to be $\pi^i$,
we can engineer a rotation-free boost of the form 
\begin{equation}\label{eq:RotFreeBpi}
\mathcal{B}_{\frac{\pi}{mc}}\equiv   \frac{1}{mc} \begin{pmatrix}
\pi^0 & \pi_j & 0\\
\pi^i & mc\,\delta^i{}_j+ \frac{\pi^i \pi_j}{\pi^0+mc} & 0\\
0 & 0 & mc 
\end{pmatrix}\ , 
\end{equation}
where  $\pi^0$ is determined from the vector data via the condition $\pi^0 \equiv  \sqrt{ m^2c^2+\pi^2}$. As we did with spatial isometries above, we can throw in the spatial rotations along with these boosts to form a $SO(D,1)$ element of the form
\begin{equation}
\frac{1}{mc} \begin{pmatrix}
\pi^0 & \pi_j & 0\\
(R\pi)^i & mc\,(R)^i{}_j+ \frac{(R\pi)^i \pi_j}{\pi^0+mc} & 0\\
0 & 0 & mc 
\end{pmatrix}\ . 
\end{equation}
where $R$ is an $SO(D)$ rotation. Thus, we get a different $SO(D,1)\subset SO(D,2)$ which has the rotations in common with the AdS spatial isometry $SO(D,1)$, but the translations are now replaced by boosts.

\subsection{Massive spinning particle in AdS}\label{ssec:spinning}
Let us now use the  action of these isometries to map out the phase space of a free, massive, spinning particle in AdS. We are especially interested in how the $\mathfrak{so}(D,2)$ charges of the particle changes across the phase space. The motivation for studying this is that, we want to eventually prove some sort of a cente of mass theorem for an arbitrary collection of interacting particles in AdS. More precisely, we want to show that, to any such collection, we can assign in a natural way, a centroid that behaves exactly like a 
free, massive, spinning particle in AdS.

With this motivation, let us begin by considering a spinning particle of mass $m$  and spin $s_{ij}$ resting at the origin of AdS. Such a particle is characterised by the following $\mathfrak{so}(D,2)$ charges : we have $\mathbb{P}_i=0=\mathbb{G}_i$, along with $\mathbb{H}=mc^2$ and $\mathbb{J}_{ij}=s_{ij}$. In the matrix notation we are using, we can package the conserved $\mathfrak{so}(D,2)$ charges  into the matrix
\begin{equation}
\mathbb J^{AB}
\equiv \Gamma_t
\begin{pmatrix}
0& -c\,\mathbb{G}^j &\dfrac{\mathbb{H}}{\omA}\\[5mm]
c\,\mathbb{G}^i & \mathbb{J}^{ij}&\dfrac{c\, \mathbb{P}^i}{\omA}\\[5mm]
-\dfrac{\mathbb{H}}{\omA} & -\dfrac{c\, \mathbb{P}^j}{\omA} &0\\[5mm] 
\end{pmatrix}\Gamma_t^{\rm T}\ .
\end{equation}
For the particle at rest at the origin, we have $\mathbb J^{AB}=\mathbb J^{AB}_{\rm Rest}$ with
\begin{equation}
\mathbb J^{AB}_{\rm Rest}\equiv \begin{pmatrix}
0& 0 &\dfrac{mc^2}{\omA}\\
0 & s^{ij}&0\\
-\dfrac{mc^2}{\omA} & 0 &0
\end{pmatrix}\ .
\end{equation}
Starting from this state, we can map out the full space of states(i.e., the phase space) by appropriate $SO(D,2)$ transformations. The set of all states are covered by performing the following transformations in sequence\footnote{What we are describing here is the construction of the phase-space through the co-adjoint orbit method.} :
\begin{enumerate}
\item We can rotate the spin at the origin : such a little group transformation would preserve the above  form of with the replacement $s^{ij}\mapsto (RsR^{\rm T})^{ij}$. We will assume that such a rotation has been taken into account already by constructing the spin $s^{ij}$ via rotation from some reference configuration(s).\footnote{A more precise statement is that choice of reference spin configuration fixes all the $SO(D)$ Casimir invariants/co-adjoint orbit that characterise the particle (more on Casimir invariants later). In the quantum theory, we will do this by specifying an irreducible representation of $SO(D)$. } From now on, we will assume that this has been done, and we already know the rotationally connected space of allowed spin configurations for the particle.
\item We can then apply a rotation-free boost of the form given in Eq.\eqref{eq:RotFreeBpi} to make the particle have any desired momentum $\vec{\pi}$. As we shall see, $\vec{\pi}$ coincides with the physical momentum $\mathbb{P}^i$ only when  the particle is at the origin. Otherwise, they differ because of corrections incurred during  parallel transport. To make this distinction clear, we will call $\pi^i$ as the `covariant momentum'.
\item We can parallel  translate such a particle with arbitrary spin/momentum sitting at the origin of AdS, to an arbitrary spatial position $\vec{r}$, by using a spatial transvection of the form given in Eq.\eqref{eq:Gammx}.
\item We can then time-evolve the particle by AdS time evolution.
\end{enumerate}
Thus a general state of the particle is characterised by the triple $(x^i,\pi^i,s^{ij})$, and the $\mathfrak{so}(D,2)$ charges in such a state is summarised by the matrix 
\begin{equation}
\mathbb J^{AB}\equiv \Gamma_t\Gamma_x\mathcal{B}_{\frac{\pi}{mc}}\begin{pmatrix}
0& 0 &\dfrac{mc^2}{\omA}\\
0 & s^{ij}&0\\
-\dfrac{mc^2}{\omA} & 0 &0
\end{pmatrix}\mathcal{B}_{\frac{\pi}{mc}}^{\rm T}\Gamma_x^{\rm T}\Gamma_t^{\rm T}\ ,
\end{equation}
obtained by transforming the $\mathfrak{so}(D,2)$ charges of the original reference state.

The above construction can be written in a slightly more intuitive form
by defining the following embedding space objects denoting the position, covariant momentum and the spin of the particle respectively:
\begin{equation}\begin{split}
X^A &\equiv \Gamma_t\Gamma_x\mathcal{B}_{\frac{\pi}{mc}}\begin{pmatrix}
0\\
0\\
-\dfrac{c}{\omA} \end{pmatrix}\ ,\quad 
\Pi^A\equiv \Gamma_t\Gamma_x\mathcal{B}_{\frac{\pi}{mc}}\begin{pmatrix}
mc\\
0\\
0\end{pmatrix}\ ,\\
S^{AB} &\equiv \Gamma_t\Gamma_x\mathcal{B}_{\frac{\pi}{mc}}\begin{pmatrix}
0& 0 &0\\
0 & s^{ij}&0\\
0 & 0 &0
\end{pmatrix}\mathcal{B}_{\frac{\pi}{mc}}^{\rm T}\Gamma_x^{\rm T}\Gamma_t^{\rm T}\ .
\end{split}\end{equation}
The above formula for the $\mathfrak{so}(D,2)$ charges can then be re-written in the suggestive form 
\begin{equation}\label{eq:JXPiSdeco}
\mathbb J^{AB}\equiv X^A \Pi^B - X^B \Pi^A +S^{AB}\ .
\end{equation}
Since we started with mutual orthonormal vectors/tensor and then did a $SO(D,2)$ transformation that  preserves the inner products, the following properties hold in a general state constructed this way
\begin{equation}\begin{split}
 X^A \Pi_A &=0\ ,\quad    X^AS_{AB}=0=\Pi^AS_{AB}\ ,\\
 X^AX_A &=-\frac{c^2}{\omA^2}\ , \quad \Pi^A\Pi_A= -m^2c^2\, \quad S^{AB}S_{AB}=s^{ij}s_{ij}\ .
\end{split}\end{equation}
The first equation in the second line says that $X^A$ is a point in the AdS hyperboloid.\footnote{In fact, the plane spanned by the pair $(X^A,\Pi^A)$ intersect the AdS hyperboloid along an AdS geodesic curve with tangent vector $\Pi^A$. This is the geometric way to reconstruct the time-like trajectory of the massive spinning particle.} The condition $X^A \Pi_A =0$ along with $\Pi^A\Pi_A= -m^2c^2$ implies $\Pi^A$ is a time-like momentum tangent to AdS satisfying the mass shell condition. The condition $X^AS_{AB}=0$ implies that the spin $S_{AB}$ is also tangent to AdS, and $\Pi^AS_{AB}=0$ implies that it is transverse also to momentum.\footnote{This shows that the bivector $S_{AB}$ contains the similar information as the Pauli-Lubanski vector in flat space. In the context of spinning particles in curved spacetime, this condition goes by the name of Tulczyjew-Dixon(TD) condition, a very common and convenient way to define spin in curved spacetimes.  } Another useful relation is 
\begin{equation}\label{eq:JXPieigen}\begin{split}
 \mathbb{J}^A{}_B\,X^B=\frac{c^2}{\omA^2} \Pi_A \ ,\quad    \mathbb{J}^A{}_B\,\Pi^B=-m^2c^2\, X^A\ ,
\end{split}\end{equation}
which tell us that $X^A$ and $\Pi^A$ are two mutually orthogonal time-like eigenvectors of the symmetric matrix $(\mathbb{J}^2)^A{}_B$ with eigenvalue $-\dfrac{m^2c^4}{\omA^2}$. For a heavy multi-particle of the type we are considering, such time-like eigenvectors always exist.

The above equations can also be checked against the explicit forms. For position, we have
\begin{equation}\begin{split}
X^A &\equiv \Gamma_t\Gamma_x\mathcal{B}_{\frac{\pi}{mc}}\begin{pmatrix}
0\\
0\\
-\dfrac{c}{\omA} \end{pmatrix}=\Gamma_t\Gamma_x\begin{pmatrix}
0\\
0\\
-\dfrac{c}{\omA} \end{pmatrix}=\begin{pmatrix} \frac{c}{\omA}\sqrt{f(r)}\sin(\omA t)\\ x^i\\ -\frac{c}{\omA}\sqrt{f(r)}\cos(\omA t)\end{pmatrix}\ ,
\end{split}\end{equation}
which we recognise as the usual description of an AdS spacetime point in the embedding space. For covariant momentum, we have
\begin{equation}\begin{split}
\Pi^A\equiv \Gamma_t\Gamma_x\mathcal{B}_{\frac{\pi}{mc}}\begin{pmatrix}
mc\\
0\\
0\end{pmatrix}=\Gamma_t\Gamma_x\begin{pmatrix}
\pi^0\\
\pi^i\\
0\end{pmatrix}=\Gamma_t\begin{pmatrix}
\pi^0\\[1mm]
\pi^i+\frac{\omA^2}{c^2}
\frac{x^i (x\cdot\pi)}{1+\sqrt{f(r)}}\\
-\frac{\omA}{c}(x\cdot\pi)\end{pmatrix}\ ,
\end{split}\end{equation}
with $\pi^0\equiv \sqrt{m^2c^2+\pi^2}$. As for the spin contribution, let us first compute the effect of the boost at the origin :
\begin{equation}\begin{split}
\mathcal{B}_{\frac{\pi}{mc}}\begin{pmatrix}
0& 0 &0\\
0 & s^{ij}&0\\
0 & 0 &0
\end{pmatrix}\mathcal{B}_{\frac{\pi}{mc}}^{\rm T}\equiv \begin{pmatrix}
0& -s^{i0}_\pi &0\\
s^{i0}_\pi & s^{ij}_\pi&0\\
0 & 0 &0
\end{pmatrix}\ ,
\end{split}\end{equation}
where we have defined the boosted spin components
\begin{equation}\begin{split}
s^{ij}_\pi
&\equiv 
\frac{
s^{ik}\pi_k
}{
mc(\pi^0+mc)
}\pi^j
-
\pi^i\frac{
s^{jk}\pi_k
}{
mc(\pi^0+mc)
}+s^{ij}
\ ,\quad s^{i0}_\pi \equiv 
\frac{s^{ik}\pi_k}{mc}\ .
\end{split}
\end{equation}
This is, in fact, identical to the corresponding flat spacetime computation relating boosted spin to the rest frame spin, assuming spin tensor $s^{ij}$ is the spatial component of an antisymmetric spacetime tensor. In fact, we note the appearence of the standard Newton-Wigner shift $\frac{
s^{ik}\pi_k
}{
mc(\pi^0+mc)
}$ for spinning particles. Another important fact to note is  the local Tulczyjew-Dixon(TD) condition satified by momentum/spin at the origin :
\begin{equation}\label{eq:localTD}\begin{split}
\pi^0 s^{i0}_\pi=\pi_js^{ij}_\pi\ .
\end{split}\end{equation}
This also has the same form as in flat spacetime. Parallel transporting this answer using the transvection $\Gamma_x$, and adding the result with the $X\wedge\Pi$ term, we get  the final $\mathfrak{so}(D,2)$  charges as
\begin{equation}\begin{split}
\frac{\mathbb{H}}{c}
&=
\sqrt{f(r)}\,\pi^0+\frac{\omA^2}{c^2}\,x_j s^{j0}_\pi
,\\[1mm]
\mathbb{P}^i
&=
\sqrt{f(r)}\,\pi^i-\frac{\omA^2}{c^2}
\frac{x_j (x\cdot\pi)}{1+\sqrt{f(r)}}-\frac{\omA^2}{c^2}
x_js_\pi^{ij},\\[1mm]
\mathbb{J}^{ij}
&=
x^i\left\{\pi^j-\frac{\omA^2}{c^2}
\frac{x_k s_\pi^{jk}}{1+\sqrt{f(r)}}\right\}-x^j\left\{\pi^i-\frac{\omA^2}{c^2}
\frac{x_ks_\pi^{ik}}{1+\sqrt{f(r)}}\right\}+s_\pi^{ij}\ ,\\
c\,\mathbb{G}^i
&=x^i\Bigl\{\pi^0 + \frac{\omA^2}{c^2}\frac{x_j s^{j0}_\pi}{1+\sqrt{f(r)}}\Bigr\}\,
+s_\pi^{i0}\ .
\end{split}\end{equation}
This then generalises the expressions in the main text given for spinless free particles.

\subsection{The TD centroid}\label{ssec:TDcen}
We can, in principle, invert the above relations to express $(x^i,\pi^i,s^{ij})$ in terms of the $\mathfrak{so}(D,2)$  charges using the local TD condition given in Eq.\eqref{eq:localTD}. Such an $x^i$ will then move along an AdS geodesic. Here is a quick geometric argument from embedding space before we see how this works algebraically. Given the $\mathbb{J}^{AB}$ of a heavy, isolated multi-particle state, we can always find the time-like embedding space eigen-plane  of the symmetric matrix $(\mathbb{J}^2)^A{}_B$. We will use the corresponding  eigenvalue $-\dfrac{m^2c^4}{\omA^2}$ to define the invariant mass $m$. On this invariant plane, we can then choose a unique time-like  eigen-vector $X^A$ such that it obeys 
\begin{equation}\begin{split}
 X^AX_A &=-\frac{c^2}{\omA^2}\ , \quad \dfrac{X^0}{\cos(\omA t)} \geq 0\ .
\end{split}\end{equation}
We automatically  then have an orthonormal time-like $\Pi^A$ which satisfies
\begin{equation}\begin{split}
 \mathbb{J}^A{}_B\,X^B=\frac{c^2}{\omA^2} \Pi_A \ ,\quad    \mathbb{J}^A{}_B\,\Pi^B=-m^2c^2\, X^A\ ,\quad  X^A \Pi_A &=0\ ,\quad \Pi^A\Pi_A= -m^2c^2\ ,
\end{split}\end{equation}
such  that $X^A$ and $\Pi^A$ span the time-like eigen-plane of the matrix $(\mathbb{J}^2)^A{}_B$. Subtracting $X\wedge \Pi$ from $\mathbb{J}$, we then get the spin bi-vector $S^{AB}$. As we mentioned before  plane spanned by $(X^A,\Pi^A)$ intersect the AdS hyperboloid along an AdS geodesic curve with tangent vector $\Pi^A$. This construction then gives a definition of a centroid that moves along the time-like geodesic.

To see how this works at the level of algebra, let us reconstruct the momentum/spin at the origin from the $\mathfrak{so}(D,2)$  charges by starting from a yet to be determined point $x^i$. In other words, we define the objects 
\begin{equation}\label{eq:Jpidef}\begin{split}
\mathbb{J}_\pi &\equiv \mathcal{B}_{\frac{\pi}{mc}}\mathbb{J}_{\rm Rest}\mathcal{B}_{\frac{\pi}{mc}}^{\rm T} = \begin{pmatrix}
0& -s^{i0}_\pi &\dfrac{c}{\omA}\pi^0\\[5mm]
s^{i0}_\pi & s^{ij}_\pi&\dfrac{c}{\omA}\pi^i\\[5mm]
-\dfrac{c}{\omA}\pi^0 & -\dfrac{c}{\omA}\pi_j &0
\end{pmatrix}\ ,
\end{split}\end{equation}
Given our previous definition, we can also write the same object as
\begin{equation}\begin{split}
\mathbb{J}_\pi&=\Gamma_x^{-1}\  \Gamma_t^{-1}\mathbb{J}\ (\Gamma_t^{-1})^{\rm T}\,(\Gamma_x^{-1})^{\rm T}=\Gamma_{-x}\ \Gamma_t^{-1}\mathbb{J}\ (\Gamma_t^{-1})^{\rm T}\, \Gamma_{-x}^{\rm T}\ .
\end{split}\end{equation}
Comparing these two expressions at the level of components, we get
\begin{equation}\begin{split}
\pi^0
&=\frac{1}{c}
\Bigl\{\sqrt{f(r)}\,\mathbb{H}-\omA^2\,x_j \mathbb{G}^j\Bigr\}
,\\[1mm]
\pi^i
&=
\sqrt{f(r)}\,\mathbb{P}^i-\frac{\omA^2}{c^2}
\frac{x_j (x\cdot\mathbb{P})}{1+\sqrt{f(r)}}+\frac{\omA^2}{c^2}
x_j\mathbb{J}^{ij},\\[1mm]
s^{ij}_\pi
&=\mathbb{J}^{ij}
-x^i\left\{\mathbb{P}^j+\frac{\omA^2}{c^2}
\frac{x_k \mathbb{J}^{jk}}{1+\sqrt{f(r)}}\right\}+x^j\left\{\mathbb{P}^i+\frac{\omA^2}{c^2}
\frac{x_k\mathbb{J}^{ik}}{1+\sqrt{f(r)}}\right\}\ ,\\
s^{i0}_\pi
&=c\,\mathbb{G}^i-\frac{x^i}{c}\Bigl\{\mathbb{H} - \omA^2\frac{x_j\,\mathbb{G}^j}{1+\sqrt{f(r)}}\Bigr\}\ .
\end{split}\end{equation}
These formulae parallel transport the $\mathfrak{so}(D,2)$  charge data from some arbitrary point to the origin.

The point $x^i$ can now be determined by substituting the above expressions back into the 
local TD condition  in Eq.\eqref{eq:localTD}.  For a fixed set of $\mathfrak{so}(D,2)$  charges, we then get $D$ non-linear equations for the D components $x^i$, which can then be solved uniquely. The uniqueness in guaranteed by the geometric construction we gave before. Algebraically, uniqueness can be checked, for example, by expanding the above equation about the flat limit.

\subsubsection*{Constructing the spin }
Once the $x^i$s are found this way, $\pi^i$ can be derived from the equation given above and the spin $s^{ij}$ can, in turn, be defined by
\begin{equation}\begin{split}
s^{ij}=s^{ij}_\pi
- 
\frac{
s^{i0}_\pi\pi^j-s^{j0}_\pi\pi^i
}{
\pi^0+mc
}
\ .
\end{split}
\end{equation}

The invariant mass $m$ that appears above as well as the spin magnitude $|s_{ij}|^2$  for a given set of charges  can be derived by computing the appropriate Casimir invariants in the enveloping algebra of $\mathfrak{so}(D,2)$. The quadratic/quartic Casimir invariants\footnote{
Our convention for p-form norm is \begin{equation}
\begin{split}
|F_{A_1A_2\ldots A_p}|^2\equiv \frac{1}{p!}F_{A_1A_2\ldots A_p}F^{A_1A_2\ldots A_p}\ ,
\end{split}
\end{equation}
where the indices are raised via the inverse of $\mathfrak{so}(D,2)$ invariant metric given in Eq.\eqref{eq:sod2eta}. All our wedge products are to be interpreted as standard exterior products with no extra factorials inserted while taking the product.}    of $\mathfrak{so}(D,2)$ are  
\begin{equation}
\begin{split}
m^2+\frac{\omA^2}{c^4} |s_{ij}|^2 &=\frac{\omA^2}{c^4} |\mathbb{J}_{AB}|^2=\frac{1}{c^4}\Biggl\{\mathbb{H}^2 - c^2 \vec{\mathbb{P}}^2\Biggr\}+\frac{\omA^2}{c^4}\, \Biggl\{ |\mathbb{J}_{ij}|^2- c^2 \vec{\mathbb{G}}^2 \Biggr\}\ ,\\
\end{split}
\end{equation}
as well as
\begin{equation}
\begin{split}
m^2 &|s_{ij}|^2 +\frac{\omA^2}{c^4} |\frac{1}{2!}(s \wedge s)_{ijkl}|^2\\
& = \frac{\omA^2}{c^4}|\frac{1}{2!}\left(\mathbb{J} \wedge \mathbb{J}\right)_{ABCD}|^2 \\
&=   \Biggl\lvert\frac{\mathbb{H}}{c^2} \mathbb{J}_{ij}+ (\mathbb{P}\wedge\mathbb{G})_{ij}\Biggr\rvert^2 -\frac{1}{c^2}|(\mathbb{P} \wedge \mathbb{J})_{ijk}|^2+\frac{\omA^2}{c^4}\Biggl\{ |\frac{1}{2!}(\mathbb{J} \wedge \mathbb{J})_{ijkl}|^2  - c^2|(\mathbb{G}\wedge \mathbb{J})_{ijk}|^2 \Biggr\}\ .
\end{split}
\end{equation}
Here we have normalised the invariants to have the correct  flat space limit, i.e., as $\omA\to 0$, the above expressions reduce to the norm squares of the spacetime momentum $(\mathbb{H}/c,\vec{\mathbb{P}})$ and the Pauli-Lubanski 3-form respectively. The relation to mass and spin comes about by evaluating the invariants on $\mathbb{J}^{AB}_{\rm Rest}$. For $D=3$, the above two equations are sufficient\footnote{after choosing the positive branch for taking the square roots.} to define $m$, since $s\wedge s=0$ in that case. For $D>3$, one should look at higher Casimir invariants associated with $\mathfrak{so}(D,2)$ to completely solve for $m$. For example, we have the sextic invariant
\begin{equation}
\begin{split}
m^2 |\frac{1}{2!}(s\wedge s)_{ijkl}|^2 +\frac{\omA^2}{c^4} |\frac{1}{3!}(s \wedge s\wedge s)_{ijklmn}|^2
& = \frac{\omA^2}{c^4}|\frac{1}{3!}\left(\mathbb{J} \wedge \mathbb{J} \wedge \mathbb{J}\right)_{ABCDEF}|^2\ ,
\end{split}
\end{equation}
and so on. This tower of invariants always truncates at any $D$ and then allows us to solve for $m$ in higher dimensions.

\subsubsection*{Dynamics of the TD centroid}
We will now discuss two important consequences of the above construction :
\begin{itemize}
\item Given the $\mathfrak{so}(D,2)$  charges for a collection of interacting particles/fields, we can assign to it a centroid $x^i$, a covariant momentum $\pi^i$ and a spin $s^{ij}$ \emph{which only depends on the $\mathfrak{so}(D,2)$  charges.}
\item Since the Poisson brackets and dynamics of  $\mathfrak{so}(D,2)$  charges is universal, the  Poisson brackets and dynamics of the triple $(x^i,\pi^i,s^{ij})$ is also universal. This then generalises the flat space result that the centre of mass degrees of freedom have a closed universal kinematics determined by the net charges.
\end{itemize}

While it is hard to solve the local TD constraint, the non-linear equation can be used to  derive exact evolution equations for all the quantities involved. For example, it can be shown using the harmonic oscillator equations for $(\mathbb{G}^i,\mathbb{P}^i)$ that 
\textbf{the point $x^i$ moves along an AdS geodesic.} It is actually easier to prove the converse, i.e., to show that if $(x^i,\pi^i,s^{ij})$ are given appropriate geodesic dynamics, the correct evolution  equations for $(\mathbb{G}^i,\mathbb{P}^i)$ follow.\footnote{Note that given the uniqueness argument before, the local TD constraint is invertible, and either of these statements imply their converse.}

The entire dynamics is quite easy to state in the language of embedding space. We begin by recalling that the oscillator-like equations for the pair $(\mathbb{G}^i,\mathbb{P}^i)$, and the conservation of $(\mathbb{H},\mathbb{J}^{ij})$ are all contained in the simple statement that $\mathbb{J}^{AB}$ is conserved. As far the embedding space triple $(X^A,\Pi^A,S^{AB})$ that appears in the decomposition of $\mathbb{J}^{AB}$, we have the following simple set of dynamical equations :
\begin{equation}
\begin{split}
\frac{d X^A}{d\tau}&\equiv \frac{\pi^0}{mc\sqrt{f(r)}}\frac{d X^A}{dt}
=
\frac{\Pi^A}{m}\ ,\\
\frac{d\Pi^A}{d\tau}&\equiv \frac{\pi^0}{mc\sqrt{f(r)}}\frac{d\Pi^A}{dt}
=
-m\,\omA^2 X^A\ ,\\
\frac{dS^{AB}}{d\tau}&\equiv \frac{\pi^0}{mc\sqrt{f(r)}}\dot S^{AB}=0\ .
\end{split}
\end{equation}
Here $\tau$ denotes the proper time along the particle trajectory, and overdot denotes the coordinate time derivative. The first two equations are essentially the equations for the AdS geodesic. The last is that statement that $S^{AB}$ is parallel transported along the geodesic. It is evident that above equations are consistent with all constraints on the triple $(X^A,\Pi^A,S^{AB})$ and imply that $\mathbb{J}^{AB}$ is conserved.

At the level of components, we have the following equation for the velocity
\begin{equation}
\begin{split}
\frac{dx^i}{d\tau} &\equiv \frac{\pi^0}{mc\sqrt{f(r)}}\,\dot x^i
=\frac{1}{m}
\left\{
\pi^i+\frac{\omA^2}{c^2}
\frac{x^ix_j\pi^j}{1+\sqrt{f(r)}}
\right\}\ .
\end{split}
\end{equation}
Further, the components of $\mathbb{J}_\pi^{AB}$ (see Eq.\eqref{eq:Jpidef}) evolve via infinitesimal Lorentz transformations of the form
\begin{equation}
\begin{split}(\Omega_\pi)^{i0}\equiv
-(\Omega_\pi)^{0i}\equiv\frac{\omA^2}{c} x^i\ ,\quad (\Omega_\pi)^{ij}\equiv  \frac{\omA^2}{c^2}
\frac{(x\wedge \dot x)^{ij}}{1+\sqrt{f(r)}}=\frac{\omA^2}{ c}
\frac{\sqrt{f(r)}(x\wedge \pi)^{ij}}{\pi^0(1+\sqrt{f(r)})}\ ,
\end{split}
\end{equation}
i.e., we have the evolution equations
\begin{equation}
\begin{split}
\dot \pi^0 &= (\Omega_\pi)^0{}_i \pi_i\ ,\quad \dot \pi^i = (\Omega_\pi)^i{}_0 \pi^0  + (\Omega_\pi)^i{}_j \pi^j\ ,\\
\dot s_\pi^{ij} &=(\Omega_\pi)^j{}_0 s_\pi^{i0} -(\Omega_\pi)^i{}_0 s_\pi^{j0}+ (\Omega_\pi)^i{}_k s_\pi^{kj}+ (\Omega_\pi)^j{}_k s_\pi^{ik}\ ,\\
\dot s_\pi^{i0} &=(\Omega_\pi)^0{}_j s_\pi^{ij} + (\Omega_\pi)^i{}_j s_\pi^{j0}\ .
\end{split}
\end{equation}
and the above equations describe how the components of $\mathbb{J}_\pi$ describing the locally measured momentum/spin of the centroid get parallel-transported along the time-like geodesic. We can also equivalently define a vector that behaves like the canonical momentum associated with the geodesic
\begin{equation}
\begin{split}
p_i
\equiv 
\pi_i
-
\frac{\omA^2}{c^2}
\frac{x_i(x\cdot\pi)}
{\sqrt{f(r)}(1+\sqrt{f(r)})} .
\end{split}
\end{equation}
We will leave it to the reader to check that they obey the evolution equations appropriate for the geodesic
\begin{equation}
\begin{split}
\frac{dx^i}{d\tau}
=\frac{1}{m}\Big[p^i+\frac{\omA^2}{c^2}\,x^i(x\cdot p)\Big],\quad
\frac{dp_i}{d\tau}
=
-\frac{\omA^2}{mc^2}
\left[
(x\cdot p)\,p_i
+\frac{(\pi^0)^2}{f(r)}\,x_i
\right]
\end{split}
\end{equation}
If desired, we can use canonical momentum to  eliminate all references to $(\pi^0,\pi^i)$ via the relations 
\begin{equation}
\begin{split}
\pi^0=\sqrt{
m^2c^2
+p^2
+\frac{\omA^2}{c^2}(x\cdot p)^2
}\ ,\quad 
\pi_i
=
p_i
+
\frac{\omA^2}{c^2}
\frac{x_i(x\cdot p)}
{1+\sqrt{f(r)}}\ .
\end{split}
\end{equation}
 
 As for the proper spin $s^{ij}$, its geodetic precession is described via 
 \begin{equation}
\begin{split}
\dot s^{ij} &= (\Omega_{\rm geo})^i{}_k s^{kj}+ (\Omega_{\rm geo})^j{}_k s^{ik}\ ,\\
\end{split}
\end{equation}
where we have defined
\begin{equation}
\begin{split}
(\Omega_{\rm geo})^{ij}\equiv-\frac{\omA^2}{ c}
\frac{\pi^0-mc\sqrt{f(r)}}{\pi^0(1+\sqrt{f(r)})(\pi^0+mc)} (x\wedge \pi)^{ij}\ ,
\end{split}
\end{equation}
Thus, the evolution of proper spin is a pure  $SO(D)$ rotation with no change in its intrinsic magnitude or in the higher  Casimirs.

\subsection{Centroid in dS}\label{ssec:dSCent}
Much of the formalism developed in this appendix carries over to dS via the analytic continuation
\begin{equation}
\omA\rightarrow \compi\,\hds\ ,\quad X^{(D+1)}_{\rm dS}\rightarrow \compi\,X^{(D+1)}_{\rm dS}\ . 
\end{equation}
For this reason, we will content ourselves by pointing out some qualitatively new features that show up under such an analytic continuation. The time translation matrix in dS turns into a boost matrix, whereas the spatial transvection turns into a translation on a sphere:   
\begin{equation}
\begin{split}
\Gamma_t^{\rm dS}
&\equiv 
\begin{pmatrix}
\cosh \hds t&0&\sinh \hds t\\
0&\delta^i{}_j&0\\
\sinh \hds t&0&\cosh \hds t
\end{pmatrix}\ ,\quad 
\Gamma_x^{\rm dS}
\equiv
\begin{pmatrix}
1&0&0\\[1mm]
0&
\displaystyle
\delta^i{}_j-
\frac{\hds^2}{c^2}
\frac{x^ix_j}{1+\sqrt{f(r)}}
&
\displaystyle \frac{\hds }c x^i\\[3mm]
0&
-\displaystyle\frac{\hds }c x_j&
\sqrt{f(r)}
\end{pmatrix}.
\end{split}
\end{equation}
The boost matrices on the other hand do not change. As in AdS, the $\mathfrak{so}(D+1,1)$ charges of dS can be put into an anti-symmetric matrix, and then brought to the canonical form corresponding to the particle at rest at the origin, i.e.,
\begin{equation}
\mathbb J^{AB}
\equiv \Gamma_t^{\rm dS}
\begin{pmatrix}
0& -c\,\mathbb{G}^j &-\dfrac{\mathbb{H}}{\hds}\\[5mm]
c\,\mathbb{G}^i & \mathbb{J}^{ij}&-\dfrac{c\, \mathbb{P}^i}{\hds}\\[5mm]
\dfrac{\mathbb{H}}{\hds} & \dfrac{c\, \mathbb{P}^j}{\hds} &0\\[5mm] 
\end{pmatrix}(\Gamma_t^{\rm dS})^{\rm T}=\Gamma_t^{\rm dS}\Gamma_x^{\rm dS}\mathcal{B}_{\frac{\pi}{mc}}\begin{pmatrix}
0& 0 &-\dfrac{mc^2}{\hds}\\
0 & s^{ij}&0\\
\dfrac{mc^2}{\hds} & 0 &0
\end{pmatrix}(\Gamma_t^{\rm dS}\Gamma_x^{\rm dS}\mathcal{B}_{\frac{\pi}{mc}})^{\rm T}\ .
\end{equation}
The matrices needed can be constructed purely from embedding space arguments.
Given the $\mathbb{J}^{AB}$ of a heavy, isolated multi-particle state, we can always find the  eigen-plane  in  embedding space where the symmetric matrix $(\mathbb{J}^2)^A{}_B$ has positive eigenvalue $\dfrac{m^2c^4}{\hds^2}$. Crucial in this regard is the observation that spin part of $(\mathbb{J}^2)^A{}_B$, being a square of an anti-symmetric matrix, is negative semi-definite. On the positive eigen-plane, we can then choose a unique space-like  eigen-vector $X^A$ and a time-like eigen-vector $\Pi^A$ such that they obey
\begin{equation}\begin{split}
 X^AX_A &=\frac{c^2}{\hds^2}\ , \quad X^0\geq 0,\quad  X^A \Pi_A =0\ ,\quad \Pi^A\Pi_A= -m^2c^2\ ,\\
\mathbb{J}^A{}_B\,X^B&=-\frac{c^2}{\hds^2} \Pi_A \ ,\quad    \mathbb{J}^A{}_B\,\Pi^B=-m^2c^2\, X^A\ . 
\end{split}\end{equation}
Subtracting $X\wedge \Pi$ from $\mathbb{J}$, we then get the spin bi-vector $S^{AB}$. The transformations required to bring $\mathbb{J}$ matrix to the canonical form
can then be read off from
\begin{equation}\begin{split}
X^A &\equiv 
\Gamma_t^{\rm dS} 
\Gamma_x^{\rm dS} \begin{pmatrix} 0\\ 0\\ \frac{c}{\hds}\end{pmatrix}=\Gamma_t^{\rm dS}\begin{pmatrix} 0\\ x^i\\ \frac{c}{\hds}\sqrt{f(r)}\end{pmatrix}=\begin{pmatrix} \frac{c}{\hds}\sqrt{f(r)}\sinh \hds t\\ x^i\\ \frac{c}{\hds}\sqrt{f(r)}\cosh \hds t\end{pmatrix}\ ,\\
\Pi^A &\equiv
\Gamma_t^{\rm dS}\Gamma_x^{\rm dS}\mathcal{B}_{\frac{\pi}{mc}}\begin{pmatrix}
mc\\[1mm]
\displaystyle
0\\[3mm]
0\end{pmatrix}=
\Gamma_t^{\rm dS}\Gamma_x^{\rm dS}\begin{pmatrix}
\pi^0\\[1mm]
\displaystyle
\pi^i\\[3mm]
0\end{pmatrix}=
\Gamma_t^{\rm dS}\begin{pmatrix}
\pi^0\\[1mm]
\displaystyle
\pi^i-\frac{\hds ^2}{c^2}
\frac{x^i(\mathbf x\cdot\boldsymbol\pi)}
{1+\sqrt{f(r)}}\\[3mm]
\displaystyle
-\frac{\hds }{c}(\mathbf x\cdot\boldsymbol\pi)
\end{pmatrix}.
\end{split}
\end{equation}
With these, it is clear that the rest of the AdS construction goes through for dS with appropriate changes.

\subsection{Related constructions in the literature }\label{ssec:WigDisc}
We will conclude by discussing how our construction above relates to various related constructions in the literature. The existing constructions can be classified into three types : 
\begin{itemize}
 \item Authors who construct models for classical particles in AdS (often via co-adjoint orbit),
 \item Authors who study quantum particles in AdS ala Wigner,
\item Authors who approach the AdS particle problem from the dual CFT perspective. The issue here is to establish bulk localisability via purely CFT objects.
\end{itemize}
Our construction,  of course, is classical and is directly related to the first type of constructions.

At the classical level, spinning
particles in AdS can be described as co-adjoint orbits of the AdS isometry group. Our construction bears close similarity to the   massive spinning-particle models in AdS$_4$ by Kuzenko,
Lyakhovich, Segal and Sharapov \cite{Kuzenko:1995aq}, as well as to recent co-adjoint orbit constructions in AdS\cite{Basile:2023vyg,Joung:2024akb}. The latter classification includes, besides the generic massive spinning orbits we consider here, various special
orbits describing massless and conformal particles, including the
singletons, and other  more exotic possibilities \cite{Joung:2024akb}. Our focus here is  on heavy states which also admit a non-relativistic limit, and as in the rest of this work, we have retained all our $c$'s and $\omA$'s to enable flat/non-relativistic limit. Our final goal  below is
somewhat the reverse of these constructions : our goal is not to construct a
prescribed one-particle representation or co-adjoint orbit, but to consider the net $\mathfrak{so}(D,2)$ charges of an arbitrary collection of interacting
particles and ask whether they can be interpreted in terms of a single massive spinning particle.\footnote{The standard co-adjoint orbit constructions are of course helpful in answering this question.}  In what follows we restrict to the regular massive
sector, and assume that the total charges are sufficiently generic that
the special or degenerate representations just mentioned do not arise.

Quantum models of spinning particles can be constructed  by extending Wigner's description of quantum particles in terms of spacetime symmetries
\cite{Wigner:1939cj,Weinberg:1995mt} to AdS. The original group-theoretic programme was developed by Fronsdal long ago, who described massive quantum particles in AdS by their spin and a complex ``pseudo-momentum''\cite{FRONSDAL:1965nuq,Fronsdal:1974ew}.  The Fronsdal's programme has been explored in more detail recently by authors of \cite{Nguyen:2026rwt,Iacobacci:2024laa,Nguyen:2026yho}. Authors of \cite{Berenstein:2025qhb,Berenstein:2025tts} provide a complementary
construction directly in terms of Lorentzian AdS wavefunctions. Fronsdal's pseudo-momentum becomes real in the flat limit, but the  a direct physical interpretation away from flat limit seems elusive in such Fronsdal-like constructions.

Unfortunately, the classical construction that we present here does not  provide a simple remedy since it is hard to quantise the dynamics of Tulczyjew--Dixon centroid even in the flat limit. For the quantum problem, we need a generalisaton of the  canonical Newton-Wigner centroid, which can presumably be done by redefinitions starting from the construction presented here.

Finally, while it is not our primary goal, our construction below has implications for the dual CFT, and especially to the question of studying bulk localisation via boundary CFT data.  The standard HKLL construction represents local bulk fields in terms of non-local operators in the boundary CFT \cite{Hamilton:2006az}.  However, closer in spirit to the discussion here, are constructions which use the global conformal symmetry to associate CFT states or operators with points in AdS\cite{Nakayama:2015mva,Guo:2022syo}.  There is also a closely related geometric
description of conformal blocks in terms of geodesic Witten diagrams
\cite{Hijano:2015zsa}. What we show here is that, rather than reconstructing a local bulk
operator from the detailed CFT data, we can ask whether the finite set of total $\mathfrak{so}(D,2)$ charges of a state is sufficient to associate with it a preferred bulk worldline. We shall find that, in the massive sector, these
charges do naturally determine a time-like AdS geodesic which may be interpreted
as the trajectory of a centroid.

\section{Free scalar fields in (A)dS}\label{app:scalar-ads-ds}
In this appendix, we review various facts about free scalar fields in AdS/dS that are relevant for this work. Like appendix~\ref{app:AdS}, the material here has quite some overlap with lecture notes on AdS/CFT\cite{Aharony:1999ti,DHoker:2002nbb,Kaplan:AdSCFT,Penedones:2016voo}. Our purpose here is to set the notation and emphasise some aspects which do not seem to have appeared elsewhere.

This appendix is organised as follows: after a brief introduction, we move to the central result of this appendix-- the nature of the classical field in various zones relative to the source. The basic idea is familiar from flat spacetime. For slowly moving point sources of the type we are interested in, we
first have a \textbf{near zone} defined as the region near the source (usually within a wavelength of the emitted wave) where we can understand the behaviour of the field starting from statics and adding corrections. In other words, we can start with the approximation that the near zone field is determined at every instant by solving the static problem with the instantaneous position of the sources, and then deviations from this assumption can be taken into account systematically. For a multipole of rank $l$, we have two standard static behaviours for massless/light fields
near the origin: a regular behaviour going as $r^l$ and a singular behaviour varying $r^{-(l+D-2)}$, the latter appearing only if a point source is sitting at the origin. These leading  power laws near origin  continue to hold even when we move  to AdS or dS since the behaviour near origin is unaffected by the curvature. The sub-leading curvature corrections can then be taken into account. In AdS or dS, the exact solutions which have regular/singular behaviour can be written down in terms of hypergeometric functions, as we discuss in \S\S\ref{ssec:NearZone}. 

In flat spacetime, the next zone of interest is the \textbf{radiation zone}, where the field is far from being static, and is closer to being a wave with the temporal variations being comparable to spatial variation. In flat spacetime, this is the behaviour near  light-like infinities, where we see purely outgoing or purely incoming waves. In the dS static patch, once go near the future cosmological horizon, we get a behaviour very similar to what happens at light-like infinity of flat spacetime.  Hence, there is a sense in which we can extend the definition of radiation zone, as well as the definition of outgoing and incoming waves to dS. As we shall see below, the AdS case is very different: an outgoing wave in AdS can only stay outgoing for an AdS period, after which it hits the boundary and bounces back into an ingoing wave. Nevertheless, one can define waves in AdS which for a while  look like outgoing/incoming waves. In fact, we will see that the easiest way to understand these solutions is to think about them as obtained by analytically continuing outgoing/incoming solutions from dS. In either case, we will call these incoming/outgoing pairs as Jost solutions, following  a common nomenclature in scattering theory\cite{Newton:1982qc,Rakityansky2022Jost}. The construction of these Jost solutions is the goal of \S\S\ref{ssec:RadZone}. 

There is actually a third zone which is peculiar to cosmological spacetimes like AdS/dS. We will call it the \textbf{cosmic zone}, where the effect of cosmological constant on the waves can no more be neglected. For dS, this is true for a wave which has long exited the static patch horizon and its behaviour at late times is governed by completely different power laws than in near zone or radiation zone. For light fields, one mode grows relative to the other, thus leading to the `growing/decaying mode' distinction familiar from cosmological perturbation theory.  

For AdS, we have characteristic falloffs as we approach the time-like conformal boundary, which is again far from any flat spacetime behaviour. This is the  normalisable vs. non-normalisable  mode distinction well-known in AdS/CFT, especially because the dual CFT correlation functions can be read off easily from this `cosmic zone' behaviour. Another characteristic feature in the cosmic zone is the perfect reflection of the waves reaching the boundary which then interfere with original waves, giving rise eventually to a discrete spectrum of standing waves. The physics of this `AdS echo' affects the dynamics of AdS sources over cosmologically long times. Both these dS as well as AdS behaviours at the cosmic zone is the subject of 
\S\S\ref{ssec:CosmZone}.

\subsection{The AdS scalar as a dynamical system}
\label{subsec:scalar-dynamics}
We begin with the ordinary massive Klein--Gordon field on global AdS$_{D+1}$,
\begin{equation}
\begin{split}
 ds^{2}
 &=-f(r)c^{2}dt^{2}
   +\frac{dr^{2}}{f(r)}
   +r^{2}d\Omega_{D-1}^{2}.
\end{split}
\label{eq:ads-global-metric}
\end{equation}
with the action
\begin{equation}
\begin{split}
\mathcal{S}^{\phi}
 &=-\frac12\int d^{D+1}x\sqrt{-g}
 \left[g^{\mu\nu}\partial_{\mu}\phi\partial_{\nu}\phi
       +m_{\phi}^{2}c^{2}\phi^{2}\right]\ .
       \end{split}
\label{eq:ads-action}
\end{equation}
The equation of motion is $(\Box_{\rm AdS}-m_{\phi}^{2}c^{2})\phi=0$ and its energy-momentum tensor is given by 
\begin{equation}
\begin{split}
 T_{\mu\nu}
 &=\partial_{\mu}\phi\,\partial_{\nu}\phi
 -\frac12 g_{\mu\nu}\left[(\partial\phi)^2+m_{\phi}^{2}c^{2}\phi^{2}\right].
\end{split}
\label{eq:scalar-stress-tensor}
\end{equation}

We will find it convenient to characterise the mass appearing here by a  parameter $\mu$ as well as the exponents $\Delta_\pm\equiv \dfrac{D}{2}\pm \mu$. These parameters are defined via 
\begin{equation}
\begin{split}
 m_{\phi}^{2}
 &\equiv\frac{\omA ^{2}}{c^{4}}
   \left(\mu^{2}-\frac{D^{2}}{4}\right)\equiv\frac{\omA ^{2}}{c^{4}}
  \Delta_{\pm}(\Delta_{\pm}-D)\ .
\end{split}
\label{eq:ads-action-mass}
\end{equation}
 This parametrisation is useful because the two roots $\Delta_{\pm}$ appear directly as the two exponents at the AdS conformal boundary. At fixed $\mu$, the physical mass goes to zero as $\omA \to0$. To instead obtain a finite massive flat-space limit, one takes $\mu\sim m_{\phi}c^{2}/\omA $ while sending the curvature scale to zero. All the formulae above can be analytically continued to dS under $\omA \to \compi\hds $. 
For a massive scalar field, this gives
\begin{equation}
\begin{split}
 m_{\phi}^{2}
 &\equiv\frac{\hds ^{2}}{c^{4}}
 \left(\frac{D^{2}}4-\mu^{2}\right)\equiv-\frac{\hds ^{2}}{c^{4}}
 \Delta_{\pm}(\Delta_{\pm}-D)\ .
\end{split}
\label{eq:ds-mass-mu}
\end{equation}
For heavy fields with $m_{\phi}c^{2}>\frac{D}{2}\hds $, we can  write $\mu\equiv\compi\rho$, with $\rho\in\mathbb R$, and
\begin{equation}
\begin{split}
 \Delta_{\pm}\equiv\frac D2\pm \compi\,\rho .
\end{split}
\label{eq:principal-series}
\end{equation}
The case with imaginary $\mu$ in dS is usually referred to as the \emph{principal series}, whereas cosmologically light fields have real $\mu$ and belong to the \emph{complementary series}.

We would like to repeat for the scalar field our discussion of conserved quantities for free particles given in \S\ref{sec:FPartAdS}. As described around Eq.~\eqref{eq:noether-charge}, if $\xi^{\mu}$ is an AdS Killing vector, there is a charge $\mathbb{Q}_\xi$ defined by
\begin{equation}
\begin{split}
\mathbb{Q}_{\xi}\equiv\compi\,\int_{\Sigma_t}T^{0}{}_{\mu}\xi^{\mu}\ .
\end{split}
\label{eq:noether-chargeII}
\end{equation}
Further, the conservation law $\nabla_{\mu}T^{\mu\nu}=0$, together with the Killing equation, gives $\nabla_{\mu}(T^{\mu}{}_{\nu}\xi^{\nu})=0$. This is the local statement behind the conservation of $\mathbb{Q}_{\xi}$. In dS, however, the current can be locally conserved but energy and other charges restricted to one static patch change because of flux through the cosmological horizon. In other words, unlike AdS, the static observer sees an open subsystem, characterised by a cosmological influence phase\cite{Loganayagam:2023pfb}.

The action of isometry on a scalar field is via a Lie drag
\begin{equation}
\begin{split}
 \compi\,\delta_{\xi}\phi\equiv\compi\,[\phi,\mathbb{Q}_\xi]_{\rm PB}={\mathscr L}_{\xi}\phi=\xi^{\mu}\partial_{\mu}\phi\ .
\end{split}
\label{eq:lie-drag}
\end{equation}
Since $[\Box_{\rm AdS},{\mathscr L}_{\xi}]=0$, such a Lie drag  maps a solution of the wave equation to another solution. 

To understand the action of isometries on the solution, we begin by  separating variables to write
\begin{equation}
\begin{split}
 \phi(t,r,\Omega)
 &=e^{-\compi\omega t}Y_{ l \bar m}\left(D;\hat{\Omega}\right)\,\Xi_{ l \omega}(r),\\
 \nabla_{S^{D-1}}^{2}Y_{ l \bar m}
 &=- l ( l +D-2)Y_{ l \bar m}.
\end{split}
\label{eq:separation}
\end{equation}
Here $Y_{ l \bar m}\left(D;\hat{\Omega}\right)$ are orthonormal spherical harmonics on $\mathbb{S}^{D-1}$ with the label $\bar m$ collectively denoting all the admissible labels for a given $ l $. We can also use an STF (symmetric trace-free) basis, in which $Y_{ l \bar m}$ is represented by a symmetric, traceless rank-$ l $ tensor contracted with $\hat x^i$s, i.e.,
\begin{equation}
\begin{split}
 Y_{ l \bar m}(\hat x)
 =\frac{1}{ l !}\,\mathcal Y^{( l \bar m)}_{\langle i_1i_2\ldots i_ l \rangle}
 \hat x^{i_1}\hat x^{i_2}\ldots \hat x^{i_l}.
\end{split}
\label{eq:stf-harmonics}
\end{equation}
The angular bracket here signifies the STF projection.Such a basis of STF tensors is the natural language for discussing multipole expansion.

The action of AdS isometries (see Eqs.\eqref{eq:AdSKilling-glob} and \eqref{eq:AdSKilling-globII}) on such a variable-separated solution is now easy to see. The solutions above are eigen-functions of the Killing vector $H \equiv \compi\frac{\partial}{\partial t}$ with eigenvalue $\omega$, whereas the $J_{ij}$ acts to rotate the spherical harmonic pre-factor without changing $l$. The only non-trivial action is from the Killing vector fields
\begin{equation}
\begin{split}
\frac{1}{c}J^{i\mp}\equiv   \frac{e^{\pm\compi\,\omA t}}{\sqrt{f(r)}}
\left(\compi\,\frac{x_i}{c^2}\frac{\partial}{\partial t}\pm \frac{f(r)}{\omA} \frac{\partial}{\partial x^i}\right)\ .
\end{split}
\end{equation}
We see right away that, these operators shift the frequency $\om$ up and down in steps of $\omA$. Further being a vector differential operator, they shift $l$ by one unit up or down when they act. Similar statements hold for dS, except that the shifts in $\om$ are along the imaginary axis in steps of $\compi\,\hds$.

\subsection{Near zone multipoles}
\label{ssec:NearZone}
Let us now examine the variable separated solutions in more detail. The radial equation after variable separation is
\begin{equation}
\begin{split}
 0={}&\frac1{r^{D-1}}\partial_r\!\left[r^{D-1}f(r)\,\partial_r\Xi\right]
 +\frac{\omega^{2}}{c^{2}f(r)}\Xi
 -\frac{ l ( l +D-2)}{r^{2}}\Xi
 -m_{\phi}^{2}c^{2}\Xi .
\end{split}
\label{eq:ads-radial-equation}
\end{equation}
After a change of variable $u_{\rm AdS}\equiv -\omA ^{2}r^{2}/c^{2}$, this equation becomes a  hypergeometric equation with three regular singular points $ u_{\rm AdS}=0,1,\infty$ corresponding to 
\begin{equation}
\begin{split}
 r=0,&\qquad r=\pm \compi\frac{c}{\omA },\qquad r=\infty 
\end{split}
\label{eq:ads-three-singular-points}
\end{equation}
respectively. The solutions of the hypergeometric equation can hence be written with the last hypergeometric variable being either $u,f\equiv 1-u$ or $\frac{1}{u}$ depending on around which singular point we are power series expanding our solution. In the dS static patch, the same three points move to $u_{\rm dS}=0,1,\infty$, with $u_{\rm dS}=1$ denoting the real cosmological horizon. The three choices for  the singular point neatly map on to the three zones described in the beginning of this appendix:
\begin{itemize}
\item Near the origin $u=0$, we have a basis of regular and singular solutions $\{\Xi_{\rm Reg},\Xi_{\rm Sing}\}$ which are well-suited to describing the  near-zone behaviour.
\item Near $u=1$, we have the Jost basis $\{\Xi_+,\Xi_-\}$ describing outgoing/incoming solutions in the radiation-zone. 
\item Near $u=\infty$, we have the asymptotic basis $\{\Xi_{\rm Dom},\Xi_{\rm Sub}\}$ describing the cosmic-zone behaviour.
\end{itemize}
Thus the three bases  are the Frobenius bases adapted to the three singular points. 

Let us begin with  the origin-adapted basis. We define
\begin{equation}
\begin{split}
 u_{\rm AdS}\equiv -\frac{\omA ^{2}r^{2}}{c^{2}},
 \qquad
 \widehat\omega\equiv \frac{\omega}{\omA }\ ,\ \nu\equiv l+\frac{D}{2}-1
\end{split}
\end{equation}
in terms of which the regular/singular  solutions at the origin are given as
\begin{equation}
\begin{split}
 \Xi_{\rm Reg}^{\rm AdS}
 &\equiv \frac{r^{ l }}{ 2\nu}f(r)^{-\widehat\omega/2}
 {}_2F_1\!\left(
 \frac{1+\nu-\mu-\widehat\omega}{2},
 \frac{1+\nu+\mu-\widehat\omega}{2};
 1+\nu;u_{\rm AdS}\right),\\
 \Xi_{\rm Sing}^{\rm AdS}
 &\equiv r^{-( l +D-2)}f(r)^{-\widehat\omega/2}
 {}_2F_1\!\left(
 \frac{1-\nu+\mu-\widehat\omega}{2},
 \frac{1-\nu-\mu-\widehat\omega}{2};
 1-\nu;u_{\rm AdS}\right).
\end{split}
\label{eq:ads-reg-sing}
\end{equation}
At integer $\nu$ (i.e., for even $D$) the second function becomes proportional to the first and this basis degenerates; in that case, an appropriate amount of regular solution has to be subtracted from the singular solution at a general $\nu$ before we take the limit of $\nu$ tending to an integer.

The above solutions differ by their near origin power laws:
\begin{equation}
\begin{split}
 \Xi_{\rm Reg}^{\rm AdS}\sim\frac{r^{ l }}{2\nu},
 \qquad
 \Xi_{\rm Sing}^{\rm AdS}\sim r^{-( l +D-2)},
 \qquad r\to0\ .
\end{split}
\label{eq:origin-normalisation}
\end{equation}
which indeed matches the power-laws for static solutions in flat spacetime.

The corresponding solutions in dS static patch are obtained by analytic continuation $\omA\to\compi\,\hds$. In terms of the variables
\begin{equation}
\begin{split}
 u_{\rm dS}\equiv \frac{\hds ^{2}r^{2}}{c^{2}},
 \qquad
 \varpi\equiv \frac{\omega}{\hds },
\end{split}
\label{eq:y-varpi}
\end{equation}
we have
\begin{equation}
\begin{split}
 \Xi_{\rm Reg}^{\rm dS}
 &\equiv \frac{r^{ l }}{2\nu}f(r)^{-\compi\varpi/2}
 {}_2F_1\!\left(
 \frac{1+\nu+\mu-\compi\varpi}{2},
 \frac{1+\nu-\mu-\compi\varpi}{2};
 1+\nu;u_{\rm dS}\right),\\
 \Xi_{\rm Sing}^{\rm dS}
 &\equiv r^{-( l +D-2)}f(r)^{-\compi\varpi/2}
 {}_2F_1\!\left(
 \frac{1-\nu+\mu-\compi\varpi}{2},
 \frac{1-\nu-\mu-\compi\varpi}{2};
 1-\nu;u_{\rm dS}\right).
\end{split}
\label{eq:ds-reg-sing}
\end{equation}
The same two powers occur at the observer worldline $r=0$, and their relative coefficient in outgoing solution encodes the radiation reaction\cite{Loganayagam:2023pfb}.

We can now examine the flat limit of these solutions by examining the limit of either $\omA \to0$ or $\hds \to0$ at fixed $(r,\omega,\nu,\mu)$.
In this limit, hypergeometric functions   appearing above have first two parameters large while their argument tends to zero inversely as the product of the large parameters.The relevant confluent limit
is
\begin{equation}
\begin{split}
{}_2F_1
\left(
a_\lambda,b_\lambda;c;z_\lambda
\right)
\longrightarrow
{}_0F_1
\left(
;c;\zeta
\right),
\qquad
a_\lambda b_\lambda z_\lambda
\longrightarrow
\zeta ,
\end{split}
\label{eq:2F1-to-0F1}
\end{equation}
provided $a_\lambda,b_\lambda\rightarrow\infty$ and
$z_\lambda\rightarrow0$ in the correlated manner above. We further note that $f(r)$ power prefactors in the solutions tend to unity in this limit. In either case, the argument of ${}_0F_1$ is $\zeta=-\frac{k^2r^2}{4}$ with $k\equiv \frac{\om}{c}$. This then leads to the limit
\begin{equation}
\begin{split}
 \Xi_{\rm Reg}^{\rm flat}
 &\equiv \frac{r^{ l }}{2\nu}
 {}_0F_1\!\left(;1+\nu;-\frac{k^{2}r^{2}}{4}\right),\quad
 \Xi_{\rm Sing}^{\rm flat}
 \equiv r^{-( l +D-2)}
 {}_0F_1\!\left(;1-\nu;-\frac{k^{2}r^{2}}{4}\right).
\end{split}
\label{eq:flat-reg-sing-0f1}
\end{equation}
If instead of holding $\mu$ fixed, we keep the physical mass $m_\phi$ fixed as we take the limit, we get the same result after replacing $k\equiv \omega/c$ by 
\begin{equation}
\begin{split}
 k^{2}\equiv \frac{\omega^{2}}{ c^{2}}-m_{\phi}^{2}c^{2}.
\end{split}
\label{eq:flat-k}
\end{equation}
Using the identity 
\begin{equation}
\begin{split}
{}_0F_1
\left(
;1+\nu;
-\frac{x^2}{4}
\right)
=
\Gamma(1+\nu)
\left(
\frac{2}{x}
\right)^\nu
J_\nu(x),
\end{split}
\label{eq:0F1-Bessel}
\end{equation}
we can can recognise the above solutions as the standard spherical Bessel eigen-functions of the flat Helmholtz operator.
For non-integer $\nu$, $J_\nu$ and $J_{-\nu}$ form a convenient pair of
independent solutions. At integer $\nu$ (i.e., for even $D$) this basis degenerates, and an appropriate amount of regular solution has to be subtracted from the singular solution at a general $\nu$ before we take the limit of $\nu$ tending to an integer, anf obtain a solution proportional to  Bessel $Y_\nu$ function.

\subsection{Radiation zone: Jost solutions}
\label{ssec:RadZone}
Next we now turn to the \emph{Jost solutions} of the (A)dS radial problem. Let us first recall their role in ordinary scattering theory. For a radial Schr\"odinger problem with a sufficiently short-ranged potential, one distinguishes the solution regular at the origin from two solutions which, at large radius, carry purely outgoing or incoming wave behaviour. These are the Jost solutions. The coefficients relating the regular solution to the Jost pair\footnote{or equivalently Wronskian with Jost pairs} are the Jost functions. The scattering matrix is given by a ratio of these coefficients, while their zeros and poles locate bound states and resonances. This gives a common language for the scattering and bound-state problems. Standard references are Newton's textbook on scattering\cite{Newton:1982qc} and the monograph on Jost functions by Rakityansky \cite{Rakityansky2022Jost}.

Schematically, let $\varphi_{ l }$ be the regular radial solution and $f_{ l }^{(\pm)}$ the asymptotic Jost solutions. The scattering amplitude $S_{ l }(k)$ is then given by the ratio of the Jost functions $A_{ l }^{(\pm)}$, i.e., the coefficient functions in the expansion of $\varphi_{ l }$ in terms of $f_{ l }^{(\pm)}$:
\begin{equation}
\begin{split}
 \varphi_{ l }(k,r)
 =A_{ l }^{(-)}(k)f_{ l }^{(-)}(k,r)
  +A_{ l }^{(+)}(k)f_{ l }^{(+)}(k,r),
 \qquad
 S_{ l }(k)=\frac{A_{ l }^{(+)}(k)}{ A_{ l }^{(-)}(k)},
\end{split}
\label{eq:qm-jost-schematic}
\end{equation}
for an appropriate normalisation of the two asymptotic waves. The basic idea is to choose a basis adapted to the second end of the radial problem and use its connection coefficients with the origin basis to separate local propagation from the global boundary condition.

Our hypergeometric problem has the same structure, with one important difference.  In the dS static patch the second regular singular point is a genuine Lorentzian horizon, so the two Jost solutions are literally outgoing and incoming waves.  In global AdS the corresponding singular point lies at the imaginary radius $r=\pm \compi c/\omA $ with no actual  flux surface there.  The useful AdS Jost pair is instead obtained by analytically continuing the physical dS outgoing and incoming solutions through conjugate paths in the curvature plane.  This gives two Schwarz-conjugate boundary values on the physical AdS radial axis, naturally adapted to the real, perfectly reflecting AdS cavity.

To see how this works, we begin with dS, where this distinction does not arise.  With
\begin{equation}
\begin{split}
 u_{\rm dS}\equiv \frac{\hds ^{2}r^{2}}{ c^{2}},
 \qquad
 f(r)\equiv 1-u_{\rm dS},
 \qquad
 \varpi\equiv \frac{\omega}{ \hds },
\end{split}
\label{eq:ds-jost-vars}
\end{equation}
the outgoing solution at the future cosmological horizon is chosen with unit coefficient of $\Xi_{\rm Sing}^{\rm dS}$ at the origin,
\begin{equation}
\begin{split}
 \Xi_{+}^{\rm dS}
 &\equiv \Xi_{\rm Sing}^{\rm dS}-K_{+}^{\rm dS}\Xi_{\rm Reg}^{\rm dS}\nonumber\\
 &\equiv \left(\frac{\hds }{ c}\right)^{2\nu}
 \frac{\Gamma\!\left(\frac{1+\nu+\mu-\compi\varpi}{2}\right)
  \Gamma\!\left(\frac{1+\nu-\mu-\compi\varpi}{2}\right)
  }{
  \Gamma(1-\compi\varpi)\Gamma(\nu)}
 r^{ l }f(r)^{-\compi\varpi/2}\nonumber\\
 &\hspace{0.5cm}\times
 {}_2F_1\!\left(
 \frac{1+\nu+\mu-\compi\varpi}{2},
 \frac{1+\nu-\mu-\compi\varpi}{2};
 1-\compi\varpi;f(r)\right).
\end{split}
\label{eq:ds-jost-solutions}
\end{equation}
The incoming solution is its frequency reverse,
\begin{equation}
\begin{split}
 \Xi_{-}^{\rm dS}(\omega,r)=\Xi_{+}^{\rm dS}(-\omega,r),
 \qquad
 \Xi_{-}^{\rm dS}
 =\Xi_{\rm Sing}^{\rm dS}-K_{-}^{\rm dS}\Xi_{\rm Reg}^{\rm dS}.
\end{split}
\label{eq:ds-jost-frequency-reversal}
\end{equation}
Near the horizon,
\begin{equation}
\begin{split}
 \Xi_{+}^{\rm dS}\sim f(r)^{-\compi\varpi/2},
 \qquad
 \Xi_{-}^{\rm dS}\sim f(r)^{+\compi\varpi/2}.
\end{split}
\label{eq:ds-horizon-behaviour}
\end{equation}
The $+$ solution is analytic in the upper half-plane of complex $\varpi$ and is retarded/outgoing, whereas the $-$ solution is advanced/incoming. For real $\omega,\mu,\nu$, the two are also complex conjugates.  

The corresponding Jost functions are
\begin{equation}
\begin{split}
 K_{+}^{\rm dS}
 &\equiv 2\left(\frac{\hds }{ c}\right)^{2\nu}
 \frac{\Gamma(1-\nu)}{\Gamma(\nu)}
 \frac{\Gamma\!\left(\frac{1+\nu+\mu-\compi\varpi}{2}\right)
  \Gamma\!\left(\frac{1+\nu-\mu-\compi\varpi}{2}\right)
  }{
  \Gamma\!\left(\frac{1-\nu+\mu-\compi\varpi}{2}\right)
  \Gamma\!\left(\frac{1-\nu-\mu-\compi\varpi}{2}\right)},\\
 K_{-}^{\rm dS}(\omega)&\equiv K_{+}^{\rm dS}(-\omega).
\end{split}
\label{eq:ds-Kpm}
\end{equation}
For real parameters, $K_{-}^{\rm dS}=K_{+}^{\rm dS\,*}$.  In the notation of \cite{Loganayagam:2023pfb}, $K_{+}^{\rm dS}$ is the outgoing static-patch response coefficient after converting their Eddington--Finkelstein radial function to the static-time convention used here.  We will therefore interpret it as an observer-centred response or radiation-reaction kernel.  As emphasised in that work, one should first simplify the complete Gamma-function ratio before assigning significance to individual apparent poles; in many massless sectors, the answer reduces to a finite polynomial in frequency.

We now define the AdS pair by analytically continuing these two physical dS solutions. The important point is that a real dS solution has a definite Schwarz-conjugation property.  If the outgoing solution is continued through the lower half of the curvature plane,
\begin{equation}
\begin{split}
 \hds \mapsto e^{-\compi\pi/2}\omA ,
\end{split}
\label{eq:lower-curvature-rotation}
\end{equation}
then its complex conjugate is naturally continued through the conjugate path
\begin{equation}
\begin{split}
 \hds \mapsto e^{+\compi\pi/2}\omA .
\end{split}
\label{eq:upper-curvature-rotation}
\end{equation}
We therefore define
\begin{equation}
\begin{split}
 \Xi_{+}^{\rm AdS}
 &\equiv
 \left.\Xi_{+}^{\rm dS}\right|_{\hds \to-\compi\omA },
 \\
 \Xi_{-}^{\rm AdS}
 &\equiv
 \left.\Xi_{-}^{\rm dS}\right|_{\hds \to+\compi\omA }.
\end{split}
\label{eq:ads-jost-continuation}
\end{equation}
These are boundary values on the two opposite sheets associated with the imaginary-horizon singularity.  They are not obtained from one another simply by $\omega\to-\omega$ at fixed AdS sheet.  Instead, for real $\omega,\mu,\nu$,
\begin{equation}
\begin{split}
 \Xi_{-}^{\rm AdS}(\omega,r)
 =\Xi_{+}^{\rm AdS}(\omega,r)^{*}.
\end{split}
\label{eq:ads-jost-conjugacy}
\end{equation}
This property will be useful for interpreting the AdS cavity.

We retain the same origin normalisation,
\begin{equation}
\begin{split}
 \Xi_{+}^{\rm AdS}
 &=\Xi_{\rm Sing}^{\rm AdS}-K_{+}^{\rm AdS}\Xi_{\rm Reg}^{\rm AdS},\\
 \Xi_{-}^{\rm AdS}
 &=\Xi_{\rm Sing}^{\rm AdS}-K_{-}^{\rm AdS}\Xi_{\rm Reg}^{\rm AdS}.
\end{split}
\label{eq:ads-jost-origin-def}
\end{equation}
The  functions appearing here are
\begin{equation}
\begin{split}
 K_{+}^{\rm AdS}
 &\equiv 2e^{-\compi\pi\nu}\left(\frac{\omA }{ c}\right)^{2\nu}
 \frac{\Gamma(1-\nu)}{\Gamma(\nu)}
 \frac{\Gamma\!\left(\frac{1+\nu+\mu+\widehat\omega}{2}\right)
  \Gamma\!\left(\frac{1+\nu-\mu+\widehat\omega}{2}\right)
  }{
  \Gamma\!\left(\frac{1-\nu+\mu+\widehat\omega}{2}\right)
  \Gamma\!\left(\frac{1-\nu-\mu+\widehat\omega}{2}\right)},\\
 K_{-}^{\rm AdS}
 &\equiv 2e^{+\compi\pi\nu}\left(\frac{\omA }{ c}\right)^{2\nu}
 \frac{\Gamma(1-\nu)}{\Gamma(\nu)}
 \frac{\Gamma\!\left(\frac{1+\nu+\mu+\widehat\omega}{2}\right)
  \Gamma\!\left(\frac{1+\nu-\mu+\widehat\omega}{2}\right)
  }{
  \Gamma\!\left(\frac{1-\nu+\mu+\widehat\omega}{2}\right)
  \Gamma\!\left(\frac{1-\nu-\mu+\widehat\omega}{2}\right)}.
\end{split}
\label{eq:ads-KRA}
\end{equation}
Thus $K_{-}^{\rm AdS}=K_{+}^{\rm AdS\,*}$ for real parameters.  The notation $\pm$ signifies that, \emph{after taking the flat limit}, these solutions reduce to retarded/outgoing and advanced/incoming waves.  The above solutions should not be confused with the fully dressed global-AdS retarded and advanced Green functions, which also require the AdS boundary condition and a prescription for the normal-mode poles.

Let us now check the flat limit.  At fixed $\mu$ and fixed non-zero $\omega$, we have the Stirling approximation
\begin{equation}
\begin{split}
 \frac{\Gamma\!\left(\frac{1+\nu+\mu+\widehat\omega}{2}\right)
  \Gamma\!\left(\frac{1+\nu-\mu+\widehat\omega}{2}\right)
  }{
  \Gamma\!\left(\frac{1-\nu+\mu+\widehat\omega}{2}\right)
  \Gamma\!\left(\frac{1-\nu-\mu+\widehat\omega}{2}\right)}
 \sim \left(\frac{\widehat\omega}{2}\right)^{2\nu},
 \qquad
 \widehat\omega\to\infty.
\end{split}
\label{eq:gamma-ratio-flat-asymptotic}
\end{equation}
Hence
\begin{equation}
\begin{split}
 K_{+}^{\rm AdS}&\longrightarrow
 K_{+}^{\rm flat}
 \equiv 2e^{-\compi\pi\nu}\frac{\Gamma(1-\nu)}{\Gamma(\nu)}
 \left(\frac{\omega}{2c}\right)^{2\nu},\\
 K_{-}^{\rm AdS}&\longrightarrow
 K_{-}^{\rm flat}
 \equiv 2e^{+\compi\pi\nu}\frac{\Gamma(1-\nu)}{\Gamma(\nu)}
 \left(\frac{\omega}{2c}\right)^{2\nu}.
\end{split}
\label{eq:flat-jost-def}
\end{equation}
The two characteristic phases $e^{\mp \compi\pi\nu}$ are therefore already explicit in the two AdS sheets.  In the same limit
\begin{equation}
\begin{split}
 \Xi_{+}^{\rm AdS}\longrightarrow\Xi_{+}^{\rm flat},
 \qquad
 \Xi_{-}^{\rm AdS}\longrightarrow\Xi_{-}^{\rm flat},
\end{split}
\label{eq:ads-flat-jost-limit}
\end{equation}
where, using the origin basis of Eq.~\eqref{eq:flat-reg-sing-0f1},
\begin{equation}
\begin{split}
 \Xi_{\pm}^{\rm flat}
 \equiv {}&r^{-( l +D-2)}
 {}_0F_1\!\left(;1-\nu;-\frac{k^{2}r^{2}}{4}\right)\\
 &-e^{\mp \compi\pi\nu}\frac{\Gamma(1-\nu)}{\Gamma(1+\nu)}
 \left(\frac{\omega}{2c}\right)^{2\nu}r^{ l }
 {}_0F_1\!\left(;1+\nu;-\frac{k^{2}r^{2}}{4}\right)\ ,
\end{split}
\label{eq:flat-jost-0f1}
\end{equation}
with $k\equiv \om/c$. We recognise these as the outgoing and incoming Hankel functions.  The dS pair has the same limit directly,
\begin{equation}
\begin{split}
 \Xi_{+}^{\rm dS}\longrightarrow\Xi_{+}^{\rm flat},
 \qquad
 \Xi_{-}^{\rm dS}\longrightarrow\Xi_{-}^{\rm flat},
 \qquad \hds \to0^{+}\ .
\end{split}
\label{eq:ds-flat-jost-limit}
\end{equation}

We stress that the word ``Jost'' refers here to a choice of radial basis, and not to a new boundary condition. In dS, the Jost basis is the physical out/in basis at the cosmological horizon. In AdS, it is the two-sheet analytic continuation of this basis and gives a travelling-wave decomposition of a globally standing-wave problem. As we shall see shortly, the AdS boundary condition, and hence the normal-mode spectrum, enters only when this local Jost pair is recombined into the dominant or subdominant solutions at infinity.

\subsection{Cosmic zone : Dominant vs subdominant solutions}
\label{ssec:CosmZone}

We now turn to the third regular singular point of the hypergeometric equation.  This is the AdS conformal boundary, while in dS it becomes future timelike infinity after analytic continuation through the static horizon.  The same two Frobenius exponents therefore have different physical interpretations in the two geometries. We write
\begin{equation}
\begin{split}
 \sigma=-1&:\quad {\rm Dom},
 \qquad
 \sigma=+1:\quad {\rm Sub},
 \qquad
 \Delta_{\sigma}\equiv \frac{D}{2}+\sigma\mu.
\end{split}
\label{eq:sigma-dom-sub}
\end{equation}
For $\Re\mu>0$, the dominant branch carries the slower $\Delta_-$ falloff and the subdominant branch the faster $\Delta_+$ falloff(and hence the labels).  When $\mu\equiv \compi\rho$ in the dS principal series, neither branch dominates in magnitude and the words Dom/Sub should simply be regarded as analytic labels.

We first define the asymptotic solutions with unit coefficient of $\Xi_{\rm Sing}$ at the origin,
\begin{equation}
\begin{split}
 \widehat\Xi_{\rm Dom,Sub}
 =\Xi_{\rm Sing}-K_{\rm Dom,Sub}\Xi_{\rm Reg}.
\end{split}
\label{eq:hatXi-sigma-origin}
\end{equation}
For AdS, we have
\begin{equation}
\begin{split}
 K_{\rm Dom,Sub}^{\rm AdS}
 \equiv {}&2\left(\frac{\omA }{ c}\right)^{2\nu}
 \frac{\Gamma(1-\nu)}{\Gamma(\nu)}
 \frac{\Gamma\!\left(\frac{1+\nu+\sigma\mu+\widehat\omega}{2}\right)
  \Gamma\!\left(\frac{1+\nu+\sigma\mu-\widehat\omega}{2}\right)
  }{
  \Gamma\!\left(\frac{1-\nu+\sigma\mu+\widehat\omega}{2}\right)
  \Gamma\!\left(\frac{1-\nu+\sigma\mu-\widehat\omega}{2}\right)}\ ,
\end{split}
\label{eq:ads-Kdomsub}
\end{equation}
where we take $\sigma=\mp 1$ for the Dom/Sub solutions, respectively.
This can be derived using a boundary-adapted solution, whose last hypergeometric argument vanishes at $r=\infty$, i.e.,
\begin{equation}
\begin{split}
 \widehat\Xi_{\rm Dom,Sub}^{\rm AdS}
 \equiv {}&\left(\frac{\omA }{ c}\right)^{2\nu}
 \frac{\Gamma\!\left(\frac{1+\nu+\sigma\mu-\widehat\omega}{2}\right)
  \Gamma\!\left(\frac{1+\nu+\sigma\mu+\widehat\omega}{2}\right)
  }{\Gamma(1+\sigma\mu)\Gamma(\nu)}\\
 &\times r^{ l }f(r)^{-\widehat\omega/2}
 (-u_{\rm AdS})^{-\frac{1+\nu+\sigma\mu-\widehat\omega}{2}}\\
 &\times{}_2F_1\!\left(
 \frac{1+\nu+\sigma\mu-\widehat\omega}{2},
 \frac{1-\nu+\sigma\mu-\widehat\omega}{2};
 1+\sigma\mu;\frac{1}{ u_{\rm AdS}}\right).
\end{split}
\label{eq:ads-hatXi-domsub}
\end{equation}
Thus, we have the falloffs
\begin{equation}
\begin{split}
 \widehat\Xi_{\rm Dom}^{\rm AdS}\sim r^{-\Delta_-},
 \qquad
 \widehat\Xi_{\rm Sub}^{\rm AdS}\sim r^{-\Delta_+}\ ,
\end{split}
\label{eq:ads-domsub-falloff}
\end{equation}
Re-expanding this solution in terms of the origin-adapted hypergeometrics gives the $K_{\rm Dom,Sub}^{\rm AdS}$ quoted above. The same expansion fixes the normalisation of $\widehat\Xi_{\rm Dom,Sub}^{\rm AdS}$ by requiring the coefficient of $\Xi_{\rm Sing}$ to be unity. This differs from the usual AdS/CFT normalisation, but is convenient for studying near-origin physics, the PN expansion, and the flat limit. For standard AdS quantisation, the two solutions are, up to this non-standard normalisation, respectively the source/non-normalisable and response/normalisable falloffs.\footnote{  In the alternate-quantisation window, the admissible choice of source and response can be exchanged, but the two radial subspaces themselves are unchanged.}

To interpret $K_{\rm Dom,Sub}^{\rm AdS}$, it is useful to move away from the unit-singular normalisation in Eq.\eqref{eq:hatXi-sigma-origin}. We instead define solutions with unit coefficient for $\Xi_{+}^{\rm AdS}$:
\begin{equation}
\begin{split}
 \Xi_{\rm Dom,Sub}^{\rm AdS}
 &\equiv(1+\mathfrak R_{\rm Dom,Sub}^{\rm AdS})
 \widehat\Xi_{\rm Dom,Sub}^{\rm AdS}
 =\Xi_{+}^{\rm AdS}
  +\mathfrak R_{\rm Dom,Sub}^{\rm AdS}\Xi_{-}^{\rm AdS}.
\end{split}
\label{eq:ads-Xi-sigma-jost}
\end{equation}
We will interpret $R_{\rm Dom,Sub}^{\rm AdS}$ as the AdS reflection coefficient: it gives the amount of the incoming echo $\Xi_{-}^{\rm AdS}$ that must be added to the outgoing wave $\Xi_{+}^{\rm AdS}$ to obtain the Dom/Sub solutions.

To compute $R_{\rm Dom,Sub}^{\rm AdS}$, we proceed as follows.
since both $\Xi_\pm$  have unit $\Xi_{\rm Sing}$ coefficient, comparison at the origin gives
\begin{equation}\label{eq:domsubReExp}
\begin{split}
 \Xi_{\rm Dom,Sub}^{\rm AdS}
 &=(1+\mathfrak R_{\rm Dom,Sub}^{\rm AdS})\Xi_{\rm Sing}^{\rm AdS}
 -\left(K_{+}^{\rm AdS}
       +\mathfrak R_{\rm Dom,Sub}^{\rm AdS}K_{-}^{\rm AdS}\right)
  \Xi_{\rm Reg}^{\rm AdS}\ .
\end{split}
\end{equation}
Normalising this solution to have a unit coefficient for $\Xi_{\rm Sing}^{\rm AdS}$, we obtain
\begin{equation}
\begin{split}  
 K_{\rm Dom,Sub}^{\rm AdS}
 &=\frac{K_{+}^{\rm AdS}
   +\mathfrak R_{\rm Dom,Sub}^{\rm AdS}K_{-}^{\rm AdS}
   }{
   1+\mathfrak R_{\rm Dom,Sub}^{\rm AdS}}\ \implies\ 
 \mathfrak R_{\rm Dom,Sub}^{\rm AdS}
 \equiv \frac{K_{\rm Dom,Sub}^{\rm AdS}-K_{+}^{\rm AdS}
   }{ K_{-}^{\rm AdS}-K_{\rm Dom,Sub}^{\rm AdS}}.
\end{split}
\label{eq:ads-R-K-crossratio}
\end{equation}
Substituting the expressions for $K_{\rm Dom,Sub}^{\rm AdS}$ and $K_{\pm}^{\rm AdS}$, the Gamma functions simplify by the reflection formula and give
\begin{equation}
\begin{split}
 \boxed{\displaystyle
 \mathfrak R_{\rm Dom,Sub}^{\rm AdS}
 \equiv \exp\!\left[\compi\pi
 \left(\widehat\omega-\nu-\sigma\mu\right)\right].}
\end{split}
\label{eq:ads-pure-reflection-phase}
\end{equation}
For real frequency,
\begin{equation}
\begin{split}
 |\mathfrak R_{\rm Dom,Sub}^{\rm AdS}|=1.
\end{split}
\label{eq:ads-unit-reflection}
\end{equation}
This is the expected result for global AdS: the boundary is lossless, and a wave sent out from the central region returns with a phase but no attenuation.

The frequency-dependent part of this phase has a simple geometrical interpretation:
\begin{equation}
\begin{split}
 \mathfrak R_{\rm Dom,Sub}^{\rm AdS}
 =e^{\compi\omega T_{\rm echo}}
  e^{-\compi\pi(\nu+\sigma\mu)},
 \qquad
 T_{\rm echo}=\frac{\pi}{\omA }.
\end{split}
\label{eq:echo-time}
\end{equation}
A radial null ray takes $\pi/(2\omA )$ to travel from the centre to the boundary, so $T_{\rm echo}$ is the round-trip time.  The second factor is a frequency-independent monodromy phase determined by the radial index and the boundary falloff.  Thus the AdS reflection coefficient separates the propagation across the cavity from the phase acquired in matching to the chosen asymptotic branch.

We can now read off the normal-mode quantisation.  The coefficient of the singular origin solution in Eq.~\eqref{eq:ads-R-K-crossratio} is $1+\mathfrak R_{\rm Dom,Sub}^{\rm AdS}$.  Hence the boundary-selected solution is regular at the origin precisely when
\begin{equation}
\begin{split}
 1+\mathfrak R_{\rm Dom,Sub}^{\rm AdS}=0.
\end{split}
\label{eq:ads-normal-mode-standing-wave}
\end{equation}
At such a frequency the remaining regular coefficient in Eq.\eqref{eq:domsubReExp} is proportional to $K_--K_+$ and is generically non-zero so the solution does not vanish. Thus the singular pieces of the two conjugate Jost waves cancel, leaving a regular solution.  For the standard subdominant branch,
\begin{equation}
\begin{split}
 \widehat\omega_{n l }
 \equiv 1+\nu+\mu+2n
 =\Delta_++ l +2n,
 \qquad n=0,1,2,\ldots,
\end{split}
\label{eq:ads-normal-spectrum}
\end{equation}
whereas the dominant branch gives
\begin{equation}
\begin{split}
 \widehat\omega_{n l }^{(-)}
 \equiv 1+\nu-\mu+2n
 =\Delta_-+ l +2n.
\end{split}
\label{eq:alternate-spectrum}
\end{equation}
The latter is the alternate-quantisation spectrum when that boundary condition is admissible. The same frequencies can equivalently be obtained from the poles of $K_{\rm Sub}^{\rm AdS}$ or $K_{\rm Dom}^{\rm AdS}$, respectively. With the standard $\Delta_+$ boundary condition,  the normalisable radial bulk-to-bulk Green function is given as
\begin{equation}
\begin{split}
 G_{ l \omega}(r_1,r_2)
 &\equiv 
 \Xi_{\rm Reg}(r_<)\,\widehat\Xi_{\rm Sub}(r_>)\\
 &=
 \Xi_{\rm Reg}(r_<)\Xi_{\rm Sing}(r_>)
 -K_{\rm Sub}^{\rm AdS}(\omega)
  \Xi_{\rm Reg}(r_1)\Xi_{\rm Reg}(r_2)\ .
\end{split}
\end{equation}
Hence the same normal-mode poles also appear in the normalisable bulk-to-bulk Green function.  

Let us now turn to dS.  The asymptotic point $u_{\rm dS}=\infty$ lies beyond the static horizon and corresponds, in the expanding Poincare description, to future timelike infinity.  The solutions with unit coefficient for $\Xi_{\rm Sing}$ are\footnote{There is a branch cut ambiguity here in this analytic continuation. We will work with $\text{Arg}(-u_{\rm dS})=-\pi$ branch.}
\begin{equation}
\begin{split}
 \widehat\Xi_{\rm Dom,Sub}^{\rm dS}
 \equiv {}&e^{-\compi\pi\nu}\left(\frac{\hds }{ c}\right)^{2\nu}
 \frac{\Gamma\!\left(\frac{1+\nu+\sigma\mu-\compi\varpi}{2}\right)
  \Gamma\!\left(\frac{1+\nu+\sigma\mu+\compi\varpi}{2}\right)
  }{\Gamma(1+\sigma\mu)\Gamma(\nu)}\\
 &\times r^{ l }f(r)^{-\compi\varpi/2}
 (-u_{\rm dS})^{-\frac{1+\nu+\sigma\mu-\compi\varpi}{2}}\\
 &\times{}_2F_1\!\left(
 \frac{1+\nu+\sigma\mu-\compi\varpi}{2},
 \frac{1-\nu+\sigma\mu-\compi\varpi}{2};
 1+\sigma\mu;\frac{1}{ u_{\rm dS}}\right).
\end{split}
\label{eq:ds-hatXi-domsub}
\end{equation}
They correspond to the late-time expansion
\begin{equation}
\begin{split}
 \phi\sim(-\eta)^{\Delta_-}\phi_-(\vec y)
       +(-\eta)^{\Delta_+}\phi_+(\vec y).
\end{split}
\label{eq:ds-late-time}
\end{equation}
For a light field these are the growing/dominant and decaying/subdominant cosmological branches.  For $\mu\equiv \compi\rho$ they are the two oscillatory principal-series branches, neither one more normalisable than the other at a timelike boundary. We can write, as in AdS,
\begin{equation}
\begin{split}
 \widehat\Xi_{\rm Dom,Sub}
 =\Xi_{\rm Sing}-K_{\rm Dom,Sub}\Xi_{\rm Reg}\ ,
\end{split}
\label{eq:hatXi-sigma-origin-dS}
\end{equation}
with
\begin{equation}
\begin{split}
 K_{\rm Dom,Sub}^{\rm dS}
 \equiv {}&2e^{-\compi\pi\nu}\left(\frac{\hds }{ c}\right)^{2\nu}
 \frac{\Gamma(1-\nu)}{\Gamma(\nu)}
 \frac{\Gamma\!\left(\frac{1+\nu+\sigma\mu-\compi\varpi}{2}\right)
  \Gamma\!\left(\frac{1+\nu+\sigma\mu+\compi\varpi}{2}\right)
  }{
  \Gamma\!\left(\frac{1-\nu+\sigma\mu-\compi\varpi}{2}\right)
  \Gamma\!\left(\frac{1-\nu+\sigma\mu+\compi\varpi}{2}\right)}.
\end{split}
\label{eq:ds-Kdomsub}
\end{equation}
Re-expanding late-time solutions in terms of  the  horizon Jost pair, we can write 
\begin{equation}
\begin{split}
 \Xi_{\rm Dom,Sub}^{\rm dS}
 &\equiv(1+\mathfrak R_{\rm Dom,Sub}^{\rm dS})\widehat\Xi_{\rm Dom,Sub}^{\rm dS}
 =\Xi_{+}^{\rm dS}+\mathfrak R_{\rm Dom,Sub}^{\rm dS}\Xi_{-}^{\rm dS},
\end{split}
\label{eq:ds-Xi-sigma-jost}
\end{equation}
with
\begin{equation}
\begin{split}
 \mathfrak{R}_{\rm Dom,Sub}^{\rm dS}
 \equiv -e^{-\pi\varpi}
 \frac{\sin\!\left[\frac{\pi}{2}(1-\nu+\sigma\mu-\compi\varpi)\right]
  }{
  \sin\!\left[\frac{\pi}{2}(1-\nu+\sigma\mu+\compi\varpi)\right]}.
\end{split}
\label{eq:R-ds-causal}
\end{equation}
For real $\mu,\nu,\omega$ the sine ratio is a phase and therefore
\begin{equation}
\begin{split}
 |\mathfrak{R}_{\rm Dom}^{\rm dS}|=|\mathfrak{R}_{\rm Sub}^{\rm dS}|
 =e^{-\pi\omega/\hds },
 \qquad
 |\mathfrak{R}_{\rm Dom,Sub}^{\rm dS}|^{2}=e^{-2\pi\omega/\hds }.
\end{split}
\label{eq:thermal-magnitude}
\end{equation}
This coefficient measures the mixture of incoming and outgoing static-horizon waves needed to construct a pure late-time cosmological solution. Its magnitude is the thermal Boltzmann factor, while its phase depends on the choice of the $\Delta_{\sigma}$ branch. In AdS, by contrast, $\mathfrak{R}_{\rm Dom,Sub}^{\rm AdS}$ is a unit-modulus reflection phase in a conservative cavity.

There is also a useful analytic relation between the two cases.  The frequency-dependent part of the AdS phase is
\begin{equation}
\begin{split}
 e^{\compi\pi\omega/\omA }.
\end{split}
\end{equation}
Under the Wick rotation of the curvature scale this becomes
\begin{equation}
\begin{split}
 e^{\compi\pi\omega/\omA }
 \longrightarrow e^{-\pi\omega/\hds }.
\end{split}
\label{eq:reflection-to-thermal}
\end{equation}
Up to the frequency-independent monodromy phase, the oscillatory AdS echo factor therefore continues to the dS thermal attenuation. In this sense, reflection from a lossless AdS boundary and Hawking mixing in dS are related by analytic continuation.\footnote{See \cite{Sleight:2021plv} for a detailed exploration of how such an analytic continuation works in the Mellin space.}

The two interpretations of the asymptotic basis may be summarised as follows:
\begin{center}
\begin{tabular}{@{}lll@{}}
\toprule
 & AdS & dS \\
\midrule
$\Xi_{\rm Dom}$ & $r^{-\Delta_-}$, source/non-normalisable & growing or late-time dominant branch \\
$\Xi_{\rm Sub}$ & $r^{-\Delta_+}$, response/normalisable & decaying or late-time subdominant branch \\
middle point & imaginary Jost point & cosmological horizon \\
global real $\omega$ & reflecting cavity & continuous static-patch response \\
\bottomrule
\end{tabular}
\end{center}
The table is phrased at the level of classical solutions.  In AdS the standard boundary condition promotes the subdominant regular solutions to a positive-energy Hilbert-space basis.  In dS the same special-function branches are more naturally thought of as cosmological data which a static observer resolves into incoming and outgoing horizon components.

Let us finally discuss the flat limit in the two cases.  In dS, for fixed positive real $\omega$,
\begin{equation}
\begin{split}
 |\mathfrak{R}_{\rm Dom,Sub}^{\rm dS}|
 =e^{-\pi\omega/\hds }\longrightarrow0,
 \qquad
 \Xi_{\rm Dom}^{\rm dS},\Xi_{\rm Sub}^{\rm dS}
 \longrightarrow\Xi_{+}^{\rm flat}.
\end{split}
\label{eq:ds-domsub-flat}
\end{equation}
The cosmological horizon is pushed to infinite distance and the incoming component is exponentially suppressed.  In AdS the reflection amplitude remains of unit modulus on the real frequency axis,
\begin{equation}
\begin{split}
 \mathfrak {R}_{\sigma}^{\rm AdS}
 =e^{\compi\omega T_{\rm echo}}e^{-\compi\pi(\nu+\sigma\mu)},
 \qquad T_{\rm echo}\to\infty,
\end{split}
\end{equation}
so there is no pointwise real-frequency limit of the global standing wave as the normal-mode poles become dense.   With a retarded prescription,  we can keep $\operatorname{Im}\omega>0$  fixed as $\omA \to0$, 
\begin{equation}
\begin{split}
 \left|e^{\compi(\omega+\compi\epsilon)T_{\rm echo}}\right|
 =e^{-\epsilon T_{\rm echo}}\longrightarrow0,
\end{split}
\label{eq:ads-echo-flat-suppression}
\end{equation}
and the boundary echoes are exponentially suppressed. The same outgoing flat solution is recovered, i.e., $\Xi_{+}\to\Xi_{+}^{\rm flat}$ in this limit. The advanced prescription, with $\operatorname{Im}\omega <0$ fixed as $\omA \to0$, similarly gives the incoming branch $\Xi_{-}\to\Xi_{-}^{\rm flat}$. 
 
Thus dS reaches the causal flat problem by thermally suppressing the unwanted horizon component, whereas AdS does so by pushing the reflected echo to infinite time. At the level of the radial ODE, the three regular singular points of the finite-curvature hypergeometric equation coalesce in the flat limit into a regular singular point at the origin and an irregular point at infinity.  The origin basis becomes Eq.~\eqref{eq:flat-reg-sing-0f1} in the near-zone, while the Jost pair becomes Eq.~\eqref{eq:flat-jost-0f1} of the radiation-zone.  The distinction between Dom and Sub, which is specific to the cosmic-zone, disappears in this limit. Thus both the discrete reflecting cavity of global AdS and the absorbing horizon of dS reduce to the ordinary causal radiation problem when the curvature scale is removed.

We end with two standard qualifications.  When $\nu\in\mathbb Z$ (i.e., for even $D$) the origin Frobenius basis is degenerate and the expressions containing $\Gamma(1-\nu)$ are understood by analytic continuation and the corresponding local renormalisation; logarithms can then appear\cite{Loganayagam:2023pfb}. When $2\mu\in\mathbb Z$, the two asymptotic exponents are resonant and the boundary basis can likewise acquire logarithmic terms.  We will therefore understand the connection formulae as first derived for generic $(\nu,\mu)$ and only then continued to the degenerate limit.  Neither complication changes the organisation described above: the origin basis controls local multipoles, the Jost basis controls causal response, and the Dom/Sub basis controls the global or cosmological completion of the same two-dimensional radial solution space.

\section{Multipole resummation and PN numerators}\label{app:Scalar bulk-bulk Green's function in (A)dS}

\subsection{Proof of multipole resummation formulae}\label{app:multProof}
In this section, we present a proof of the two formulae Eq.\eqref{eq:MultFunI} and \eqref{eq:MultFunII} we used to resum the multipole expansion. The two identities under question are valid for $0\leq\tau <1$, $n$ a non-negative integer, and odd $D$:: 
\begin{equation}\label{eq:MultFunIalt}\begin{split}
&\frac{\Gamma\left(\frac{D}{2}-1-n\right)}{\Gamma\left(\frac{D}{2}\right)\left(1+\tau^{2}-2z\tau\right)^{\frac{D}{2}-1-n}}=  \sum_{l}\mathcal{N}_{\text{HH}}\left(D,l\right)\mathcal{P}_{l}\left(D;z\right)\tau^l \frac{\Gamma (\nu -n) }{\Gamma (\nu +n+1)}\, n!\ \mathcal{P}^{(\nu,-2n-1)}_n(1-2\tau^2)
\ ,
\end{split}\end{equation}
as well as
\begin{equation}\label{eq:MultFunIIalt}\begin{split}
\frac{\Gamma\left(\frac{D}{2}+n\right)(1-\tau^2)^{2n+1}}{\Gamma\left(\frac{D}{2}\right)\left(1+\tau^{2}-2z\tau\right)^{\frac{D}{2}+n}}
=&  \sum_{l}\mathcal{N}_{\text{HH}}\left(D,l\right)\mathcal{P}_{l}\left(D;z\right) \tau^{l} \ n!\ \mathcal{P}^{(\nu,-2n-1)}_{n}(1-2\tau^2)\ .
\end{split}\end{equation}

To this end, we use the following identity to  rewrite the expansion formulae we want to prove:
\begin{equation}\begin{split}
\mathcal{P}_n^{\left(\nu, -1-2n\right)}\left(1-2 \tau^2\right)= (1-\tau^2)^n \mathcal{P}_n^{\left(\nu, -\nu\right)}\left(\frac{1+ \tau^2}{1-\tau^2}\right)\ .
\end{split}\end{equation}
We will now prove these identities then in the following form:
\begin{equation}\label{eq:MultNNeg}\begin{split}
&\frac{\Gamma\left(\frac{D}{2}-1-n\right)(1-\tau^2)^{-n}}{n!\ \Gamma\left(\frac{D}{2}\right)\left(1+\tau^{2}-2z\tau\right)^{\frac{D}{2}-1-n}}\\
&\quad =  \sum_{l}\mathcal{N}_{\text{HH}}\left(D,l\right)\mathcal{P}_{l}\left(D;z\right)\tau^l \frac{\Gamma (\nu -n) }{\Gamma (\nu +n+1)}\,\mathcal{P}_n^{\left(\nu, -\nu\right)}\left(\frac{1+ \tau^2}{1-\tau^2}\right)
\ ,
\end{split}\end{equation}
as well as
\begin{equation}\label{eq:MultNPos}\begin{split}
\frac{\Gamma\left(\frac{D}{2}+n\right)(1-\tau^2)^{n+1}}{n!\ \Gamma\left(\frac{D}{2}\right)\left(1+\tau^{2}-2z\tau\right)^{\frac{D}{2}+n}}
=&  \sum_{l}\mathcal{N}_{\text{HH}}\left(D,l\right)\mathcal{P}_{l}\left(D;z\right) \tau^{l} \,\mathcal{P}_n^{\left(\nu, -\nu\right)}\left(\frac{1+ \tau^2}{1-\tau^2}\right)\ .
\end{split}\end{equation}

In order to prove Eq.\eqref{eq:MultNPos}, we consider the following binomial series with a dummy  variable $s$ which sums to
\begin{equation}\begin{split}
\sum_{n=0}^\infty s^n\frac{ \Gamma\left(\frac{D}{2}+n\right)(1-\tau^2)^{n+1}}{n!\ \Gamma\left(\frac{D}{2}\right)\left(1+\tau^{2}-2z\tau\right)^{\frac{D}{2}+n}}
=& \frac{1-\tau^2}{\left(1+\tau^{2}+s(\tau^2-1)-2z\tau\right)^{\frac{D}{2}}}\ .
\end{split}\end{equation}
Here we have assumed $s$ to be appropriately small to assure convergence of this binomial series. Next, we introduce a new variable $t$ implicitly via $\tau=t\Lambda$ and $1+\tau^{2}+s(\tau^2-1)=(1+t^2)\Lambda$. An explicit form for $\Lambda$ can be given by solving the above equalities : 
\begin{equation}\begin{split}
\Lambda &\equiv \frac{1}{2}(1-\tau^2)\Bigl\{\frac{1+\tau^2}{1-\tau^2}-s+\sqrt{1-2s\frac{1+\tau^2}{1-\tau^2}+s^2}\Bigr\} \\
=&\frac{1-s+\sqrt{1-2s\frac{1+\tau^2}{1-\tau^2}+s^2}}{1+s+\sqrt{1-2s\frac{1+\tau^2}{1-\tau^2}+s^2}}\ .
\end{split}\end{equation}
We will choose positive square root in what follows.

We can then rewrite our original binomial series as
\begin{equation}\begin{split}
\sum_{n=0}^\infty \frac{s^n \Gamma\left(\frac{D}{2}+n\right)(1-\tau^2)^{n+1}}{n!\ \Gamma\left(\frac{D}{2}\right)\left(1+\tau^{2}-2z\tau\right)^{\frac{D}{2}+n}}
=&\frac{1-\tau^2}{\Lambda^{\frac{D}{2}}(1-t^2)}\times \frac{1-t^2}{\left(1+t^{2}-2zt\right)^{\frac{D}{2}}}\\
=&\frac{1-\tau^2}{\Lambda^{\frac{D}{2}}(1-t^2)}\times\sum_{l}\mathcal{N}_{\text{HH}}\left(D,l\right)\mathcal{P}_{l}\left(D;z\right)\ t^l \\
=&\frac{\Lambda(1-\tau^2)}{\Lambda^2-\tau^2}\times \sum_{l}\mathcal{N}_{\text{HH}}\left(D,l\right)\mathcal{P}_{l}\left(D;z\right)\ \frac{\tau^l}{\Lambda^\nu}\\
=&\frac{1}{\sqrt{1-2s\frac{1+\tau^2}{1-\tau^2}+s^2}}\sum_{l}\mathcal{N}_{\text{HH}}\left(D,l\right)\mathcal{P}_{l}\left(D;z\right) \frac{\tau^l }{\Lambda^\nu}\\
=&\sum_{n=0}^\infty s^n\sum_{l}\mathcal{N}_{\text{HH}}\left(D,l\right)\mathcal{P}_{l}\left(D;z\right) \tau^{l} \  \mathcal{P}_n^{\left(\nu, -\nu\right)}\left(\frac{1+ \tau^2}{1-\tau^2}\right)
\ .
\end{split}\end{equation}
In the second line, we have used the  Poisson identity (see it proved below) to do a multipole expansion. In the third line, we have substituted back for variable $t$ in terms of $\tau$, and in the last line, we have used the standard generating function of Jacobi polynomials.
Comparing the coefficients of $s^n$ in LHS and RHS then proves Eq.\eqref{eq:MultNPos}.

To prove the Poisson identity, we begin by recalling the standard Gegenbauer expansion is given by 
\begin{equation}
\frac{\Gamma\left(\frac{D}{2}-1\right)}{\Gamma\left(\frac{D}{2}\right)\left(1+\tau^{2}-2z\tau\right)^{\frac{D}{2}-1}}=  \sum_{l}\mathcal{N}_{\text{HH}}\left(D,l\right)\mathcal{P}_{l}\left(D;z\right) \frac{\tau^l}{l+\frac{D}{2}-1}
\end{equation}
This generalises the well-known generating function for Legendre polynomials in $D=3$.
The Poisson identity we used above can then be obtained by applying the operator
\begin{equation}
\frac{1}{\tau^{\frac{D}{2}-1}}\frac{\partial}{\partial \tau} \tau^{\frac{D}{2}-1} \end{equation}
applied to both sides of the above identity. This yields
\begin{equation}
\frac{1-\tau^2}{\left(1+\tau^{2}-2z\tau\right)^{\frac{D}{2}}}=\sum_{l}\mathcal{N}_{\text{HH}}\left(D,l\right)\mathcal{P}_{l}\left(D;z\right)\ \tau^l\ . \end{equation}

We now turn to the proof of  Eq.\eqref{eq:MultNNeg}. We begin by noting that, apart from a factor of $n!$ in LHS, the LHS of Eq.\eqref{eq:MultNNeg} can be thought of as analytic continuation of LHS of  Eq.\eqref{eq:MultNPos} with $n\mapsto -n-1$. The $n!$ of course has poles on non-negative integers and cannot be analytically continued this way. On RHS, the standard definition of Jacobi polynomials also does not admit its degree to be  negative. But, fortunately these two problems cancel each other as can be seen from the hyper-geometric representation 
\begin{equation}\begin{split}
m!\,\mathcal{P}_m^{\left(\nu, -\nu\right)}\left(\frac{1+ \tau^2}{1-\tau^2}\right) =\frac{\Gamma(\nu+m+1)}{\Gamma(\nu+1)} \, _2F_1\left(-m,m+1;\nu +1;-\frac{\tau ^2}{1-\tau^2}\right) \ .
\end{split}\end{equation}
Setting $m=-n-1$ in the RHS, we get an analytic continuation of LHS as
\begin{equation}\begin{split}
\lim_{m\to -n-1} &m!\,\mathcal{P}_m^{\left(\nu, -\nu\right)}\left(\frac{1+ \tau^2}{1-\tau^2}\right) =\frac{\Gamma(\nu-n)}{\Gamma(\nu+1)} \, _2F_1\left(n+1,-n;\nu +1;-\frac{\tau ^2}{1-\tau^2}\right)\\
=&\frac{\Gamma(\nu-n)}{\Gamma(\nu+n+1)} \frac{\Gamma(\nu+n+1)}{\Gamma(\nu+1)}\, _2F_1\left(-n,n+1;\nu +1;-\frac{\tau ^2}{1-\tau^2}\right)\\
=& \frac{\Gamma(\nu-n)}{\Gamma(\nu+n+1)} \times n!\,\mathcal{P}_n^{\left(\nu, -\nu\right)}\left(\frac{1+ \tau^2}{1-\tau^2}\right)\ .
\end{split}\end{equation}
Under this analytic continuation, Eq.\eqref{eq:MultNNeg} follows from  Eq.\eqref{eq:MultNPos}. 

\subsection{Explicit expressions for PN numerators}\label{app:An,Bn}

\textbf{0PN :}
\begin{equation}\label{eq:AB0PN}\begin{split}
\mathbb{A}_0(r_1,r_2) =&\frac{1}{2}-\frac{\omA^2}{2^3c^2}(r_1^2+r_2^2)\\
&+\frac{\omA^2}{2^5c^4}(\om^2-\mu^2\omA^2)(r_1^2-r_2^2)^2+\frac{\omA^4}{2^6c^4}(5r_1^4+2 r_1^2 r_2^2+5r_2^4)\\
&-\frac{\omA^4}{2^7c^6}(r_1^2+r_2^2)\Bigl\{
(5\om^2-3\mu^2\omA^2)(r_1^2-r_2^2)^2+\omA^2(7r_1^4-4 r_1^2 r_2^2+7r_2^4) \Bigr\}+\text{O}(c^{-8})\\
\end{split}\end{equation}
\\~\\
\textbf{1PN :}
\begin{equation}\label{eq:AB1PN}\begin{split}
\mathbb{B}_0(r_1,r_2) =&\frac{1}{2^3}-\frac{3\omA^2}{2^5c^2}(r_1^2+r_2^2)\\
&+\frac{\omA^2}{2^8c^4}(\om^2-\mu^2\omA^2)(r_1^2-r_2^2)^2+\frac{\omA^4}{2^8c^4}(19\ r_1^4+22\  r_1^2 r_2^2+19\ r_2^4)\\
&-\frac{\omA^4}{2^{11}c^6}(r_1^2+r_2^2)\Bigl\{
2(7\om^2-5\mu^2\omA^2)(r_1^2-r_2^2)^2+5\omA^2(25\ r_1^4+6\ r_1^2 r_2^2+25\ r_2^4) \Bigr\}\\
&+\text{O}(c^{-8})\ ,\\
\mathbb{A}_1(r_1,r_2) =&\frac{1}{2^3c^2}(\om^2-\mu^2\omA^2)+\frac{\omA^2}{2^6c^4}\left[-6\om^2+(1+2\mu^2)\omA^2\right](r_1^2+r_2^2)\\
&+\frac{\omA^2}{2^{12}c^6}\Biggl\{
2(\om^2-\mu^2\omA^2)^2(r_1^2-r_2^2)^2+ \om^2\omA^2(19\ r_1^4+22\  r_1^2 r_2^2+19\ r_2^4) \Biggr\}\\
&-\frac{\omA^6}{2^{8}c^6}\Biggl\{3(r_1^2+r_2^2)^2+\mu^2(
7\ r_1^4-2\ r_1^2r_2^2+7\ r_2^4)\Biggr\}\\
&-\frac{\omA^4}{2^{11}c^8}(r_1^2+r_2^2)\Biggl\{2(r_1^2-r_2^2)^2(7\om^4-10\mu^2\om^2\omA^2+3\mu^4\omA^4)\Biggr.\\
&\quad\Biggl.\quad+4\om^2\omA^2(31\ r_1^4+ 8\ r_1^2 r_2^2+31\ r_2^4) \Biggr\}\\
&+\frac{\omA^8}{2^{11}c^8}(r_1^2+r_2^2)\Biggl\{ 3(
7\ r_1^4+6\ r_1^2r_2^2+7\ r_2^4)
+40\mu^2(r_1^4- r_1^2 r_2^2+r_2^4) \Biggr\}+\text{O}(c^{-10})\ .
\end{split}\end{equation}
\\~\\
\textbf{2PN :}
\begin{equation}\label{eq:AB2PN}\begin{split}
\mathbb{B}_1(r_1,r_2) =&\frac{\omA^2}{2^6c^2}-\frac{5\omA^4}{2^8c^4}(r_1^2+r_2^2)\\
&+\frac{\omA^4}{3\times 2^{10}c^6}(\om^2-\mu^2\omA^2)(r_1^2-r_2^2)^2+\frac{\omA^6}{2^{11}c^6}(41\ r_1^4+58\  r_1^2 r_2^2+41\ r_2^4)+\text{O}(c^{-8})\ ,\\
\mathbb{A}_2(r_1,r_2) =&\frac{1}{2^6c^4}[\om^2-(\mu-1)^2\omA^2][\om^2-(\mu+1)^2\omA^2]\\
&+\frac{\omA^2}{2^8c^6}\left[-5\om^4+(11+6\mu^2)\om^2\omA^2-(\mu^4-\mu^2+3)\omA^4\right](r_1^2+r_2^2)+\text{O}(c^{-8})\\
\end{split}\end{equation}
\\~\\
\textbf{3PN :}
\begin{equation}\label{eq:AB3PN}\begin{split}
\mathbb{B}_2(r_1,r_2) =&\frac{\omA^4}{3\times 2^8c^4}-\frac{7\omA^6}{3\times 2^{10}c^6}(r_1^2+r_2^2)+\text{O}(c^{-8})\ ,\\
\mathbb{A}_3(r_1,r_2) =&\frac{1}{3\times 2^8c^6}[\om^2-(\mu-2)^2\omA^2][\om^2-(\mu+2)^2\omA^2][\om^2-\mu^2\omA^2]+\text{O}(c^{-8})\ .
\end{split}\end{equation}

\section{Form invariance of translation generator till 2PN}\label{app:momInv}

In this appendix, we will argue for an important property of the contact transformation in Eq.\eqref{eq:deltax} that was used to order reduce the Lagrangian: under such a transformation,  
canonical form of the AdS Noether momentum remains invariant through 2PN order. More precisely, let $\widetilde{\mathbb P}_i$ denote the translation
generator obtained by transforming the Noether charge of the
original higher-derivative action to the reduced phase space.
We will show that
\begin{equation}
\widetilde{\mathbb P}_i
=\sum_a\sqrt{f(r_a)}\,p_{ai}+\mathcal O(c^{-6}),
\qquad
f(r)=1+\frac{\omA^2r^2}{c^2},
\end{equation}
where $\vec p_a=\partial\mathcal L/\partial\vec v_a$
are the canonical momenta of the reduced Lagrangian.

Consider an infinitesimal AdS spatial translation with
parameter $\vec\varepsilon$, evaluated on the reference time
slice for which the corresponding Killing vector has no
time component. Its action on the original particle
positions is
\begin{equation}
\delta_\varepsilon\vec r_a^{\,\mathrm{old}}
=\sqrt{f(r_a^{\,\mathrm{old}})}\,\vec\varepsilon.
\end{equation}
At Newtonian order this reduces to an ordinary spatial
translation,
\begin{equation}
\delta_\varepsilon\vec r_a^{\,\mathrm{old}}
=\vec\varepsilon+\mathcal O(c^{-2}),
\qquad
\delta_\varepsilon\vec v_a^{\,\mathrm{old}}
=\mathcal O(c^{-2}).
\end{equation}
The contact shifts $\delta\vec x_a$ are of $\mathcal{O}(c^{-4})$
and depend on the particle positions only through
$\vec r_{12}=\vec r_1-\vec r_2$ and
$\hat n_{12}$. Consequently, we have
\begin{equation}
\sum_{b=1}^2
\frac{\partial\delta\vec x_a}{\partial\vec r_b}
=0.
\end{equation}
Furthermore, since
\begin{equation}
\begin{split}
\delta_\varepsilon\vec r_{12}
&=\left(\sqrt{f(r_1)}-\sqrt{f(r_2)}\right)
\vec\varepsilon+\mathcal O(c^{-6})\\
&=\frac{\omA^2}{2c^2}
(r_1^2-r_2^2)\vec\varepsilon
+\mathcal O(c^{-4}),
\end{split}
\end{equation}
the induced variation of either contact shift satisfies
\begin{equation}
\begin{split}
\delta_\varepsilon(\delta\vec x_a)
={}&
\sum_b
\frac{\partial\delta\vec x_a}{\partial\vec r_b}
\cdot\delta_\varepsilon\vec r_b
+
\sum_b
\frac{\partial\delta\vec x_a}{\partial\vec v_b}
\cdot\delta_\varepsilon\vec v_b
\\
={}&\mathcal O(c^{-6}).
\end{split}
\end{equation}
Here the leading common translation cancels because
the shifts depend only on relative positions, while
the first curvature corrections and velocity variations
are already of $\mathcal{O}(c^{-2})$.

Using
$\vec r_a^{\,\mathrm{old}}
=\vec r_a+\delta\vec x_a$,
we therefore obtain
\begin{equation}
\begin{split}
\delta_\varepsilon\vec r_a
&=\delta_\varepsilon\vec r_a^{\,\mathrm{old}}
-\delta_\varepsilon(\delta\vec x_a)=\sqrt{f(r_a^{\,\mathrm{old}})}\,\vec\varepsilon
+\mathcal O(c^{-6})=\sqrt{f(r_a)}\,\vec\varepsilon
+\mathcal O(c^{-6}).
\end{split}
\end{equation}
In the last step we used
\begin{equation}
\sqrt{f(r_a^{\,\mathrm{old}})}
-\sqrt{f(r_a)}
=
\frac{\omA^2}{c^2}
\frac{\vec r_a\cdot\delta\vec x_a}
{\sqrt{f(r_a)}}
+\mathcal O(c^{-8})
=\mathcal O(c^{-6}).
\end{equation}

Since the reduced variables are canonical, the
transformed Noether charge must generate this
transformation through Poisson brackets:
\begin{equation}
\left[r_a^i,\widetilde{\mathbb P}_j\right]_{\rm PB}
=\frac{\partial\widetilde{\mathbb P}_j}
{\partial p_{ai}}
=\delta_{ij}\sqrt{f(r_a)}
+\mathcal O(c^{-6}).
\end{equation}
Integrating with respect to the canonical momenta gives
\begin{equation}
\widetilde{\mathbb P}_i
=\sum_a\sqrt{f(r_a)}\,p_{ai}
+F_i(\vec r_1,\vec r_2)
+\mathcal O(c^{-6}),
\end{equation}
where $F_i$ is independent of the momenta.

The remaining ambiguity is fixed by time reversal.
The effective action is invariant under
time reversal, and the contact shifts are quadratic in
velocities, so the redefinition preserves this symmetry.
The AdS translation charge is odd under time reversal,
as is $\sum_a\sqrt{f(r_a)}\,p_{ai}$. A function depending
only on positions, however, is even under time reversal.
It follows that $F_i=0$ through the retained order.
Thus
\begin{equation}
\widetilde{\mathbb P}_i
=\sum_a\sqrt{f(r_a)}\,p_{ai}
+\mathcal O(c^{-6}).
\end{equation}
Hence, we conclude that the contact transformation Eq.\eqref{eq:deltax} does not introduce any
interaction-dependent corrections to the canonical
realisation of AdS Noether momentum through
2PN order.

\section{Explicit expressions for 2PN Fokker Lagrangian and equations of motion}\label{app:2PN action and equations of motion}

\subsection{Effective potential before order reduction}

In this subsection, we give the two-particle effective potential through 2PN order before order reduction, following the notation introduced in \S\ref{sec:Effective Fokker action and equations of motion}. The contribution associated with $\mathcal{G}_{p,q}$ at total PN order $n$ is denoted by $V^{(n)}_{p,q}$:
\begin{equation}
W^{(n)}_{p,q}=-\int_t V^{(n)}_{p,q},
\qquad
V_{\text{eff}}^{(n)}
=\sum_{p=0}^{n}\sum_{q=0}^{p}V^{(n)}_{p,q}.
\end{equation}
When the second subscript is omitted, we sum over all curvature contributions at the specified Green's-function order:
\begin{equation}
V^{(n)}_p=\sum_{q=0}^{p}V^{(n)}_{p,q}.
\end{equation}
In particular, the combined 2PN contribution from $\mathcal{G}^{(1)}$ is
$V^{(2)}_1=V^{(2)}_{1,0}+V^{(2)}_{1,1}$.

\medskip
\noindent\textbf{0PN expression.}

\begin{equation}\label{eq:0PN V}
V_{\text{eff}}^{(0)}=V^{(0)}_{0,0}
=-\gsc\frac{m_1m_2}{r_{12}^{D-2}}.
\end{equation}

\pagebreak
\noindent\textbf{1PN expressions.}

\begin{equation}\label{eq:1PN V}
\begin{aligned}
V^{(1)}_{0,0}
&=-\gsc\frac{m_1m_2}{r_{12}^{D-2}}
\left[
\frac{r_1^2\omA^2-v_1^2}{2c^2}
+\frac{r_2^2\omA^2-v_2^2}{2c^2}
\right],\\[3mm]
V^{(1)}_{1,0}
&=-\frac{\gsc}{c^2}\frac{m_1m_2}{r_{12}^{D-2}}
\left[
\frac{1}{2}\,\vec{v}_1.\vec{v}_2
-\left(\frac{D}{2}-1\right)
(\hat{n}_{12}.\vec{v}_1)(\hat{n}_{12}.\vec{v}_2)
\right],\\[3mm]
V^{(1)}_{1,1}
&=-\omA^2\frac{\gsc}{c^2}\frac{m_1m_2}{r_{12}^{D-2}}
\left[
-\frac{\mu^2r_{12}^2}{2(D-4)}
-\frac{r_1^2+r_2^2}{4}
+\frac{D-2}{8r_{12}^2}(r_1^2-r_2^2)^2
\right].
\end{aligned}
\end{equation}
\medskip
\noindent\textbf{2PN expressions.}

\begin{subequations}\label{eq:2PN V}
\begin{equation}
\begin{split}
V^{(2)}_{0,0}
&=-\gsc\frac{m_1m_2}{r_{12}^{D-2}}
\Bigg[
\frac{(r_1^2\omA^2-v_1^2)(r_2^2\omA^2-v_2^2)}{4c^4}\\
&\qquad
+\frac{(\vec{r}_1.\vec{v}_1)^2\omA^2}{2c^4}
+\frac{(\vec{r}_2.\vec{v}_2)^2\omA^2}{2c^4}\\
&\qquad
-\frac{(r_1^2\omA^2-v_1^2)^2}{8c^4}
-\frac{(r_2^2\omA^2-v_2^2)^2}{8c^4}
\Bigg].
\end{split}
\end{equation}

\begin{equation}
\begin{split}
V^{(2)}_1
&=\frac{\gsc}{c^4}\frac{m_1m_2}{4r_{12}^{D-2}}
\Bigg[
2r_{12}(\hat{n}_{12}.\vec{v}_2)
\left(\vec{v}_1.\vec{a}_1-\omA^2(\vec{r}_1.\vec{v}_1)\right)\\
&\qquad
-2r_{12}(\hat{n}_{12}.\vec{v}_1)
\left(\vec{v}_2.\vec{a}_2-\omA^2(\vec{r}_2.\vec{v}_2)\right)\\
&\qquad
+\left(v_1^2+v_2^2-(r_1^2+r_2^2)\omA^2\right)\\
&\qquad\quad\times\Bigg\{
\left(
\frac{(D-2)(r_1^2-r_2^2)^2}{4r_{12}^2}
-\frac{r_1^2+r_2^2}{2}
-\frac{\mu^2r_{12}^2}{D-4}
\right)\omA^2\\
&\qquad\qquad
+\vec{v}_1.\vec{v}_2
-(D-2)(\hat{n}_{12}.\vec{v}_1)(\hat{n}_{12}.\vec{v}_2)
\Bigg\}
\Bigg].
\end{split}
\end{equation}

\begin{equation}
\begin{split}
V^{(2)}_{2,0}
&=-\frac{\gsc}{c^4}
\frac{m_1m_2}{8(D-4)r_{12}^{D-2}}
\Bigg[
r_{12}^2
\Big(
\vec{a}_1.\vec{a}_2
-(D-4)(\hat{n}_{12}.\vec{a}_1)(\hat{n}_{12}.\vec{a}_2)
\Big)\\
&\qquad
-2(D-4)r_{12}
\Big(
(\hat{n}_{12}.\vec{v}_1)(\vec{v}_1.\vec{a}_2)
-(\hat{n}_{12}.\vec{v}_2)(\vec{v}_2.\vec{a}_1)
\Big)\\
&\qquad
+(D-4)(D-2)r_{12}
\Big(
(\hat{n}_{12}.\vec{a}_2)(\hat{n}_{12}.\vec{v}_1)^2
-(\hat{n}_{12}.\vec{a}_1)(\hat{n}_{12}.\vec{v}_2)^2
\Big)\\
&\qquad
-(D-4)r_{12}
\Big(
v_1^2(\hat{n}_{12}.\vec{a}_2)
-v_2^2(\hat{n}_{12}.\vec{a}_1)
\Big)\\
&\qquad
-4(D-4)(D-2)
(\hat{n}_{12}.\vec{v}_1)(\hat{n}_{12}.\vec{v}_2)
(\vec{v}_1.\vec{v}_2)
+2(D-4)(\vec{v}_1.\vec{v}_2)^2\\
&\qquad
+D(D-4)(D-2)
(\hat{n}_{12}.\vec{v}_1)^2(\hat{n}_{12}.\vec{v}_2)^2
+(D-4)v_1^2v_2^2\\
&\qquad
-(D-4)(D-2)
\Big(
v_1^2(\hat{n}_{12}.\vec{v}_2)^2
+v_2^2(\hat{n}_{12}.\vec{v}_1)^2
\Big)
\Bigg].
\end{split}
\end{equation}

\begin{equation}
\begin{split}
V^{(2)}_{2,1}
&=-\omA^2\frac{\gsc}{c^4}
\frac{m_1m_2}{32r_{12}^{D-2}}
\Bigg[
-\frac{2D(D-2)(r_1^2-r_2^2)^2}{r_{12}^2}
(\hat{n}_{12}.\vec{v}_1)(\hat{n}_{12}.\vec{v}_2)\\
&\qquad
+\frac{2(D-2)(r_1^2-r_2^2)^2}{r_{12}^2}
(\vec{v}_1.\vec{v}_2)
+\frac{8(D-2)(r_1^2-r_2^2)}{r_{12}}
(\vec{r}_1.\vec{v}_1)(\hat{n}_{12}.\vec{v}_2)\\
&\qquad
+\frac{8(D-2)(r_1^2-r_2^2)}{r_{12}}
(\hat{n}_{12}.\vec{v}_1)(\vec{r}_2.\vec{v}_2)
+12(D-2)(r_1^2+r_2^2)
(\hat{n}_{12}.\vec{v}_1)(\hat{n}_{12}.\vec{v}_2)\\
&\qquad
-16(\vec{r}_1.\vec{v}_1)(\vec{r}_2.\vec{v}_2)
-12(r_1^2+r_2^2)(\vec{v}_1.\vec{v}_2)\\
&\qquad
-24r_{12}(\vec{r}_1.\vec{v}_1)(\hat{n}_{12}.\vec{v}_2)
+24r_{12}(\hat{n}_{12}.\vec{v}_1)(\vec{r}_2.\vec{v}_2)\\
&\qquad
+8(1+\mu^2)r_{12}^2
(\hat{n}_{12}.\vec{v}_1)(\hat{n}_{12}.\vec{v}_2)
-\frac{8(1+\mu^2)}{D-4}r_{12}^2
(\vec{v}_1.\vec{v}_2)
\Bigg].
\end{split}
\end{equation}

\begin{equation}
\begin{split}
V^{(2)}_{2,2}
&=-\omA^4\frac{\gsc}{c^4}
\frac{m_1m_2}{r_{12}^{D-2}}
\Bigg[
\frac{(\mu^2-1)^2r_{12}^4}{8(D-6)(D-4)}
+\frac{(r_1^2+r_2^2)(1+2\mu^2)r_{12}^2}{16(D-4)}\\
&\qquad
-\frac{(2\mu^2-5)r_1^4
-2(1+2\mu^2)r_1^2r_2^2
+(2\mu^2-5)r_2^4}{32}\\
&\qquad
-\frac{3(D-2)(r_1^2-r_2^2)^2(r_1^2+r_2^2)}
{32r_{12}^2}
+\frac{D(D-2)(r_1^2-r_2^2)^4}{128r_{12}^4}
\Bigg].
\end{split}
\end{equation}
\end{subequations}

\pagebreak
\subsection{Lagrangian}
In this subsection, we give the two-particle Fokker Lagrangian $\mathcal{L}$ through 2PN order, as described in \S\ref{sec:Effective Fokker action and equations of motion}. We organise its contributions according to their PN order, interaction order, and curvature dependence:
\begin{equation*}
\mathcal{L}
=\sum_{\mathit{n}_1,\mathit{n}_2,\mathit{n}_3}
\left(\frac{1}{c^2}\right)^{\mathit{n}_1}
\left(\gsc m_1m_2\right)^{\mathit{n}_2}
\left(\omA^2\right)^{\mathit{n}_3}
L^{(\mathit{n}_1,\mathit{n}_2,\mathit{n}_3)}.
\end{equation*}
Here $\mathit{n}_1$ labels the total PN order, with $\mathit{n}_1=0,1,2$ corresponding to Newtonian, 1PN, and 2PN order, respectively. The index $\mathit{n}_2$ counts powers of the interaction factor $\gsc m_1m_2$, while $\mathit{n}_3$ counts powers of $\omA^2$. Thus, $\mathit{n}_2=0$ identifies the free-particle contributions, and $\mathit{n}_3=0$ identifies the terms that survive the flat-space limit. The coefficients $L^{(\mathit{n}_1,\mathit{n}_2,\mathit{n}_3)}$ contain the remaining dependence on the particle masses and worldline variables, with the displayed expansion factors extracted.
\\~\\
{\bf -1PN Expression:}
\\~\\
\begin{align}\label{eq:-1PN action}
L^{(-1,0,0)}
=& -(m_1 + m_2).
\end{align}
\\~\\
{\bf 0PN Expressions:}
\\~\\
\begin{equation}\begin{split}\label{eq:0PN action}
L^{(0,0,0)}
= \frac{1}{2}\left(m_1 v_1^2 + m_2 v_2^2\right),
\qquad
L^{(0,1,0)}
= \frac{1}{r_{12}^{D-2}},
\qquad
L^{(0,0,1)}
= -\frac{1}{2}\left(m_1 r_1^2 + m_2 r_2^2\right).
\end{split}\end{equation}
\pagebreak
\\~\\
{\bf 1PN Expressions:}
\\~\\
\begin{equation}\begin{split}\label{eq:1PN action}
L^{(1,0,0)}
=& \frac{1}{8}\left(m_1 v_1^4 + m_2 v_2^4\right).
\\[3mm]
L^{(1,0,1)}
=& -\frac{1}{4}m_1
\left[
2\left(\vec{v}_1.\vec{r}_1\right)^2
+ r_1^2 v_1^2
\right]
- \frac{1}{4}m_2
\left[
2\left(\vec{v}_2.\vec{r}_2\right)^2
+ r_2^2 v_2^2
\right].
\\[3mm]
L^{(1,0,2)}
=& \frac{1}{8}\left(m_1 r_1^4 + m_2 r_2^4\right).
\\[3mm]
L^{(1,1,0)}
=& -\frac{1}{2r_{12}^{D-2}}
\left[
\vec{v}_1.\vec{v}_2
+ (D-2)
\left(\hat{n}_{12}.\vec{v}_1\right)
\left(\hat{n}_{12}.\vec{v}_2\right)
+ v_{12}^2
\right].
\\[3mm]
L^{(1,1,1)}
=& \frac{1}{2r_{12}^{D-2}}
\left[
r_{\mathrm{av}}^2
+ (D-2)\left(\hat{n}_{12}.\vec{r}_{\mathrm{av}}\right)^2
+ \left(\frac{1}{4}-\frac{\mu^2}{D-4}\right)r_{12}^2
\right].
\end{split}\end{equation}
\\~\\
{\bf 2PN Expressions:}
\\~\\
\begin{subequations}\label{eq:2PN action}
\begin{align}
L^{(2,0,0)}
=& \frac{1}{16}\left(m_1 v_1^6 + m_2 v_2^6\right).
\end{align}
\begin{align}
L^{(2,0,1)}
=& -\frac{1}{16}m_1v_1^2
\left[
4\left(\vec{v}_1.\vec{r}_1\right)^2
+ 3v_1^2r_1^2
\right]
- \frac{1}{16}m_2v_2^2
\left[
4\left(\vec{v}_2.\vec{r}_2\right)^2
+ 3v_2^2r_2^2
\right].
\end{align}
\begin{align}
L^{(2,0,2)}
=& \frac{3}{16}m_1r_1^2
\left[
r_1^2v_1^2
+ 4\left(\vec{v}_1.\vec{r}_1\right)^2
\right]
+ \frac{3}{16}m_2r_2^2
\left[
r_2^2v_2^2
+ 4\left(\vec{v}_2.\vec{r}_2\right)^2
\right].
\end{align}
\begin{align}
L^{(2,0,3)}
=& -\frac{1}{16}\left(m_1 r_1^6 + m_2 r_2^6\right).
\end{align}
\begin{align}
L^{(2,1,0)}
=& \frac{1}{8r_{12}^{D-2}}
\Bigg[
-\left(v_1^4+v_2^4\right)
+ 3v_1^2v_2^2
+ 2\left(\vec{v}_1.\vec{v}_2\right)^2
- 2\left(\vec{v}_1.\vec{v}_2\right)
\left(v_1^2+v_2^2\right)
\nonumber\\
&\quad
+ D(D-2)
\left(\hat{n}_{12}.\vec{v}_1\right)^2
\left(\hat{n}_{12}.\vec{v}_2\right)^2
- (D-2)
\left[
\left(\hat{n}_{12}.\vec{v}_1\right)^2v_2^2
+ \left(\hat{n}_{12}.\vec{v}_2\right)^2v_1^2
\right]
\nonumber\\
&\quad
+ 2(D-2)
\left(\hat{n}_{12}.\vec{v}_1\right)
\left(\hat{n}_{12}.\vec{v}_2\right)
v_{12}^2
\Bigg].
\end{align}
\begin{align}
L^{(2,1,1)}
=& \frac{1}{4r_{12}^{D-2}}
\Bigg[
r_{\mathrm{av}}^2
\Big[
\vec{v}_1.\vec{v}_2
+ (D-2)
(\hat{n}_{12}.\vec{v}_1)
(\hat{n}_{12}.\vec{v}_2)
+ v_{12}^2
\Big]
\nonumber\\
&\quad
- (D-2)
(\hat{n}_{12}.\vec{r}_{\mathrm{av}})^2
\Big[
-\vec{v}_1.\vec{v}_2
+ D(\hat{n}_{12}.\vec{v}_1)
(\hat{n}_{12}.\vec{v}_2)
+ v_1^2+v_2^2
\Big]
\nonumber\\
&\quad
+ 2(\vec{v}_{12}.\vec{r}_{\mathrm{av}})^2
+ 2(\vec{v}_1.\vec{r}_{\mathrm{av}})
(\vec{v}_2.\vec{r}_{\mathrm{av}})
- (D-2)
(\hat{n}_{12}.\vec{r}_{\mathrm{av}})
(\hat{n}_{12}.\vec{v}_{12})
(\vec{v}_{12}.\vec{r}_{\mathrm{av}})
\nonumber\\
&\quad
+ (D-2)
(\hat{n}_{12}.\vec{r}_{\mathrm{av}})
\left[
\hat{n}_{12}.(\vec{v}_1+\vec{v}_2)
\right]
\left[
(\vec{v}_1+\vec{v}_2).\vec{r}_{\mathrm{av}}
\right]
\Bigg]
\nonumber\\
&\quad
+ \frac{1}{2r_{12}^{D-2}}
\Big[
(v_1^2-v_2^2)
\vec{r}_{12}.\vec{r}_{\mathrm{av}}
+ (\vec{v}_{12}.\vec{r}_{\mathrm{av}})
(\vec{v}_1+\vec{v}_2).\vec{r}_{12}
\Big]
\nonumber\\
&\quad
- \frac{1}{32r_{12}^{D-2}}
\Bigg[
\left(
\frac{1}{2}-\frac{2\mu^2}{D-4}
\right)r_{12}^2
(\vec{v}_1+\vec{v}_2).(\vec{v}_1+\vec{v}_2)
\nonumber\\
&\quad
-\left(\frac{3D}{2}+2\mu^2\right)
\left[
(\vec{v}_1+\vec{v}_2).\vec{r}_{12}
\right]^2
-\left(
\frac{5}{2}+\frac{6\mu^2}{D-4}
\right)r_{12}^2v_{12}^2
\nonumber\\
&\quad
+\left(
\frac{3D}{2}-4+2\mu^2
\right)
(\vec{r}_{12}.\vec{v}_{12})^2
\Bigg].
\end{align}
\begin{align}
L^{(2,1,2)}
=& -\frac{1}{8r_{12}^{D-2}}
\left[
r_{\mathrm{av}}^4
- D(D-2)\left(\hat{n}_{12}.\vec{r}_{\mathrm{av}}\right)^4
+ 2(D-2)r_{\mathrm{av}}^2
\left(\hat{n}_{12}.\vec{r}_{\mathrm{av}}\right)^2
\right]
\nonumber\\
&\quad
- \frac{1}{8r_{12}^{D-2}}
\left[
\frac{1}{2}
\left(\hat{n}_{12}.\vec{r}_{\mathrm{av}}\right)^2
\left(D+4\mu^2\right)
+ r_{\mathrm{av}}^2
\left(
\frac{1}{2}+\frac{2\mu^2}{D-4}
\right)
\right]r_{12}^2
\nonumber\\
&\quad
- \frac{1}{8r_{12}^{D-2}}
\left[
\frac{5}{16}
+\frac{\mu^2-1}{2(D-6)}
-\frac{\mu^2(\mu^2-1)}{(D-4)(D-6)}
\right]r_{12}^4.
\end{align}
\begin{align}
L^{(2,2,0)}
=& \frac{1}{16r_{12}^{2(D-2)}}
\left(\frac{1}{m_1}+\frac{1}{m_2}\right)
\Bigg[
-(D-2)(v_1^2+v_2^2)
-\left(3+\frac{2}{D-4}\right)v_{12}^2
\nonumber\\
&\quad
+ (D-2)D
\left[
\left(\hat{n}_{12}.\vec{v}_1\right)^2
+\left(\hat{n}_{12}.\vec{v}_2\right)^2
\right]
+\left(
\frac{8}{D-4}-2(D-3)D
\right)
\left(\hat{n}_{12}.\vec{v}_{12}\right)^2
\Bigg]
\nonumber\\
&\quad
- \frac{1}{16r_{12}^{2(D-2)}}
\left(\frac{1}{m_1}-\frac{1}{m_2}\right)
\Bigg[
-\left(1+\frac{2}{D-4}\right)(v_1^2-v_2^2)
\nonumber\\
&\quad
+\left(
\frac{8}{D-4}-(D-4)D
\right)
\left[
\left(\hat{n}_{12}.\vec{v}_1\right)^2
-\left(\hat{n}_{12}.\vec{v}_2\right)^2
\right]
\Bigg].
\end{align}
\begin{align}
L^{(2,2,1)}
=& \frac{1}{8r_{12}^{2(D-2)}}
\frac{D-5}{D-4}(D-2)
\Bigg[
\left(\frac{1}{m_1}-\frac{1}{m_2}\right)
\vec{r}_{12}.\vec{r}_{\mathrm{av}}
- \frac{1}{2}
\left(\frac{1}{m_1}+\frac{1}{m_2}\right)r_{12}^2
\Bigg].
\end{align}
\begin{align}
L^{(2,3,0)}
=& -\frac{1}{r_{12}^{3(D-2)}}
\frac{(D-5)(D-2)^2}{8m_1m_2(D-4)}.
\end{align}
\end{subequations}

\pagebreak
\subsection{Force}
In this subsection, we give the force on each particle of the binary through 2PN order, as described in \S\ref{sec:Effective Fokker action and equations of motion}. We organise its contributions according to their PN order, interaction order, and curvature dependence:
\begin{equation*}
\vec{\mathcal{F}}_a
=\sum_{\mathit{n}_1,\mathit{n}_2,\mathit{n}_3}
\left(\frac{1}{c^2}\right)^{\mathit{n}_1}
\left(\gsc m_1m_2\right)^{\mathit{n}_2}
\left(\omA^2\right)^{\mathit{n}_3}
\vec{F}^{(\mathit{n}_1,\mathit{n}_2,\mathit{n}_3)}_a.
\end{equation*}
Here $a=1,2$ labels the particles, and $\mathit{n}_1$ labels the total PN order, with $\mathit{n}_1=0,1,2$ corresponding to Newtonian, 1PN, and 2PN order, respectively. The index $\mathit{n}_2$ counts powers of the interaction factor $\gsc m_1m_2$, while $\mathit{n}_3$ counts powers of $\omA^2$. Thus, $\mathit{n}_2=0$ identifies the interaction-independent contributions, and $\mathit{n}_3=0$ identifies the terms that survive the flat-space limit. The coefficients $\vec{F}^{(\mathit{n}_1,\mathit{n}_2,\mathit{n}_3)}_a$ contain the remaining dependence on the particle masses and worldline variables, with the displayed expansion factors extracted.
\\~\\
{\bf 0PN Expressions:}
\\~\\
\begin{equation}\begin{split}\label{eq:0PN Force}
\vec{F}_{1}^{\,(0,0,1)}
= -m_1 \vec{r}_1,
\qquad
\vec{F}_{1}^{\,(0,1,0)}
=& -\frac{D-2}{r_{12}^{D}}\,\vec{r}_{12}.
\end{split}\end{equation}
\\~\\
{\bf 1PN Expressions:}
\\~\\
\begin{equation}\begin{split}\label{eq:1PN Force}
\vec{F}_{1}^{\,(1,0,1)}
=& m_1\left[
2\left(\vec{r}_1.\vec{v}_1\right)\vec{v}_1
+ v_1^2\,\vec{r}_1
\right].
\\[3mm]
\vec{F}_{1}^{\,(1,0,2)}
=& -m_1 r_1^2\,\vec{r}_1.
\\[3mm]
\vec{F}_{1}^{\,(1,1,0)}
=& \frac{1}{2}\frac{D-2}{r_{12}^{D}}\,\vec{r}_{12}
\left[
D\left(\hat{n}_{12}.\vec{v}_2\right)^2
+ 2v_1^2
\right]
+ \frac{D-2}{r_{12}^{D}}\,\vec{v}_{12}
\left(\vec{r}_{12}.\vec{v}_2\right).
\\[3mm]
\vec{F}_{1}^{\,(1,1,1)}
=& -\frac{D-2}{2r_{12}^{D-2}}\,\vec{r}_1
- \frac{D-2}{r_{12}^{D}}\,\vec{r}_{12}
\left[
r_{\mathrm{av}}^2
+ \frac{D}{2}\left(\hat{n}_{12}.\vec{r}_{\mathrm{av}}\right)^2
+ \vec{r}_{12}.\vec{r}_{\mathrm{av}}
+ \frac{1-\mu^2}{2(D-2)}r_{12}^2
\right].
\\[3mm]
\vec{F}_{1}^{\,(1,2,0)}
=& \frac{D-2}{r_{12}^{2(D-1)}}\,\vec{r}_{12}
\left(
\frac{D-3}{2m_2}
-\frac{1}{m_1}
\right).
\end{split}\end{equation}
\\~\\
{\bf 2PN Expressions:}
\\~\\
\begin{subequations}\label{eq:2PN Force}
\begin{align}
\vec{F}_{1}^{\,(2,0,2)}
=& -m_1
\Big[
\left(\vec{r}_1.\vec{v}_1\right)^2\vec{r}_1
+ 2r_1^2
\left(\vec{r}_1.\vec{v}_1\right)\vec{v}_1
\Big].
\end{align}
\begin{align}
\vec{F}_{1}^{\,(2,1,0)}
=& \frac{D-2}{4r_{12}^{D-1}}\,
\vec{v}_1
\Bigg[
2D\left(\hat{n}_{12}.\vec{v}_1\right)^2
\left(\hat{n}_{12}.\vec{v}_2\right)
-5D\left(\hat{n}_{12}.\vec{v}_1\right)
\left(\hat{n}_{12}.\vec{v}_2\right)^2
+D\left(\hat{n}_{12}.\vec{v}_2\right)^3
\nonumber\\
&\quad
-6\left(\hat{n}_{12}.\vec{v}_2\right)v_1^2
-4\left(\hat{n}_{12}.\vec{v}_1\right)
\left(\vec{v}_1.\vec{v}_2\right)
+10\left(\hat{n}_{12}.\vec{v}_2\right)
\left(\vec{v}_1.\vec{v}_2\right)
\nonumber\\
&\quad
+5\left(\hat{n}_{12}.\vec{v}_1\right)v_2^2
-5\left(\hat{n}_{12}.\vec{v}_2\right)v_2^2
\Bigg]
\nonumber\\
&\quad
+ \frac{D-2}{4r_{12}^{D-1}}\,
\vec{v}_2
\Bigg[
-D\left(\hat{n}_{12}.\vec{v}_1\right)^2
\left(\hat{n}_{12}.\vec{v}_2\right)
+3D\left(\hat{n}_{12}.\vec{v}_1\right)
\left(\hat{n}_{12}.\vec{v}_2\right)^2
\nonumber\\
&\quad
+5\left(\hat{n}_{12}.\vec{v}_2\right)v_1^2
+2\left(\hat{n}_{12}.\vec{v}_1\right)
\left(\vec{v}_1.\vec{v}_2\right)
-6\left(\hat{n}_{12}.\vec{v}_2\right)
\left(\vec{v}_1.\vec{v}_2\right)
\nonumber\\
&\quad
-3\left(\hat{n}_{12}.\vec{v}_1\right)v_2^2
+2\left(\hat{n}_{12}.\vec{v}_2\right)v_2^2
\Bigg]
\nonumber\\
&\quad
+ \frac{D-2}{8r_{12}^{D-1}}\,
\hat{n}_{12}
\Bigg[
D(D+2)\left(\hat{n}_{12}.\vec{v}_1\right)^2
\left(\hat{n}_{12}.\vec{v}_2\right)^2
-2D(D+2)\left(\hat{n}_{12}.\vec{v}_1\right)
\left(\hat{n}_{12}.\vec{v}_2\right)^3
\nonumber\\
&\quad
-5D v_1^2\left(\hat{n}_{12}.\vec{v}_2\right)^2
-D\left(\hat{n}_{12}.\vec{v}_1\right)^2v_2^2
-4D\left(\hat{n}_{12}.\vec{v}_1\right)
\left(\hat{n}_{12}.\vec{v}_2\right)
\left(\vec{v}_1.\vec{v}_2\right)
\nonumber\\
&\quad
+6D\left(\hat{n}_{12}.\vec{v}_1\right)
\left(\hat{n}_{12}.\vec{v}_2\right)v_2^2
+6D\left(\vec{v}_1.\vec{v}_2\right)
\left(\hat{n}_{12}.\vec{v}_2\right)^2
\nonumber\\
&\quad
-2D\left(\hat{n}_{12}.\vec{v}_2\right)^2v_2^2
+2\left(\vec{v}_1.\vec{v}_2\right)^2
+v_1^2v_2^2
-6\left(\vec{v}_1.\vec{v}_2\right)v_2^2
+3v_2^4
\Bigg].
\end{align}
\begin{align}
\vec{F}_{1}^{\,(2,1,1)}
=& \frac{D(D-2)}{2r_{12}^{D}}
\left(\hat{n}_{12}.\vec{r}_{\mathrm{av}}\right)^2
\Bigg[
\vec{r}_{12}
\left(
v_1^2
+ \frac{D+2}{2}
\left(\hat{n}_{12}.\vec{v}_2\right)^2
\right)
+ \vec{v}_{12}
\left(\vec{r}_{12}.\vec{v}_2\right)
\Bigg]
\nonumber\\
&\quad
+ \frac{D-2}{2r_{12}^{D}}
\vec{r}_{12}
\left(\vec{r}_{\mathrm{av}}.\vec{r}_{12}\right)
\Bigg[
v_2^2
-\vec{v}_1.\vec{v}_2
+2D\left(\hat{n}_{12}.\vec{v}_2\right)^2
+D\left(\hat{n}_{12}.\vec{v}_1\right)
\left(\hat{n}_{12}.\vec{v}_2\right)
\nonumber\\&\quad-\frac{2D}{r_{12}^2}
\left(\vec{r}_{\mathrm{av}}.\vec{v}_2\right)
\left(\vec{r}_{12}.\vec{v}_2\right)
\Bigg]
+ \frac{D-2}{r_{12}^{D}}
\vec{v}_1
\left(\vec{r}_{\mathrm{av}}.\vec{r}_{12}\right)
\left(
\vec{r}_{12}.\vec{v}_1
-\vec{r}_{\mathrm{av}}.\vec{v}_2
\right)
\nonumber\\
&\quad
- \frac{D-2}{2r_{12}^{D}}
\vec{v}_2
\left(\vec{r}_{\mathrm{av}}.\vec{r}_{12}\right)
\left(
\vec{r}_{12}.\vec{v}_1
+3\vec{r}_{12}.\vec{v}_2
-2\vec{r}_{\mathrm{av}}.\vec{v}_2
\right)
\nonumber\\
&\quad
- \frac{D-2}{2r_{12}^{D}}\vec{r}_{12}
\Bigg[
2\left(\vec{r}_{\mathrm{av}}.\vec{v}_1\right)^2
+2\left(\vec{r}_{\mathrm{av}}.\vec{v}_1\right)
\left(\vec{r}_{12}.\vec{v}_1\right)
-2\left(\vec{r}_{\mathrm{av}}.\vec{v}_2\right)
\left(\vec{r}_{12}.\vec{v}_2\right)
\nonumber\\
&\quad\quad
+\left(\vec{r}_{\mathrm{av}}.\vec{v}_1\right)
\left(\vec{r}_{12}.\vec{v}_2\right)
+\left(\vec{r}_{\mathrm{av}}.\vec{v}_2\right)
\left(\vec{r}_{12}.\vec{v}_1\right)
\Bigg]
\nonumber\\
&\quad
+ \frac{\vec{r}_1}{4r_{12}^{D-2}}
\Bigg[
(D-2\mu^2)v_1^2
+\left(8-\frac{3D}{2}\right)v_2^2
+(D-6)\vec{v}_1.\vec{v}_2
+(2-D)\left(\hat{n}_{12}.\vec{v}_1\right)^2
\nonumber\\
&\quad\quad
+(D-2)(D-8-\mu^2)
\left(\hat{n}_{12}.\vec{v}_2\right)^2
-(D-6)(D-2)
\left(\hat{n}_{12}.\vec{v}_1\right)
\left(\hat{n}_{12}.\vec{v}_2\right)
\Bigg]
\nonumber\\
&\quad
+ \frac{\vec{r}_2}{4r_{12}^{D-2}}
\Bigg[
(D-4+2\mu^2)v_1^2
+\frac{3}{2}(D-4)v_2^2
+(4-D)\vec{v}_1.\vec{v}_2
+(D-2)\left(\hat{n}_{12}.\vec{v}_1\right)^2
\nonumber\\
&\quad\quad
+(D-2)(2+\mu^2)
\left(\hat{n}_{12}.\vec{v}_2\right)^2
+(D-4)(D-2)
\left(\hat{n}_{12}.\vec{v}_1\right)
\left(\hat{n}_{12}.\vec{v}_2\right)
\Bigg]
\nonumber\\
&\quad
+ \frac{\vec{v}_1}{4r_{12}^{D-2}}
\Bigg[
(4-2D)\vec{r}_1.\vec{v}_1
+2(D-3-\mu^2)\vec{r}_1.\vec{v}_2
+2(D-4)\vec{r}_2.\vec{v}_1
+2(6-D+\mu^2)\vec{r}_2.\vec{v}_2
\Bigg]
\nonumber\\
&\quad
+ \frac{\vec{v}_2}{4r_{12}^{D-2}}
\Bigg[
(D-2)\vec{r}_1.\vec{v}_1
+2\mu^2\vec{r}_1.\vec{v}_2
+(4-D)\vec{r}_2.\vec{v}_1
-2(2+\mu^2)\vec{r}_2.\vec{v}_2
\Bigg].
\end{align}
\begin{align}
\vec{F}_{1}^{\,(2,1,2)}
=& \frac{\vec{r}_1}{128r_{12}^{D-2}}
\Bigg[
-\frac{16D(D^2-4)}{r_{12}^2}
\left(\hat{n}_{12}.\vec{r}_{\mathrm{av}}\right)^4
-\frac{64(D-2)D}{r_{12}^4}
\left(\vec{r}_{12}.\vec{r}_{\mathrm{av}}\right)^3
\nonumber\\
&\quad
-32(1+D-2\mu^2)r_{\mathrm{av}}^2
-16(D-2)\bigl(D-2(1+\mu^2)\bigr)
\left(\hat{n}_{12}.\vec{r}_{\mathrm{av}}\right)^2
\nonumber\\
&\quad
+32(7-2D+2\mu^2)
\vec{r}_{12}.\vec{r}_{\mathrm{av}}
+\frac{4\left[
88-26D+D^2+8(D-2)\mu^2-4\mu^4
\right]}{D-4}r_{12}^2
\Bigg]
\nonumber\\
&\quad
+ \frac{\vec{r}_2}{128r_{12}^{D-2}}
\Bigg[
\frac{16D(D^2-4)}{r_{12}^2}
\left(\hat{n}_{12}.\vec{r}_{\mathrm{av}}\right)^4
+\frac{64(D-2)D}{r_{12}^4}
\left(\vec{r}_{12}.\vec{r}_{\mathrm{av}}\right)^3
\nonumber\\
&\quad
-32(-5+D+2\mu^2)r_{\mathrm{av}}^2
-16(D-2)(-2+D+2\mu^2)
\left(\hat{n}_{12}.\vec{r}_{\mathrm{av}}\right)^2
\nonumber\\
&\quad
-32(1+2\mu^2)
\vec{r}_{12}.\vec{r}_{\mathrm{av}}
-\frac{4\left[
24-10D+D^2+16\mu^2-4\mu^4
\right]}{D-4}r_{12}^2
\Bigg].
\end{align}
\begin{align}
\vec{F}_{1}^{\,(2,2,0)}
=& \frac{1}{8r_{12}^{2(D-2)+1}}
\Bigg[
-\frac{8(D-3)^2(D-2)}{(D-4)m_2}
\left(\hat{n}_{12}.\vec{v}_1\right)\vec{v}_1
+ \frac{2(D-2)^3}{(D-4)m_2}
\left(\hat{n}_{12}.\vec{v}_1\right)\vec{v}_2
\nonumber\\
&\quad
+ \frac{4(D-2)}{D-4}
\left(
\frac{(D-4)(D-1)}{m_1}
+\frac{18-8D+D^2}{m_2}
\right)
\left(\hat{n}_{12}.\vec{v}_2\right)\vec{v}_1
\nonumber\\
&\quad
+ \frac{2(D-2)}{D-4}
\left(
-\frac{(D-4)D}{m_1}
+\frac{2(2-5D+D^2)}{m_2}
\right)
\left(\hat{n}_{12}.\vec{v}_2\right)\vec{v}_2
\Bigg]
\nonumber\\
&\quad
+ \frac{\hat{n}_{12}}{8r_{12}^{2(D-2)+1}}
\Bigg[
\frac{2(D-2)}{D-4}
\left(
\frac{4(D-4)}{m_1}
-\frac{22-9D+D^2}{m_2}
\right)v_1^2
\nonumber\\
&\quad
+ \frac{2(D-2)}{D-4}
\left(
\frac{5(D-4)}{m_1}
-\frac{2(17-8D+D^2)}{m_2}
\right)v_2^2
\nonumber\\
&\quad
+ \frac{2(D-2)}{D-4}
\left(
-\frac{4(D-4)}{m_1}
+\frac{28-10D+D^2}{m_2}
\right)
\vec{v}_1.\vec{v}_2
\nonumber\\
&\quad
-\frac{4(D-5)(D-2)^2(D-1)}{(D-4)m_2}
\left(\hat{n}_{12}.\vec{v}_1\right)^2
\nonumber\\
&\quad
-\frac{4(D-2)}{D-4}
\left(
\frac{(D-4)(D-3)}{m_1}
+\frac{6+D-5D^2+D^3}{m_2}
\right)
\left(\hat{n}_{12}.\vec{v}_2\right)^2
\nonumber\\
&\quad
+\frac{4(D-2)}{D-4}
\left(
\frac{2(D-4)(D-2)}{m_1}
+\frac{12-10D+D^2}{m_2}
\right)
\left(\hat{n}_{12}.\vec{v}_1\right)
\left(\hat{n}_{12}.\vec{v}_2\right)
\Bigg].
\end{align}
\begin{align}
\vec{F}_{1}^{\,(2,2,1)}
=& \frac{\vec{r}_{\mathrm{av}}}{32r_{12}^{2(D-2)}}
\left[
\frac{8(D-2)\left[(D-5)m_1-2m_2\right]}{m_1m_2}
\right]
\nonumber\\
&\quad
+ \frac{\vec{r}_{12}}{32r_{12}^{2(D-1)}}
\Bigg[
\frac{16(D-2)\left[(D-3)m_1-2m_2\right]}{m_1m_2}
r_{\mathrm{av}}^2
\nonumber\\
&\quad
+ \frac{16(D-2)(D-1)\left[(D-3)m_1-2m_2\right]}
{m_1m_2}
\left(\vec{r}_{\mathrm{av}}.\hat{n}_{12}\right)^2
+ \frac{16(D-2)}{m_2}
\vec{r}_{\mathrm{av}}.\vec{r}_{12}
\nonumber\\
&\quad
+ \frac{4}{(D-4)m_1m_2}
\Big[
-4(D-4)(D-1)m_2
+8(D-3)m_2\mu^2
\nonumber\\
&\quad\quad
+m_1\left(
-24+20D-3D^2
-4\left(11+(D-7)D\right)\mu^2
\right)
\Big]r_{12}^2
\Bigg].
\end{align}
\begin{align}
\vec{F}_{1}^{\,(2,3,0)}
=& -\frac{D-2}{8(D-4)}
\frac{\vec{r}_{12}}{r_{12}^{3(D-2)+2}}
\Bigg[
\frac{2D^3-19D^2+59D-66}{m_2^2}
+ \frac{3D^3-28D^2+87D-98}{m_1m_2}
+ \frac{8(D-4)}{m_1^2}
\Bigg].
\end{align}
\end{subequations}
\pagebreak
\section{Explicit expressions for 2PN Hamiltonian and symmetry generators}\label{app:2PN Hamiltonian and conserved charges}

\subsection{Conjugate momentum}
In this subsection, we give the conjugate momentum of each particle through 2PN order, as described in \S\ref{sec:Hamiltonian and conformal algebra}. We organise its contributions according to their PN order, interaction order, and curvature dependence:
\begin{equation}\label{eq:conjugate-momentumExpAlt}
\begin{split}
\vec{p}_a
=\frac{\partial}{\partial\vec{v}_a}\mathcal{L}=\sum_{\mathit{n}_1,\mathit{n}_2,\mathit{n}_3}
\left(\frac{1}{c^2}\right)^{\mathit{n}_1}
\left(\gsc m_1m_2\right)^{\mathit{n}_2}
\left(\omA^2\right)^{\mathit{n}_3}
\vec{p}^{\,(\mathit{n}_1,\mathit{n}_2,\mathit{n}_3)}.
\end{split}
\end{equation}
Here $a=1,2$ labels the particles, and $\mathit{n}_1$ labels the total PN order, with $\mathit{n}_1=0,1,2$ corresponding to Newtonian, 1PN, and 2PN order, respectively. The index $\mathit{n}_2$ counts powers of the interaction factor $\gsc m_1m_2$, while $\mathit{n}_3$ counts powers of $\omA^2$. Thus, $\mathit{n}_2=0$ identifies the free-particle contributions, and $\mathit{n}_3=0$ identifies the terms that survive the flat-space limit. The coefficients $\vec{p}^{\,(\mathit{n}_1,\mathit{n}_2,\mathit{n}_3)}$ contain the remaining dependence on the particle masses and worldline variables, with the displayed expansion factors extracted and the particle label $a$ understood.
\\~\\
{\bf 0PN Expression:}
\\~\\
\begin{subequations}\label{eq:0PN conjugate momentum}
\begin{align}
\vec{p}_{1}^{\,(0,0,0)}
=& m_1 \vec{v}_1.
\end{align}
\end{subequations}
\pagebreak
\\~\\
{\bf 1PN Expressions:}
\\~\\
\begin{equation}\begin{split}\label{eq:1PN conjugate momentum}
\vec{p}_{1}^{\,(1,0,0)}
=& \frac{1}{2}m_1 v_1^2\,\vec{v}_1.
\\[3mm]
\vec{p}_{1}^{\,(1,1,0)}
=& -\frac{1}{2r_{12}^{D-2}}
\left[
\vec{v}_2
+ (D-2)\hat{n}_{12}
\left(\hat{n}_{12}.\vec{v}_2\right)
+ 2\vec{v}_{12}
\right].
\\[3mm]
\vec{p}_{1}^{\,(1,0,1)}
=& -m_1\,\vec{r}_1
\left(\vec{r}_1.\vec{v}_1\right)
-\frac{1}{2}m_1 r_1^2\,\vec{v}_1.
\end{split}\end{equation}
\\~\\
{\bf 2PN Expressions:}
\\~\\
\begin{subequations}\label{eq:2PN conjugate momentum}
\begin{align}
\vec{p}_{1}^{\,(2,0,0)}
=& \frac{3}{8}m_1v_1^4\,\vec{v}_1.
\end{align}
\begin{align}
\vec{p}_{1}^{\,(2,0,1)}
=& -\frac{1}{2}m_1v_1^2
\left(\vec{v}_1.\vec{r}_1\right)\vec{r}_1
- \frac{1}{4}m_1
\left[
2\left(\vec{v}_1.\vec{r}_1\right)^2
+ 3v_1^2r_1^2
\right]\vec{v}_1.
\end{align}
\begin{align}
\vec{p}_{1}^{\,(2,0,2)}
=& \frac{3}{8}m_1r_1^2
\left[
4\left(\vec{v}_1.\vec{r}_1\right)\vec{r}_1
+ r_1^2\vec{v}_1
\right].
\end{align}
\begin{align}
\vec{p}_{1}^{\,(2,1,0)}
=& \frac{1}{4r_{12}^{D-2}}
\vec{v}_1
\Big[
-2v_1^2
+ 3v_2^2
- 2\vec{v}_1.\vec{v}_2
- (D-2)\left(\hat{n}_{12}.\vec{v}_2\right)^2
+ 2(D-2)
\left(\hat{n}_{12}.\vec{v}_1\right)
\left(\hat{n}_{12}.\vec{v}_2\right)
\Big]
\nonumber\\
&\quad
+ \frac{1}{4r_{12}^{D-2}}
\vec{v}_2
\Big[
-v_{12}^2
- 2(D-2)
\left(\hat{n}_{12}.\vec{v}_1\right)
\left(\hat{n}_{12}.\vec{v}_2\right)
\Big]
\nonumber\\
&\quad
+ \frac{D-2}{4r_{12}^{D-2}}
\hat{n}_{12}
\Big[
D\left(\hat{n}_{12}.\vec{v}_1\right)
\left(\hat{n}_{12}.\vec{v}_2\right)^2
-\left(\hat{n}_{12}.\vec{v}_1\right)v_2^2
+\left(\hat{n}_{12}.\vec{v}_2\right)v_{12}^2
\Big].
\end{align}
\begin{align}
\vec{p}_{1}^{\,(2,1,1)}
=& \frac{\vec{v}_1}{2r_{12}^{D-2}}
\Bigg[
r_{\mathrm{av}}^2
- (D-2)\left(\hat{n}_{12}.\vec{r}_{\mathrm{av}}\right)^2
+ 2\vec{r}_{12}.\vec{r}_{\mathrm{av}}
+ \left(\frac{1}{4}+\frac{\mu^2}{D-4}\right)r_{12}^2
\Bigg]
\nonumber\\
&\quad
- \frac{\vec{v}_2}{4r_{12}^{D-2}}
\Bigg[
r_{\mathrm{av}}^2
- (D-2)\left(\hat{n}_{12}.\vec{r}_{\mathrm{av}}\right)^2
+ \left(\frac{3}{4}+\frac{\mu^2}{D-4}\right)r_{12}^2
\Bigg]
\nonumber\\
&\quad
+ \frac{D-2}{4r_{12}^{D-2}}\hat{n}_{12}
\Bigg[
r_{\mathrm{av}}^2\left(\hat{n}_{12}.\vec{v}_2\right)
- D\left(\hat{n}_{12}.\vec{r}_{\mathrm{av}}\right)^2
\left(\hat{n}_{12}.\vec{v}_2\right)
\nonumber\\
&\quad\quad
+ 2\left(\hat{n}_{12}.\vec{r}_{\mathrm{av}}\right)
\left(\vec{r}_{\mathrm{av}}.\vec{v}_2\right)
\Bigg]
+ \frac{\vec{r}_{12}}{2r_{12}^{D-2}}
\Bigg[
\vec{v}_{12}.\vec{r}_{\mathrm{av}}
+ \vec{r}_{12}.
\left(
\frac{1}{2}\vec{v}_1
+ \frac{1}{8}(3D-4+4\mu^2)\vec{v}_2
\right)
\Bigg]
\nonumber\\
&\quad
+ \frac{\vec{r}_{\mathrm{av}}}{2r_{12}^{D-2}}
\Bigg[
2\vec{v}_{12}.\vec{r}_{\mathrm{av}}
+ \vec{v}_2.\vec{r}_{\mathrm{av}}
+ (D-2)
\left(\hat{n}_{12}.\vec{r}_{\mathrm{av}}\right)
\left(\hat{n}_{12}.\vec{v}_2\right)
+ \vec{r}_{12}.(\vec{v}_1+\vec{v}_2)
\Bigg].
\end{align}
\begin{align}
\vec{p}_{1}^{\,(2,2,0)}
=& \frac{\vec{v}_1}{8r_{12}^{2(D-2)}}
\Bigg[
-\left(\frac{1}{m_1}+\frac{1}{m_2}\right)
\left(1+D+\frac{2}{D-4}\right)
+\left(\frac{1}{m_1}-\frac{1}{m_2}\right)
\left(1+\frac{2}{D-4}\right)
\Bigg]
\nonumber\\
&\quad
- \frac{\vec{v}_2}{8r_{12}^{2(D-2)}}
\left(\frac{1}{m_1}+\frac{1}{m_2}\right)
\left(-3-\frac{2}{D-4}+D\right)
\nonumber\\
&\quad
+ \frac{\hat{n}_{12}}{8r_{12}^{2(D-2)}}
\left(\hat{n}_{12}.\vec{v}_1\right)
\Bigg[
\left(\frac{1}{m_1}+\frac{1}{m_2}\right)
\left(
(D-2)D+\frac{8}{D-4}-2(D-3)D
\right)
\nonumber\\
&\quad\quad
-\left(\frac{1}{m_1}-\frac{1}{m_2}\right)
\left(
\frac{8}{D-4}-(D-4)D
\right)
\Bigg]
\nonumber\\
&\quad
- \frac{\hat{n}_{12}}{8r_{12}^{2(D-2)}}
\left(\hat{n}_{12}.\vec{v}_2\right)
\left(\frac{1}{m_1}+\frac{1}{m_2}\right)
\left(
\frac{8}{D-4}-2(D-3)D
\right).
\end{align}
\end{subequations}

\pagebreak
\subsection{Hamiltonian}
In this subsection, we give the two-particle Hamiltonian through 2PN order, as described in \S\ref{sec:Hamiltonian and conformal algebra}. We organise its contributions according to their PN order, interaction order, and curvature dependence:
\begin{equation}\label{eq:HamiltonianExpII}
\begin{split}
\mathbb{H}
&=\sum_a\vec{p}_a.\,\vec{v}_a-\mathcal{L}\\
&=\sum_{\mathit{n}_1,\mathit{n}_2,\mathit{n}_3}
\left(\frac{1}{c^2}\right)^{\mathit{n}_1}
\left(\gsc m_1m_2\right)^{\mathit{n}_2}
\left(\omA^2\right)^{\mathit{n}_3}
\mathbb{H}^{\,(\mathit{n}_1,\mathit{n}_2,\mathit{n}_3)}.
\end{split}
\end{equation}
Here $\mathit{n}_1$ labels the total PN order, with $\mathit{n}_1=0,1,2$ corresponding to Newtonian, 1PN, and 2PN order, respectively. The index $\mathit{n}_2$ counts powers of the interaction factor $\gsc m_1m_2$, while $\mathit{n}_3$ counts powers of $\omA^2$. Thus, $\mathit{n}_2=0$ identifies the free-particle contributions, and $\mathit{n}_3=0$ identifies the terms that survive the flat-space limit. The coefficients $\mathbb{H}^{\,(\mathit{n}_1,\mathit{n}_2,\mathit{n}_3)}$ contain the remaining dependence on the particle masses and phase-space variables, with the displayed expansion factors extracted.
\\~\\
{\bf -1PN Expression:}
\\~\\
\begin{align}\label{eq:-1PN Hamiltonian}
\mathbb{H}^{(-1,0,0)} =&\, m_1 + m_2 .
\end{align}
\\~\\
{\bf 0PN Expressions:}
\\~\\
\begin{equation}\begin{split}\label{eq:0PN Hamiltonian}
\mathbb{H}^{(0,0,0)} = \frac{p_1^2}{2m_1} + \frac{p_2^2}{2m_2}.
\qquad
\mathbb{H}^{(0,0,1)}
= \frac{1}{2}m_1 r_1^2 + \frac{1}{2}m_2 r_2^2.
\qquad
\mathbb{H}^{(0,1,0)}
= -\frac{1}{r_{12}^{D-2}}.
\end{split}\end{equation}
\pagebreak
\\~\\
{\bf 1PN Expressions:}
\\~\\
\begin{equation}\begin{split}\label{eq:1PN Hamiltonian}
\mathbb{H}^{(1,0,0)}
=& -\frac{1}{8}\left(\frac{p_1^4}{m_1^3} + \frac{p_2^4}{m_2^3}\right).
\\[3mm]
\mathbb{H}^{(1,0,1)}
=& \frac{p_1^2}{4m_1}r_1^2
+ \frac{p_2^2}{4m_2}r_2^2
+ \frac{(\vec{p}_1.\vec{r}_1)^2}{2m_1}
+ \frac{(\vec{p}_2.\vec{r}_2)^2}{2m_2}.
\\[3mm]
\mathbb{H}^{(1,0,2)}
=& -\frac{1}{8}\left(m_1 r_1^4 + m_2 r_2^4\right).
\\[3mm]
\mathbb{H}^{(1,1,0)}
=& \frac{1}{2r_{12}^{D-2}}
\left[
\left(
\frac{\vec{p}_1}{m_1}
-\frac{\vec{p}_2}{m_2}
\right)
.
\left(
\frac{\vec{p}_1}{m_1}
-\frac{\vec{p}_2}{m_2}
\right)
+ \frac{\vec{p}_1.\vec{p}_2}{m_1m_2}
+ (D-2)
\frac{(\hat{n}_{12}.\vec{p}_1)
      (\hat{n}_{12}.\vec{p}_2)}
     {m_1m_2}
\right].
\\[3mm]
\mathbb{H}^{(1,1,1)}
=& -\frac{1}{2r_{12}^{D-2}}
\left[
r_{\mathrm{av}}^2
+ (D-2)\left(\hat{n}_{12}.\vec{r}_{\mathrm{av}}\right)^2
+ \left(\frac{1}{4}-\frac{\mu^2}{D-4}\right)r_{12}^2
\right].
\end{split}\end{equation}
\\~\\
{\bf 2PN Expressions:}
\\~\\
\begin{subequations}\label{eq:2PN Hamiltonian}
\begin{align}
\mathbb{H}^{(2,0,0)}
=& \frac{1}{16}
\left(
\frac{p_1^6}{m_1^5}
+ \frac{p_2^6}{m_2^5}
\right).
\end{align}
\begin{align}
\mathbb{H}^{(2,0,1)}
=& -\left[
\frac{p_1^4}{16m_1^3}r_1^2
+ \frac{p_1^2}{4m_1^3}
  \left(\vec{p}_1.\vec{r}_1\right)^2
+ \frac{p_2^4}{16m_2^3}r_2^2
+ \frac{p_2^2}{4m_2^3}
  \left(\vec{p}_2.\vec{r}_2\right)^2
\right].
\end{align}
\begin{align}
\mathbb{H}^{(2,0,2)}
=& -\frac{p_1^2}{16m_1}r_1^4
+ \frac{r_1^2}{4m_1}
  \left(\vec{p}_1.\vec{r}_1\right)^2
- \frac{p_2^2}{16m_2}r_2^4
+ \frac{r_2^2}{4m_2}
  \left(\vec{p}_2.\vec{r}_2\right)^2.
\end{align}
\begin{align}
\mathbb{H}^{(2,0,3)}
=& \frac{1}{16}\left(m_1 r_1^6 + m_2 r_2^6\right).
\end{align}
\begin{align}
\mathbb{H}^{(2,1,0)}
=& \frac{1}{8r_{12}^{D-2}}
\Bigg[
-3\left(\frac{p_1^4}{m_1^4}+\frac{p_2^4}{m_2^4}\right)
-3\frac{p_1^2p_2^2}{m_1^2m_2^2}
-2\frac{\left(\vec{p}_1.\vec{p}_2\right)^2}{m_1^2m_2^2}\nonumber
\\
&\quad
+4\left(\frac{p_1^2}{m_1^2}+\frac{p_2^2}{m_2^2}\right)
\frac{\vec{p}_1.\vec{p}_2}{m_1m_2}
-D(D-2)\frac{\left(\hat{n}_{12}.\vec{p}_1\right)^2}{m_1^2}
\frac{\left(\hat{n}_{12}.\vec{p}_2\right)^2}{m_2^2}
\nonumber\\
&\quad
+(D-2)\left(
\frac{p_1^2}{m_1^2}
\frac{\left(\hat{n}_{12}.\vec{p}_2\right)^2}{m_2^2}
+\frac{p_2^2}{m_2^2}
\frac{\left(\hat{n}_{12}.\vec{p}_1\right)^2}{m_1^2}
\right)
\nonumber\\
&\quad
-4(D-2)\left(
\frac{p_1^2}{m_1^2}+\frac{p_2^2}{m_2^2}
\right)
\frac{\hat{n}_{12}.\vec{p}_1}{m_1}
\frac{\hat{n}_{12}.\vec{p}_2}{m_2}
+4(D-2)
\frac{\vec{p}_1.\vec{p}_2}{m_1m_2}
\frac{\hat{n}_{12}.\vec{p}_1}{m_1}
\frac{\hat{n}_{12}.\vec{p}_2}{m_2}
\Bigg].
\end{align}
\begin{align}
\mathbb{H}^{(2,1,1)}
=& \frac{1}{4r_{12}^{D-2}}
\Bigg[
r_{\mathrm{av}}^2
\left(
\frac{\vec{p}_1.\vec{p}_2}{m_1m_2}
+ (D-2)
\frac{(\hat{n}_{12}.\vec{p}_1)(\hat{n}_{12}.\vec{p}_2)}
{m_1m_2}
+ \left(
\frac{\vec{p}_1}{m_1}
-\frac{\vec{p}_2}{m_2}
\right)^2
\right)
\nonumber\\
&\quad
+ (D-2)(\hat{n}_{12}.\vec{r}_{\mathrm{av}})^2
\left\{
\frac{\vec{p}_1.\vec{p}_2}{m_1m_2}
+ D\frac{(\hat{n}_{12}.\vec{p}_1)(\hat{n}_{12}.\vec{p}_2)}
{m_1m_2}
+ \left(
\frac{\vec{p}_1}{m_1}
-\frac{\vec{p}_2}{m_2}
\right)^2
\right\}
\nonumber\\
&\quad
+ 2\frac{(\vec{p}_1.\vec{r}_{\mathrm{av}})
(\vec{p}_2.\vec{r}_{\mathrm{av}})}{m_1m_2}
+ 2\left[
\left(
\frac{\vec{p}_1}{m_1}
-\frac{\vec{p}_2}{m_2}
\right)
.\vec{r}_{\mathrm{av}}
\right]^2
\Bigg]
\nonumber\\
&\quad
+ \frac{1}{2\,r_{12}^{D-2}}
\Bigg[
\frac{(\vec{p}_1.\vec{r}_{\mathrm{av}})
(\vec{p}_1.\vec{r}_{12})}{m_1^2}
-\frac{(\vec{p}_2.\vec{r}_{\mathrm{av}})
(\vec{p}_2.\vec{r}_{12})}{m_2^2}
+ (D-4)
\frac{
(\vec{p}_1.\vec{r}_{\mathrm{av}})
(\vec{p}_2.\vec{r}_{12})
-
(\vec{p}_2.\vec{r}_{\mathrm{av}})
(\vec{p}_1.\vec{r}_{12})
}{2\,m_1m_2}
\Bigg]
\nonumber\\
&\quad
- \frac{1}{32r_{12}^{D-2}}
\Bigg[
\left(
-\frac{3}{2}+\frac{2\mu^2}{D-4}
\right)r_{12}^2
\left(
\frac{\vec{p}_1}{m_1}
+\frac{\vec{p}_2}{m_2}
\right)^2
+ \left(
-\frac{1}{2}+\frac{6\mu^2}{D-4}
\right)r_{12}^2
\left(
\frac{\vec{p}_1}{m_1}
-\frac{\vec{p}_2}{m_2}
\right)^2
\nonumber\\
&\quad
+ \left(
-\frac{3D}{2}+2(2+\mu^2)
\right)
\left\{
\left(
\frac{\vec{p}_1}{m_1}
+\frac{\vec{p}_2}{m_2}
\right)
.\vec{r}_{12}
\right\}^2
+ \left(
\frac{3D}{2}-2(4+\mu^2)
\right)
\left\{
\left(
\frac{\vec{p}_1}{m_1}
-\frac{\vec{p}_2}{m_2}
\right)
.\vec{r}_{12}
\right\}^2
\Bigg].
\end{align}
\begin{align}
\mathbb{H}^{(2,1,2)}
=& \frac{1}{8r_{12}^{D-2}}
\left[
r_{\mathrm{av}}^4
- D(D-2)\left(\hat{n}_{12}.\vec{r}_{\mathrm{av}}\right)^4
+ 2(D-2)r_{\mathrm{av}}^2
\left(\hat{n}_{12}.\vec{r}_{\mathrm{av}}\right)^2
\right]
\nonumber\\
&\quad
+ \frac{1}{8r_{12}^{D-2}}
\left[
\frac{1}{2}
\left(\hat{n}_{12}.\vec{r}_{\mathrm{av}}\right)^2
r_{12}^2
\left(D+4\mu^2\right)
+ r_{\mathrm{av}}^2 r_{12}^2
\left(
\frac{1}{2}+\frac{2\mu^2}{D-4}
\right)
\right]
\nonumber\\
&\quad
+ \frac{1}{8r_{12}^{D-2}}
\left[
\frac{5}{16}
+\frac{\mu^2-1}{2(D-6)}
-\frac{\mu^2(\mu^2-1)}{(D-4)(D-6)}
\right]r_{12}^4.
\end{align}
\begin{align}
\mathbb{H}^{(2,2,0)}
=& \frac{1}{8r_{12}^{2(D-2)}}
\Bigg\{
\frac{1}{m_1}
\Bigg[
2\frac{D-3}{D-4}\frac{p_2^2}{m_2^2}
+ 4\frac{p_1^2}{m_1^2}
+ \frac{D^2-11D+26}{D-4}
\frac{\vec{p}_1.\vec{p}_2}{m_1m_2}\nonumber
\\
&\quad\quad
- 2\frac{D-5}{D-4}(D-2)^2
\frac{\hat{n}_{12}.\vec{p}_2}{m_2}
\hat{n}_{12}.
\left(
\frac{\vec{p}_1}{m_1}
-\frac{\vec{p}_2}{m_2}
\right)
\Bigg]
\nonumber\\
&\quad
+ \frac{1}{m_2}
\Bigg[
2\frac{D-3}{D-4}\frac{p_1^2}{m_1^2}
+ 4\frac{p_2^2}{m_2^2}
+ \frac{D^2-11D+26}{D-4}
\frac{\vec{p}_1.\vec{p}_2}{m_1m_2}\nonumber
\\
&\quad\quad
+ 2\frac{D-5}{D-4}(D-2)^2
\frac{\hat{n}_{12}.\vec{p}_1}{m_1}
\hat{n}_{12}.
\left(
\frac{\vec{p}_1}{m_1}
-\frac{\vec{p}_2}{m_2}
\right)
\Bigg]
\Bigg\}.
\end{align}
\begin{align}
\mathbb{H}^{(2,2,1)}
=& -\frac{1}{8r_{12}^{2(D-2)}}
\frac{D-5}{D-4}(D-2)
\left[
\left(\frac{1}{m_1}-\frac{1}{m_2}\right)
\vec{r}_{12}.\vec{r}_{\mathrm{av}}
-\frac{1}{2}
\left(\frac{1}{m_1}+\frac{1}{m_2}\right)r_{12}^2
\right].
\end{align}
\begin{align}
\mathbb{H}^{(2,3,0)}
=& \frac{1}{r_{12}^{3(D-2)}}
\frac{(D-5)(D-2)^2}{8m_1m_2(D-4)}.
\end{align}
\end{subequations}

\pagebreak
\subsection{Node of Inertia}
In this subsection, we give the node of inertia of the two-particle system through 2PN order, as described in \S\ref{sec:Hamiltonian and conformal algebra}. We organise its contributions according to their PN order, interaction order, and curvature dependence:
\begin{equation}
\vec{\mathbb{G}}
=\sum_{\mathit{n}_1,\mathit{n}_2,\mathit{n}_3}
\left(\frac{1}{c^2}\right)^{\mathit{n}_1}
\left(\gsc m_1m_2\right)^{\mathit{n}_2}
\left(\omA^2\right)^{\mathit{n}_3}
\vec{\mathbb{G}}^{\,(\mathit{n}_1,\mathit{n}_2,\mathit{n}_3)}.
\end{equation}
Here $\mathit{n}_1$ labels the total PN order, with $\mathit{n}_1=0,1,2$ corresponding to Newtonian, 1PN, and 2PN order, respectively. The index $\mathit{n}_2$ counts powers of the interaction factor $\gsc m_1m_2$, while $\mathit{n}_3$ counts powers of $\omA^2$. Thus, $\mathit{n}_2=0$ identifies the free-particle contributions, and $\mathit{n}_3=0$ identifies the terms that survive the flat-space limit. The coefficients $\vec{\mathbb{G}}^{\,(\mathit{n}_1,\mathit{n}_2,\mathit{n}_3)}$ contain the remaining dependence on the particle masses and phase-space variables, with the displayed expansion factors extracted.
\\~\\
{\bf 0PN Expression:}
\\~\\
\begin{align}\label{eq:0PN boost generator}
\vec{\mathbb{G}}^{\,(0,0,0)}
=& m_1 \vec{r}_1 + m_2 \vec{r}_2.
\end{align}
\\~\\
{\bf 1PN Expressions:}
\\~\\
\begin{equation}\begin{split}\label{eq:1PN boost generator}
\vec{\mathbb{G}}^{\,(1,0,0)}
= \frac{p_1^2}{2m_1}\,\vec{r}_1
+ \frac{p_2^2}{2m_2}\,\vec{r}_2.
\qquad
\vec{\mathbb{G}}^{\,(1,1,0)}
= -\frac{\vec{r}_{\mathrm{av}}}{r_{12}^{D-2}}.
\end{split}\end{equation}
\pagebreak
\\~\\
{\bf 2PN Expressions:}
\\~\\
\begin{equation}\begin{split}\label{eq:2PN boost generator}
\vec{\mathbb{G}}^{\,(2,0,0)}
=& -\frac{1}{8}
\left(
\frac{p_1^4}{m_1^3}\,\vec{r}_1
+ \frac{p_2^4}{m_2^3}\,\vec{r}_2
\right).
\\[3mm]
\vec{\mathbb{G}}^{\,(2,0,1)}
=& \frac{1}{2}
\left[
\frac{(\vec{p}_1.\vec{r}_1)^2}{m_1}\,\vec{r}_1
+ \frac{(\vec{p}_2.\vec{r}_2)^2}{m_2}\,\vec{r}_2
\right].
\\[3mm]
\vec{\mathbb{G}}^{\,(2,1,0)}
=& \frac{1}{4r_{12}^{D-2}}
\left[
\frac{\vec{r}_{12}.\vec{p}_1}{m_1m_2}\,\vec{p}_2
- \frac{\vec{r}_{12}.\vec{p}_2}{m_1m_2}\,\vec{p}_1
\right]
+ \frac{\vec{r}_{\mathrm{av}}}{2r_{12}^{D-2}}
\left[
\left(
\frac{\vec{p}_1}{m_1}
-\frac{\vec{p}_2}{m_2}
\right)^2
+ \frac{\vec{p}_1.\vec{p}_2}{m_1m_2}
+ (D-2)
\frac{(\hat{n}_{12}.\vec{p}_1)(\hat{n}_{12}.\vec{p}_2)}
{m_1m_2}
\right].
\\[3mm]
\vec{\mathbb{G}}^{\,(2,1,1)}
=& -\frac{1}{2r_{12}^{D-2}}
\left[
(D-2)\left(\hat{n}_{12}.\vec{r}_{\mathrm{av}}\right)^2
-\frac{\mu^2}{D-4}r_{12}^2
\right]\vec{r}_{\mathrm{av}}.
\\[3mm]
\vec{\mathbb{G}}^{\,(2,2,0)}
=& -
\left(\frac{D-5}{D-4}\right)\left(\frac{D-2}{8}\right)
\left(\frac{1}{m_1}-\frac{1}{m_2}\right)
\frac{\vec{r}_{12}}{r_{12}^{2(D-2)}}.
\end{split}\end{equation}

\pagebreak

\bibliography{references}
\bibliographystyle{jhepbibstyle}
\end{document}